\pdfoutput=1
\RequirePackage{ifpdf}
\ifpdf 
\documentclass[pdftex]{sigma}
\else
\documentclass{sigma}
\fi

\numberwithin{equation}{section}

\newtheorem*{Theorem*}{Theorem}

\theoremstyle{definition}

\usepackage{booktabs,integrable,stmaryrd,xparse}

\begin{document}

\allowdisplaybreaks

\newcommand{\arXivNumber}{2509.20643}

\renewcommand{\PaperNumber}{044}

\FirstPageHeading

\ShortArticleName{Non-Commutative Gauge Theory at the Beach}

\ArticleName{Non-Commutative Gauge Theory at the Beach}

\Author{Roland BITTLESTON~$^{\rm a}$, Simon HEUVELINE~$^{\rm bc}$, Surya RAGHAVENDRAN~$^{\rm d}$ \newline and David SKINNER~$^{\rm b}$}

\AuthorNameForHeading{R.~Bittleston, S.~Heuveline, S.~Raghavendran and D.~Skinner}

\Address{$^{\rm a)}$~Perimeter Institute for Theoretical Physics, 31 Caroline Street, Waterloo, Ontario, Canada}
\EmailD{\mail{rbittleston@perimeterinstitute.ca}}

\Address{$^{\rm b)}$~Department of Applied Maths \& Theoretical Physics, University of Cambridge,\\
\hphantom{$^{\rm b)}$}~Wilberforce Road, UK}
\EmailD{\mail{simonheuveline@fas.harvard.edu}, \mail{d.b.skinner@damtp.cam.ac.uk}}

\Address{$^{\rm c)}$~Center for the Fundamental Laws of Nature \& Black Hole Initiative,\\
\hphantom{$^{\rm c)}$}~Harvard University, Cambridge, USA}

\Address{$^{\rm d)}$~Department of Mathematics, Yale University, 219 Prospect St, New Haven, USA}
\EmailD{\mail{surya.raghavendran@yale.edu}}

\ArticleDates{Received October 16, 2025, in final form April 23, 2026; Published online May 05, 2026}

\Abstract{The KP equation is perhaps the most famous example of a three-dimensional integrable system. Here we show that a non-commutative five-dimensional Chern--Simons theory living on the projective spinor bundle of three-dimensional space-time compactifies to a Lagrangian formulation of the KP equation. Essential to the definition of the theory is a 2-form pulled back from minitwistor space. The dispersionless limit of the KP equation is similarly described by Poisson--Chern--Simons theory. We further show that, consistent with integrability, all tree level amplitudes vanish. The universal vertex algebra living on a two-dimensional surface defect in $5d$ is $W_{1+\infty}$, and its operator products coincide with collinear splitting functions on space-time. Taking the dispersionless limit contracts the vertex algebra to $w_{1+\infty}$.}


\Classification{17B80; 81R25; 81R60}

\section{Introduction}

The Kadomtsev--Petviashvili (KP) equation
\begin{equation} \label{eq:KP} (u_t + 6uu_x + u_{xxx})_x + 3\sigma^2u_{yy} = 0 \end{equation}
for $\sigma^2=\pm1$ is perhaps the most famous and well-studied integrable system in $2+1$ dimensions~\mbox{\cite{Ablowitz:1979evo, Kadomtsev:1970kp}}. It arises generically in physical contexts with both signs for $\sigma^2$, for example, in the propagation of shallow water waves and also plasma physics. Depending on the sign of $\sigma^2$, it admits line and lump solitons.\footnote{For $\sigma^2 = 1$, line solitons are stable but no lump solitons exist. For $\sigma^2=-1$, line solitons are unstable but lump solitons do exist \cite{Ablowitz:1991xb}.}

There exist a number of different approaches for solving the KP equation and a vast literature on the subject. We will not attempt to survey it here; nevertheless, we include here a few references for context. The associated linear problem, or Lax pair, first appeared in \cite{Dryuma:1974ast} and is the foundation of the inverse scattering transform \cite{Ablowitz:1991xb, Manakov:1981inv}. A more algebraic approach was developed by the Japanese school: it identifies the `$\tau$-function', a kind of second potential for the dynamical field $u = 2(\log\tau)_{xx}$, with a $2d$ chiral CFT correlator \cite{Jimbo:1981tov,Jimbo:1981mpt,Jimbo:1981mpf, Sato:1983se}. This $\tau$-function is the central object in Hirota's direct method \cite{Hirota:1986uc} and surprisingly can also be encountered as the partition function of the topological string \cite{Aganagic:2003qj, Dijkgraaf:1990rs,Fukuma:1990jw}.

In this work, we will understand the integrability of the KP equation from a rather different perspective: through the \emph{minitwistor correspondence}. The dispersionless KP (dKP) equation, obtained by discarding the term fourth order in derivatives from equation \eqref{eq:KP} to get
\begin{displaymath} (u_t + 6uu_x)_x + 3\sigma^2u_{yy} = 0 , \end{displaymath}
has long been known to have a minitwistor formulation. Indeed, solutions to the dKP equation yield special Einstein--Weyl spaces in $2+1$ dimensions and the Hitchin correspondence identifies these geometries with certain complex surfaces \cite{Hitchin:1982vry}. The requirement that an Einstein--Weyl space can be obtained from a solution to the dKP equation endows the associated complex surface with a holomorphic section of the dual to the fourth root of the canonical bundle,~$K^{-1/4}$, and therefore a holomorphic Poisson structure. The complex surface associated to an~Einstein--Weyl geometry is known as its \emph{minitwistor space}.

The full KP equation has proven considerably more challenging to realise in the minitwistor language. We will follow a proposal of Strachan \cite{Strachan:1994kp,Strachan:1996gx}: that the KP equation arises from a non-commutative deformation of minitwistor space quantizing the holomorphic Poisson structure. Our perspective differs in a few crucial ways from that of Strachan. The most important of these is that we will work at the level of the action; in particular we will show that a certain~$5d$ Abelian non-commutative Chern--Simons theory defined on the projective spinor bundle $\PS$ of~$\bbR^{1,2}$ descends to a $3d$ theory whose equations of motion imply the KP equation. This $5d$ Chern--Simons description of a $3d$ integrable model is a cousin of the uplift to $\PS$ \cite{Bittleston:2020hfv} of the Manakov--Zakharov--Ward model \cite{Manakov:1981kz,Ward:1988ie}. There and here the minitwistor correspondence is essential in defining the theory: $\PS$ is the correspondence space sitting above both $\bbR^{1,2}$ and its minitwistor space $\mT$. This allows holomorphic data, e.g., a holomorphic section of $K^{-1/4}$, to be pulled back to correspondence space. We will use exactly such data to define our non-commutative $5d$ Chern--Simons theory. That $\PS$ is the correct place to write actions for $3d$ integrable systems was first seen in~\cite{Adamo:2017xaf} in the context of a mixed topological-holomorphic $BF$ theory description of the Bogomol'nyi equations. The Chern--Simons theory studied here is potentially more interesting, as the Lax equation is not imposed through the equation of motion of an auxiliary Lagrange multiplier field.

While we concentrate on the KP equation in this paper, in our discussion we observe that a~wide variety of $3d$ integrable models, including novel higher genus examples \cite{Jarov:2025hgt}, are expected to be recoverable from this paradigm. (We note that there are alternative proposals for higher dimensional generalisations of $4d$ Chern--Simons, in particular the $5d$ 2-Chern--Simons of \cite{Chen:2024axr, Schenkel:2024dcd}.)

In \cite{Bittleston:2020hfv}, two of the authors argued that $4d$ Chern--Simons descriptions of $2d$ integrable models~\cite{Costello:2019tri,Delduc:2019whp,Vicedo:2019dej} could be understood as symmetry reductions of twistorial holomorphic Chern--Simons descriptions of $4d$ integrable models \cite{Costello:2021bah, Witten:2003nn}. This linked the $4d$ Chern--Simons programme to a~conjecture of Ward that all integrable partial differential equations arise as symmetry reductions of the self-dual Yang--Mills equations \cite{Mason:1991rf, Ward:1985gz}. Our $5d$ non-commutative Chern--Simons theory on correspondence does not fit directly into this framework, since it is not a~symmetry reduction of holomorphic Chern--Simons theory. Of course, it is natural to expect it should instead reduce from a $6d$ non-commutative Chern--Simons theory \cite{Bu:2022iak, Strachan:1992em,Strachan:1996gx}. A~variant of the Ward conjecture, that all integrable models should arise as symmetry reductions of twistorial theories, therefore seems plausible.

On the other hand, the KdV equation certainly arises as a symmetry reduction of the KP equation. It may therefore seem natural to seek a non-commutative deformation of $4d$ Chern--Simons theory. This is misguided: in the domain of $4d$ Chern--Simons, any bivector would have a component pointing in either a topological or anti-holomorphic direction, and translations in these directions are trivial in BRST cohomology. Vertices involving a~derivative along one of these directions can always be eliminated by a field redefinition, potentially at the cost of introducing higher-order vertices. The correspondence space of $\bbR^{1,2}$ evades this since it has two holomorphic directions. Nevertheless, it would be very interesting to understand how to recover a $4d$ Chern--Simons description of KdV by symmetry reduction.

Let us briefly summarise the content of the paper:
\begin{itemize}\itemsep=0pt
 \item Essential background material can be found in Section~\ref{sec:background}.
 Here we review how solutions to the dKP equation generate $(2+1)$-dimensional Einstein--Weyl spaces; furthermore, the Hitchin correspondence for $\bbR^{1,2}$ and more general Einstein--Weyl geometries is elucidated. We also identify an action for the potential $\phi = \p_x^{-1}u$ whose equation of motion implies the KP equation on $u$.
 \item In Section~\ref{sec:5dncCS}, we introduce mixed topological-holomorphic $5d$ Poisson--Chern--Simons theory on $\bbC^2\times\bbR$ and its non-commutative deformation. The extension of this theory to a~more general 5-manifold $X$ is described.
 \item Specialising $X$ to the projective spinor bundle of $(2+1)$-dimensional space-time, in Section~\ref{sec:PS} we define $5d$ Poisson--Chern--Simons and non-commutative Abelian Chern--Simons on correspondence space. This requires a careful analysis of the boundary conditions and (Moyal) quantizing the differential graded algebra of poly-forms.
 \item In Section~\ref{sec:off-shell-reduction}, we show how to recover Lax formulations of the dKP and KP equations by gauge fixing on correspondence space. We compactify non-commutative Chern--Simons on~$\PS$, recovering the action from Section~\ref{sec:background}.
 \item We switch gears in Section~\ref{sec:amplitudes}, and treat KP theory as a perturbative quantum field theory. Consistent with integrability, we verify that all tree level amplitudes vanish.
 \item By implementing Koszul duality on correspondence space, we show in Section~\ref{sec:3d/2d-correspondence} that \mbox{holomorphic} surface defects wrapping minitwistor lines in Poisson- and non-commuta\-tive~Chern--Simons support $w_{1+\infty}$ and $W_{1+\infty}$ vertex algebras respectively. These come equipped with a non-trivial vacuum module. We verify that this same vertex algebra structure can be recovered from the collinear splitting of form factors in the corresponding space-time theories.
\end{itemize}
We conclude in Section~\ref{sec:discussion} by discussing possible extensions of this work.

\section{Background} \label{sec:background}

In this section, we introduce a Lagrangian formulation of the dKP and KP equations, and recall the minitwistor correspondence.

\subsection{An action for the KP equation}

We begin by introducing new co-ordinates on space-time which are better suited to the correspondence space formulation. Compared to~\eqref{eq:KP}, we write
\begin{displaymath}
x^+ = x ,\qquad x^- = -12\sigma^2t
\end{displaymath}
and rescale $y\mapsto y/\sqrt{2}$, $u\mapsto - 2\sigma^2 u$. This yields
\begin{equation} \label{eq:rescaled-KP}
\biggl(u_- + uu_+ - \frac{1}{12\sigma^2} u_{+++}\biggr)_+ - \frac{1}{2}u_{yy} = 0 ,
\end{equation}
where $u_\pm=\partial_\pm u$. The dispersionless limit is obtained by relaxing the condition $\sigma^2=\pm1$ and sending $\sigma^2\to\infty$
\begin{displaymath}
(u_- + uu_+)_+ - \frac{1}{2}u_{yy} = 0 .
\end{displaymath}
We emphasise that relaxing the condition $\sigma^2 = \pm1$ by allowing any finite $\sigma^2\neq0$ does not meaningfully change the KP equation, since it can be implemented by rescaling the field and co-ordinates.

Our choice of notation here is motivated by the fact that, from the correspondence space perspective, the variables $x^+$, $x^-$ are interpreted as null co-ordinates. This is visible in the rescaled dKP equation above: the linear terms in $u$ take the form of a wave equation in $2+1$ space-time dimensions with metric $2\dif x^+\dif x^- - \dif y^2$. Of course, in physical contexts we must remember that $t = - x^-/12\sigma^2$ is identified with time.

The KP equation \eqref{eq:KP} itself is not the equation of motion of any Lagrangian. However, consider the action
\begin{equation} \label{eq:KPaction}
S_{\rm KP}[\phi] = \int_{\bbR^{1,2}}\dif^3x \biggl(\phi_+\phi_- - \frac{1}{2}\phi_y^2 + \frac{1}{12\sigma^2}\phi_{++}^2 + \frac{1}{3}\phi_+^3\biggr)
\end{equation}
for $\dif^3x = \dif x^-\dif x^+\dif y$. The equations of motion obtained by extremising $S_{\rm KP}[\phi]$ are
\begin{displaymath}
\biggl(\phi_- + \frac{1}{2}\phi_+^2 - \frac{1}{12\sigma^2}\phi_{+++}\biggr)_+ = \frac{1}{2}\phi_{yy} .
\end{displaymath}
Taking its $x^+$ derivative, this equation implies
\begin{displaymath}
\biggl(\phi_{+-} + \phi_+\phi_{++} - \frac{1}{12\sigma^2}\phi_{++++}\biggr)_+ = \frac{1}{2}\phi_{yy+} ,
\end{displaymath}
which we recognise as the KP equation~\eqref{eq:rescaled-KP} upon setting $u = \phi_+$. Thus $\phi$ is a form of first potential, sitting between $u$ and the $\tau$-function, $u = \phi_+ = -\sigma^{-2}(\log\tau)_{++}$. Writing $[\cO]$ for the mass dimension of $\cO$ under simultaneous scaling of the space and time co-ordinates, the action transforms homogeneously with $[S_{\rm KP}] = -3$ if we take $[\sigma] = 1$ and $[\phi]=-1$.\footnote{The action \eqref{eq:KPaction} has a symmetry under weighted dilations \smash{$\bigl(x^+,x^-,y\bigr)\mapsto\bigl(s^{1/2}x^+,s^{3/2}x^-,sy^2\bigr)$} whilst simultaneously scaling $\phi\mapsto s^{-1/2}\phi$ for $s\in\bbR^+$, which is perhaps a~more natural counterpart of dilations. We thank a~referee for highlighting this symmetry to us.}

A Lagrangian for the dispersionless KP equation is obtained in the limit $\sigma^2\to\infty$
\begin{equation} \label{eq:dKPaction}
S_{\rm dKP}[\phi] = \int_{\bbR^{1,2}}\dif^3x \biggl(\phi_+\phi_- - \frac{1}{2}\phi_y^2 + \frac{1}{3}\phi_+^3\biggr) .
\end{equation}
In Section~\ref{sec:off-shell-reduction}, we will see how to recover the actions \eqref{eq:KPaction} and \eqref{eq:dKPaction} from correspondence space.

\subsection[Minitwistor correspondence for {C\^{}3}]{Minitwistor correspondence for $\boldsymbol{\bbC^3}$} \label{subsec:complex-minitwistors}

Next, we briefly review the minitwistor correspondence, originally due to Hitchin \cite{Hitchin:1982vry} (see also \cite{Jones:1984phd,Jones:1985pla}). It identifies certain three-dimensional complex geometries with two-dimensional complex surfaces admitting families of rational curves of self-intersection number $2$ (or equivalently having normal bundle isomorphic to $\cO(2)$). The holomorphic metric of the complex three manifold (or more precisely its Weyl structure) is encoded entirely in the complex structure of the surface.

In the first instance, we will be interested in the case of $\bbC^3$ equipped with the standard complex metric. We fix complex co-ordinates $\bigl(x^+,x^-,y\bigr)$ in terms of which the complex metric takes the form $\eta = 2\dif x^+\dif x^- - \dif y^2$.

It will also be useful to introduce a spinor notation: We use letters from the beginning of the Greek alphabet as spinor indices; they take values in the set $\{0,1\}$. Using the isomorphism of ${\rm Spin}_3(\bbC)$ representations $S^2\bbS = \bbC^3$ \big(where $\bbS\cong\bbC^2$ is the complex two-dimensional spin representation\big), we can index points of complexified $3d$ space-time by a symmetric pair of spinor indices. In particular, we write \smash{$x^{\al\beta} = x^{(\al\beta)}$} for points in $\bbC^3$, where $(\alpha_1\dots\alpha_n)$ indicates averaging over permutations.

To pass back to the co-ordinates $\bigl(x^+,x^-,y\bigr)$, we first fix a pair of spinors $o^\al,\iota^\beta\in\bbS$ obeying \smash{$\eps_{\al\beta}o^\al\iota^\beta = \la o\iota\ra = 1$}, known as a \emph{dyad}. Here
\begin{displaymath} \eps_{\al\beta} = \begin{pmatrix} \hphantom{-}0 & 1 \\ -1 & 0 \end{pmatrix}_{\al\beta} \end{displaymath}
is the antisymmetric matrix determining the ${\rm Spin}_3(\bbC)$ invariant antisymmetric pairing $\la ab\ra = \eps_{\al\beta}a^\al b^\beta$ for $a^\al,b^\beta\in\bbS$. Spinor indices are raised and lowered using $\eps_{\al\beta}$ and its inverse $\eps^{\al\beta}$ obeying \smash{$\eps_{\al\beta}\eps^{\beta\gamma} = \delta_\al^{~\,\gamma}$}. We can then identify
\begin{displaymath} x^+ = x^{\al\beta}o_\al o_\beta ,\qquad y = \sqrt{2}x^{\al\beta}o_\al\iota_\beta ,\qquad x^- = x^{\al\beta}\iota_\al\iota_\beta . \end{displaymath}
The standard complex metric is
\begin{displaymath} \eta = \eps_{\al\gamma}\eps_{\beta\delta}\dif x^{\al\beta} \dif x^{\gamma\delta} = 2\dif x^+\dif x^- - \dif y^2 . \end{displaymath}

The minitwistor correspondence assigns the $\bigl(\bbC^3,\eta\bigr)$ the complex surface $T^{1,0}\CP^1$, or equivalently the total space of the holomorphic line bundle
\begin{displaymath} \cO(2)\to\CP^1 . \end{displaymath}
Often we will refer to this simply as \emph{minitwistor space} and denote it by $\mT$. Points in $\bbC^3$ are identified with holomorphic sections of $\mT$. Using inhomogeneous co-ordinates $(v,z)$ on the fibre and base respectively, these sections are quadratics
\begin{equation} \label{eq:inhom-inc} v = x^+ - \sqrt{2}yz + x^- z^2 . \end{equation}
This identity is the \emph{incidence relation}.

Alternatively, we can employ the homogeneous co-ordinates $(\mu,\lambda_\al)\in\bbC\times\bbC^2\setminus\{(0,0)\}$ on $\mT$, defined up to complex rescaling $(\mu,\lambda_\al)\sim\bigl(s^2\mu,s\lambda_\al\bigr)$ for $s\in\bbC^\times$. In homogeneous co-ordinates, the incidence relation is
\begin{displaymath} \mu = x^{\al\beta}\lambda_\al\lambda_\beta . \end{displaymath}
Homogeneous and inhomogeneous co-ordinates are related by $(\mu,\lambda_\al)\sim(v,o_\al + z\iota_\al)$. Throughout this work, we will often express differential forms and vector fields in terms of homogeneous co-ordinates. These methods are described in detail by Woodhouse \cite{Woodhouse:1985id}.

Points in $\bbC^3$ are null-separated if the corresponding quadratics intersect at a single point. For example, the point $x\in\bbC^3$ is on the complexified null-cone of the origin if the discriminant vanishes
\begin{displaymath} \eps_{\al\gamma}\eps_{\beta\delta}x^{\al\beta}x^{\gamma\delta} = 2x^+x^- - y^2 = 0 . \end{displaymath}
This information is sufficient to recover the conformal structure of $\bbC^3$; however, Hitchin shows that the complex structure of $\mT$ also encodes the scale of $\bbC^3$. This is because, given a tangent direction at a point in $\bbC^3$, the complex structure of $\mT$ can be used to construct a complex curve in that direction which is interpreted as a geodesic. This endows $\bbC^3$ with a projective connection which is compatible with the conformal structure. Within the projective class, there is a distinguished affine connection, which in this case is the Levi-Civita connection of the flat~metric.

We can also ask what points in $\mT$ correspond to in $\bbC^3$. We can see from the incidence relation \eqref{eq:inhom-inc} that $(v,z)$ determines a complex plane $\Pi$ in $\bbC^3$. On such a plane
\begin{displaymath} \dif x^+ = \sqrt{2}z\dif y - z^2\dif x^- \end{displaymath}
so that
\begin{displaymath} \iota^*_\Pi\eta = - \dif y^2 + 2\sqrt{2}z\dif y\dif x^- - 2z^2(\dif x^-)^2 = - \bigl(\dif y - \sqrt{2}z\dif x^-\bigr)^2 . \end{displaymath}
The complex metric $\eta$ therefore restricts to a degenerate metric, and so $\Pi$ is a null plane. (Note that a null plane is not totally null: the metric is degenerate but non-vanishing on $\Pi$.) In this way, we see that points in $\mT$ are null planes in $\bbC^3$.

The minitwistor space $\mT$ admits a natural action of $\gPSL_2(\bbC)$ acting by M\"{o}bius transformations on the base and lifting in the obvious way to the fibres. This is identified with $\gSO_3(\bbC)$, the Lorentz group of~\smash{$\bigl(\bbC^3,\eta\bigr)$}. Viewing $\mT$ as an affine bundle, $\gSO_3(\bbC)$ is extended by translations~$\bbC^3$ to the complex Poincar\'{e} group.

We can introduce an intermediate complex four-manifold $\cF = \bigl\{[(\mu,\lambda_\al)]\in\mT ,\, x^{\al\beta}\in\bbC^3 \mid \mu = x^{\al\beta}\lambda_\al\lambda_\beta\bigr\}$ consisting of points in $\mT$ and $\bbC^3$ such that the former lies on the quadratic corresponding to the latter. It is often convenient to view $\cF = \bbC^3\times\CP^1$, where the fibre co-ordinate on $\mT$ is determined by the incidence relation. From now on, we refer to $\cF$ as \emph{complex correspondence space}. By definition, $\cF$ admits a double fibration
\[
 \begin{tikzpicture}
 	\node (F) at (0,1) {$\cF$};
 	\node (mT) at (-1,0) {$\mT$};
 	\node (C3) at (1,0) {$\bbC^3$.};
 	\draw[->] (F) edge node[above right=0.25pt]{$\rho$} (C3);
		\draw[->] (F) edge node[above left=0.25pt]{$\pi$} (mT);
 \end{tikzpicture}
\]

\subsection[Minitwistor correspondence for R\^{}\{1,2\}]{Minitwistor correspondence for $\boldsymbol{\bbR^{1,2}}$} \label{subsec:Lorentzian-minitwistors}

Thus far our discussion has concerned three-dimensional complex manifolds, but we would like to restrict to a real slice. In order to do so, we pick the anti-holomorphic involution
\begin{displaymath}
C\colon\ \bbC^3\to\bbC^3 ,\qquad\bigl(x^+,x^-,y\bigr)\mapsto \bigl(\bar x^+,\bar x^-,\bar y\bigr)
\end{displaymath}
whose fixed-point locus is $\bbR^{1,2}$. We would like to distinguish those sections of $\mT$ corresponding to points in $\bbR^{1,2}$. This is achieved by choosing the anti-holomorphic involution
\begin{displaymath}
R\colon \ \mT\to\mT ,\qquad (v,z)\mapsto (\bar v,\bar z) .
\end{displaymath}
In terms of homogeneous co-ordinates, the involution acts by $R\colon (\mu,\lambda_\al)\mapsto\bigl(\bar\mu,\bar\lambda_\al\bigr)$, where $\bar\lambda_\al$ denotes component-wise conjugation. (Here we are taking the dyad $\{o,\iota\}$ to be real.) This involution picks out a subspace $\bigl\{\lambda\sim\bar\lambda\bigr\} = S^1\subset\CP^1$ on the base of $\mT$ fixed pointwise by $R$. Furthermore, the fixed points of $\mT$ form the trivial real line bundle over this $S^1$.

It is clear that those sections which are fixed under $R$ correspond to points in $\bbR^{1,2}$. (Note that we do not require the section to be fixed pointwise; indeed, this would exclude all sections.) The subgroup of $\gPSL_2(\bbC)$ acting by M\"{o}bius transformations commuting with the involution $R$ is $\gPSL_2(\bbR)$ which we recognise as the Lorentz group $\gSO^+(1,2)$. Translations compatible with $R$ are of course real, so the real Poincar\'{e} group of $\bbR^{1,2}$ survives.

We have seen that a point $[(\mu,\lambda)]\in\mT$ describes a complex null plane $\Pi\subset\bbC^3$. The intersection of this plane with the real slice $\bbR^{1,2}$ can be of three types:
\begin{itemize}\itemsep=0pt
 \item If $\lambda\not\sim\bar\lambda$, then $\Pi\cap\bbR^{1,2}$ is the timelike geodesic
 \begin{displaymath} x^{\al\beta} =x_0^{\al\beta} + sn^{\al\beta} = \frac{\mu\bar\lambda^\al\bar\lambda^\beta + \bar\mu\lambda^\al\lambda^\beta}{\big\la\lambda\bar\lambda\big\ra^2} + \sqrt{2}\im s\frac{\lambda^{(\al}\bar\lambda^{\beta)}}{\big\la\lambda\bar\lambda\big\ra} \end{displaymath}
 for $s\in\bbR$. The direction of the geodesic is determined by $n^{\al\beta} = \sqrt{2}\im\lambda^{(\al}\bar\lambda^{\beta)}/\bigl\la\lambda\bar\lambda\bigr\ra\in\bbR^{1,2}$ with $n^2 = 1$. $n$ is future pointing for $\operatorname{Im}\bigl\la\lambda\bar\lambda\bigr\ra>0$, $\operatorname{Im} z<0$, and past pointing for $\operatorname{Im}\bigl\la\lambda\bar\lambda\bigr\ra<0$, $\operatorname{Im}z>0$. Conversely, all timelike geodesics arise this way.
 \item If $(\mu,\lambda)\sim\bigl(\bar\mu,\bar\lambda\bigr)$, then $\Pi\cap\bbR^{1,2}$ is the real null plane
 \begin{displaymath} x^{\al\beta} = x_0^{\al\beta} + x^{(\al}\lambda^{\beta)} = \mu\frac{\nu^\al\nu^\beta}{\la\lambda\nu\ra^2} + \xi^{(\al}\lambda^{\beta)} . \end{displaymath}
 Here $\nu^\al\sim\bar\nu^\al\in S^1\subset\CP^1$ is a real spinor such that $\la\lambda\nu\ra\neq0$ (certainly such a spinor exists) and $\xi^\al = \bar\xi^\al\in\bbR^2$. This is the real null plane passing through $x_0$ which is orthogonal to the light ray in the direction $\lambda^\al\lambda^\beta$. Again, all real null planes arise in this way.
 \item If $\lambda\sim\bar\lambda$ but $(\mu,\lambda)\not\sim\bigl(\bar\mu,\bar\lambda\bigr)$, then $\Pi\cap\bbR^{1,2}$ is empty.
\end{itemize}

We can also define a real subspace of complex correspondence space $\PS = \bigl\{{[(\mu,\lambda_\al)]\in\mT },\allowbreak x^{\al\beta}\in\bbR^{1,2}\mid \mu = x^{\al\beta}\lambda_\al\lambda_\beta\bigr\}\subset\cF$ which coincides with the projective spinor bundle of $\bbR^{1,2}$. We will refer to $\PS$ simply as \emph{correspondence space}. It admits maps
\[
 \begin{tikzpicture}
 	\node (PS) at (0,1) {$\PS$};
 	\node (mT) at (-1,0) {$\mT$};
 	\node (R12) at (1,0) {$\bbR^{1,2}$.};
 	\draw[->] (PS) edge node[above right=0.25pt] {$\rho$} (R12);
		\draw[->] (PS) edge node[above left=0.25pt] {$\pi$} (mT);
 \end{tikzpicture}
\]
The preimage of a point in $\bbR^{1,2}$ is a $\CP^1$; however, the preimage of a point in $\mT$ differs substantially depending on which of the above three classes it belongs to. Often we will refer to the $\CP^1_x$ fibre over a point $x\in\bbR^{1,2}$ as a \emph{minitwistor line}. The anti-holomorphic involution $R$ lifts to a map $S\colon\PS\to\PS$, $(x,\lambda)\mapsto\bigl(x,\bar\lambda\bigr)$ so that $\pi\circ S = R\circ\pi$.

\subsection{Minitwistor description of the dKP equation} \label{subsec:Hitchin}

In \cite{Hitchin:1982vry}, Hitchin shows that the minitwistor correspondence of Sections~\ref{subsec:complex-minitwistors} and \ref{subsec:Lorentzian-minitwistors} is an~instance of a more general correspondence between $3d$ Einstein--Weyl geometries and complex surfaces. While we will not directly need this more general situation in the rest of the paper, we briefly review it here because the gravitational perspective on the dKP equation appears naturally both in minitwistor space and from our non-commutative gauge theory, which is also inherently gravitational. In addition, it provides a convenient way to introduce the Lax operators of the~dKP equation; we will recover these from the Chern--Simons description in Section~\ref{subsec:dKP-Lax}.

A real $3d$ Weyl manifold is a real conformal three-manifold $(M,[h])$ admitting a torsion-free connection $\nabla$ such that
\begin{equation} \label{eq:Weyl}
\nabla h = \nu\otimes h
\end{equation}
for some representative of the conformal class $h$ and 1-form $\nu\in\Omega^1(M)$. Under a Weyl transformation $h\mapsto\Omega^2h$ so that equation \eqref{eq:Weyl} continues to hold for $\nu\mapsto\nu + 2\dif\log\Omega$. We say that $(M,[h],\nabla)$ is \emph{Einstein--Weyl} if
\begin{equation} \label{eq:Einstein--Weyl}
\mrm{Ric}(\nabla) + \mrm{Ric}(\nabla)^\mrm{T} = 2\Lambda h
\end{equation}
for some function $\Lambda$, i.e., the symmetric part of the Ricci tensor of the connection $\nabla$ is proportional to a representative of the conformal class $h$. The antisymmetric part of the Ricci tensor~is
\begin{displaymath}
\mrm{Ric}(\nabla) - \mrm{Ric}(\nabla)^\mrm{T} = - 3\dif\nu .
\end{displaymath}
If this vanishes then at least locally there is a representative metric in the conformal class, unique up to constant rescaling, for which the connection $\nabla$ is Levi-Civita. The Einstein--Weyl equations~\eqref{eq:Einstein--Weyl} then specialise to the Einstein equation for this metric. Hence $\bbR^{1,2}$, $\mrm{dS}_3$ and~$\mrm{AdS}_3$ are all examples of $3d$ real Lorentzian Einstein--Weyl spaces. In these cases, $\Lambda$ is locally constant and can be identified with the cosmological constant; however, by performing a~constant rescaling we can set $\Lambda\in\{0,\pm1\}$.

The Hitchin correspondence states that complex three-folds with holomorphic Einstein--Weyl structures are in bijection with complex surfaces admitting a rational curve with normal bundle~$\cO(2)$. Let us briefly sketch how this works, following the presentation in \cite{Dunajski:2000rf}. Suppose we are given a complex three-fold $M$ with a holomorphic Einstein--Weyl structure $([h],\nabla)$. Fixing a~representative of the conformal class $h$ let \smash{$\bigl\{{\rm e}^{\al\beta} = {\rm e}^{(\al\beta)}\bigr\}$} be dreibeins, so that \smash{$h = \eps_{\al\gamma}\eps_{\beta\delta}{\rm e}^{\al\beta}{\rm e}^{\gamma\delta}$}. Since $([h],\nabla)$ is a Weyl structure there exists a spin connection \smash{$\gamma^\gamma_{\,~\,\delta} = \gamma^\gamma_{\,~\,\delta\al\beta}{\rm e}^{\al\beta}$} obeying
\begin{displaymath}
\nabla {\rm e}^{\al\beta} + 2\gamma^{(\al}_{~\,~\delta}\wedge {\rm e}^{\beta)\delta} = 0 .
\end{displaymath}
Note that $\gamma_{\gamma\delta\alpha\beta}$ is symmetric in $\al$, $\beta$ but not necessarily in $\gamma$, $\delta$ since $\nabla$ need not be metric. Furthermore, let $\bigl\{v_{\al\beta} = v_{(\al\beta)}\bigr\}$ be the dual frame of the tangent bundle satisfying
\begin{displaymath}
v_{\al\beta}\ip {\rm e}^{\gamma\delta} = \delta_{(\al}^{~~\gamma}\delta_{\beta)}^{~~\delta} ,
\end{displaymath}
where $\ip$ denotes contraction of a vector field with a form. We can then write down a rank two distribution on the complex four-fold $M\times\CP^1$, where the second factor is parametrised by the homogeneous co-ordinate $\lambda^\al$, whose integrability is equivalent to the Einstein--Weyl equations. It is spanned by the vector fields
\begin{displaymath} L_\al = \lambda^\beta v_{\al\beta} - \gamma^\gamma_{\,~\,\delta\alpha\beta}\lambda^\beta\lambda^\delta\p_{\lambda^\gamma} . \end{displaymath}
We will sometimes refer to this as the \emph{Lax distribution}, and denote it by $\la L_0,L_1\ra$. Quotienting $M\times\CP^1$ by the distribution generated by the vector fields $L_\al$ yields a complex surface $Z$, the minitwistor space of $(M,[g],\nabla)$. Points $x\in M$ lift to a $\CP^1_x\subset M\times\CP^1$, which then descend to spheres in $Z$ with normal bundle $\cO(2)$ (see \cite{Dunajski:2000rf} for details).

Conversely, given a complex surface $Z$ admitting a rational curve with normal bundle $\cO(2)$ Kodaira theory guarantees a complex three-parameter family. Each such curve represents a~point~${x\in M}$. Any two such curves intersect at two points in $Z$, counted with multiplicity. If these two points are distinct, then the corresponding points in $M$ are non-null separated and a~non-null geodesic corresponds to a family of curves in $Z$ all meeting at the same two distinct points. If the two intersection points coincide then the points are null separated in $M$. Together these determine a compatible conformal $[h]$ and projective structure $\nabla$ on $M$.

In order to recover a $3d$ real Lorentzian Einstein--Weyl space, we require that $Z$ admits an~anti-holomorphic involution with fixed points and preserving a rational curve.

Following \cite{Dunajski:2000rf}, given a solution $u$ of the dKP equation one can build a $3d$ Lorentzian Einstein--Weyl geometry with
\begin{align}
&h = \eta - 4u\,(\dif x^-)^2 = (\eps_{\al\gamma}\eps_{\beta\delta} - 4u\,\iota_\al\iota_\beta\iota_\gamma\iota_\delta)\dif x^{\al\beta}\dif x^{\gamma\delta} ,\nonumber \\
&\nu = 8\p_+u\,\dif x^- = 8\iota^\al\iota^\beta\iota_\gamma\iota_\delta\p_{\al\beta}u\,\dif x^{\gamma\delta} .\label{eq:EW-dKP}
\end{align}
Conversely, if an Einstein--Weyl geometry admits a constant weighted vector field~$l$, then there exist co-ordinates in which it takes the form \eqref{eq:EW-dKP} with $l = \iota^\al\iota^\beta\p_{\al\beta}$. Here a constant weighted vector field obeys
\begin{displaymath}\nabla l + m\,\nu\otimes l = 0 \end{displaymath}
for some $m$. In all but the trivial case of flat space $l$ is null with weight $m=-1/2$. Different solutions of the dKP equation $u$ may yield equivalent Einstein--Weyl geometries together with weighted null vector field. Indeed, a family of diffeomorphisms and Weyl transformations depending on three independent functions of $x^-$ preserve the form of the Weyl structure \eqref{eq:EW-dKP}. These generate families of solutions to the dKP equation corresponding to the same Einstein--Weyl structure.

The minitwistor space of a Lorentzian Einstein--Weyl geometry corresponding to a solution of the dKP equation admits a global section of $K^{-1/4}$. This should be invariant under the anti-holomorphic involution determining the real structure. Pulling back to $M\times\CP^1$, this section is given explicitly by $\la\lambda\iota\ra$. We will see momentarily that this is annihilated by the distribution~$\la L_0,L_1\ra$ and so descends to minitwistor space.

To specialise the Lax distribution to an Einstein--Weyl metric of the form \eqref{eq:EW-dKP}, we will need dual orthonormal frames for the tangent and cotangent bundles. Recalling that $h = \eps_{\al\gamma}\eps_{\beta\delta}{\rm e}^{\al\beta}{\rm e}^{\gamma\delta}$, these can be computed to be
\begin{equation} \label{eq:specialise-e}
{\rm e}^{\al\beta} = \dif x^{\al\beta} - 2u\iota^\al\iota^\beta\iota_\gamma\iota_\delta\dif x^{\gamma\delta} ,\qquad v_{\al\beta} = \p_{\al\beta} + 2u\iota_\al\iota_\beta\iota^\gamma\iota^\delta\p_{\gamma\delta} .
\end{equation}
From the former, we can determine the spin connection \smash{$\omega^\al_{~\,\beta} = \omega^\al_{~\,\beta\gamma\delta}{\rm e}^{\gamma\delta}$} using the torsion free condition
\begin{equation}
\label{eq:torsion-free} \dif {\rm e}^{\al\beta} + 2\omega^{(\al}_{~\,~\gamma}\wedge {\rm e}^{\beta)\gamma} = 0 .
\end{equation}
We emphasise that $\omega_{\al\beta\gamma\delta}$ is symmetric in both $\al$, $\beta$ and $\gamma$, $\delta$ since the spin connection is metric. In order to solve equation \eqref{eq:torsion-free} for $\omega$ in terms of~$u$, it is useful to decompose the dreibeins and spin connection into 1-form components according to
\begin{alignat*}{4}
&{\rm e}^+ = {\rm e}^{\al\beta}o_\al o_\beta ,\qquad&& {\rm e}^y = \sqrt{2}{\rm e}^{\al\beta}o_\al\iota_\beta ,\qquad&& {\rm e}^- = {\rm e}^{\al\beta}\iota_\al\iota_\beta ,& \\
&\omega^+ = \omega^{\al\beta}o_\al o_\beta ,\qquad&& \omega^y = \sqrt{2}\omega^{\al\beta}o_\al\iota_\beta ,\qquad&& \omega^- = \omega^{\al\beta}\iota_\al\iota_\beta .&
\end{alignat*}
In terms of these components, the metric is $h = 2{\rm e}^+{\rm e}^- - ({\rm e}^y)^2$ and the torsion free condition becomes
\begin{displaymath} \begin{aligned}
&\dif {\rm e}^- = \sqrt{2}(\omega^y\wedge {\rm e}^- - \omega^-\wedge {\rm e}^y) ,\qquad \dif {\rm e}^y = \sqrt{2}\bigl(\omega^+\wedge {\rm e}^- - \omega^-\wedge {\rm e}^+\bigr) , \\
&\dif {\rm e}^+ = \sqrt{2}\bigl(\omega^+\wedge {\rm e}^y - \omega^y\wedge {\rm e}^+\bigr) .
\end{aligned} \end{displaymath}
Specialising the dreibeins according to equation \eqref{eq:specialise-e}, we have
\begin{displaymath} \begin{aligned}
&\omega^y\wedge {\rm e}^- = \omega^-\wedge {\rm e}^y ,\qquad \omega^+\wedge {\rm e}^- = \omega^-\wedge {\rm e}^+ , \\
&\omega^+\wedge {\rm e}^y - \omega^y\wedge {\rm e}^+ = - \sqrt{2}\dif u\wedge\dif x^- = - \sqrt{2}\bigl(\p_+u{\rm e}^+ + 2\p_yu{\rm e}^y\bigr)\wedge {\rm e}^-\,
\end{aligned} \end{displaymath}
which are solved by $\omega^- = 0$, \smash{$\omega^y = - \sqrt{2}\p_+u {\rm e}^-$} and \smash{$\omega^+ = \sqrt{2}\p_yu {\rm e}^-$}. Equivalently, in spinor notation,
\begin{displaymath}
\omega^\gamma_{\,~\,\delta\al\beta} = \bigl(\iota^\gamma o_\delta\iota^\eps\iota^\zeta\p_{\eps\zeta}u + o^\gamma \iota_\delta\iota^\eps\iota^\zeta\p_{\eps\zeta}u - 2\iota^\gamma\iota_\delta o^\eps\iota^\zeta\p_{\eps\zeta}u\bigr)\iota_\al\iota_\beta .
\end{displaymath}
The spin connection $\gamma$ associated to $\nabla$ is related to $\omega$ by
\begin{displaymath} \gamma^\gamma_{\,~\,\delta\al\beta} = \omega^\gamma_{\,~\,\delta\al\beta} - \frac{1}{2}\delta^\gamma_{~\,(\al}\nu_{\beta)\delta} \end{displaymath}
since
\begin{displaymath} \nabla_{\al\beta}\eps_{\gamma\delta}
= \gamma_{\delta\gamma\al\beta} - \gamma_{\gamma\delta\al\beta} = - \eps_{\gamma\delta}\gamma^\eps_{\,~\,\eps\al\beta} = \frac{1}{2}\nu_{\al\beta}\eps_{\gamma\delta} \end{displaymath}
so that $\nabla g = \nu\otimes g$. Therefore,
\begin{displaymath} \gamma^\gamma_{\,~\,\delta\al\beta} = \bigl(\iota^\gamma o_\delta\iota^\eps\iota^\zeta\p_{\eps\zeta}u - 2\iota^\gamma\iota_\delta o^\eps\iota^\zeta\p_{\eps\zeta}u - 3o^\gamma \iota_\delta\iota^\eps\iota^\zeta\p_{\eps\zeta}u\bigr)\iota_\al\iota_\beta + 4\iota^\gamma\iota_\delta\iota^\eps\iota^\zeta\p_{\eps\zeta}uo_{(\al}\iota_{\beta)} \end{displaymath}
and the Lax distribution specialises to~\cite{Dunajski:2000rf}
\begin{align}
L_\al &{}= \lambda^\beta v_{\al\beta} - \gamma^\gamma_{\,~\,\delta\al\beta}\lambda^\beta\lambda^\delta\p_{\lambda^\gamma} \sim \lambda^\beta v_{\al\beta} - \frac{\gamma_{\gamma\delta\al\beta}\lambda^\beta\lambda^\gamma\lambda^\delta}{\la\lambda\iota\ra}\iota\cdot\p_\lambda \nonumber\\
&{}= \lambda^\beta\p_{\al\beta} + 2\iota_\al\la\lambda\iota\ra u\iota^\eps\iota^\zeta\p_{\eps\zeta} + 2\iota_\al\la\lambda\iota\ra^2o^\eps\iota^\zeta\p_{\eps\zeta}u\iota\cdot\p_\lambda - 2o_\al\la\lambda\iota\ra^2\iota^\eps\iota^\zeta\p_{\eps\zeta}u\iota\cdot\p_\lambda .\label{eq:dKP-Lax-dist}
\end{align}
Here $\sim$ indicates that we are working modulo the homogeneity operator $\Gamma = \lambda\cdot\p_\lambda$ and up to scale. It is immediate that $L_\al$ in equation \eqref{eq:dKP-Lax-dist} preserve the ideal generated by $\la\lambda\iota\ra$. In com\-po\-nent~form,
\begin{displaymath} L_0 \sim \p_- - \frac{z}{\sqrt{2}}\p_y + 2u\p_+ - \sqrt{2}\p_yu\p_z ,\qquad L_1\sim - \frac{1}{\sqrt{2}}\p_y + z\p_+ + 2\p_+u\p_z . \end{displaymath}
Integrability of the distribution spanned by the $L_\al$ requires
\[ \begin{aligned}
{}[L_0,L_1] &{}\sim \bigl(2\p_+\p_-u - \p_y^2 + 4u\p_+^2u\bigr)\p_z + \sqrt{2}\p_+u\p_y - 2z\p_+u\p_+ \\
&{}= \bigl(2\p_+\p_-u - \p_y^2u + 4\p_+(u\p_+u)\bigr)\p_z - 2\p_+u L_1 \in \la L_0,L_1\ra
\end{aligned}
\]
so that \smash{$2\p_+\p_-u - \p_y^2u + 4\p_+(u\p_+u) = 0$}, which we recognise as the dKP equation.

\section[Non-commutative 5d Chern--Simons theory]{Non-commutative $\boldsymbol{5d}$ Chern--Simons theory}
\label{sec:5dncCS}

The main goal of this work is to show that the KP theory discussed above is equivalent to a certain non-commutative $5d$ Chern--Simons theory \big(with Abelian\footnote{For us Abelian vs non-Abelian always refers to the gauge group, commutative vs non-commutative to whether multiplication of co-ordinate functions is graded commutative and quantum usually indicates the inclusion of loop effects. The one exception is that we occasionally refer to quantization of a Poisson bracket to get a non-commutative structure.} gauge group\big) which can naturally be defined on correspondence space. This can be viewed as a generalization of the program of Costello--Witten--Yamazaki~\cite{Costello:2017dso,Costello:2018gyb,Costello:2019tri} to recover $2d$ integrable systems from a $4d$ Chern--Simons theory. The $5d$ theory we will consider was originally introduced in \cite{Costello:2016nkh} as the conjectural twist of $11d$ supergravity in an $\Omega$ background, and has been studied extensively in~\mbox{\cite{Costello:2017fbo,Gaiotto:2019wcc,Gaiotto:2020dsq,Oh:2020hph}}. We begin this section by introducing this theory on $\bbC^2\times\bbR$, before addressing some of the subtleties which arise when trying to define it on $\PS$. In this section, we will abuse notation by writing $(v,z,t)$ for co-ordinates on $\bbC^2\times\bbR$.

\subsection[The theory on C\^{}2 times R]{The theory on $\boldsymbol{\bbC^2\times\bbR}$} \label{subsec:C2xR}

The dynamical field of the theory (with an Abelian gauge group) is a partial connection
\begin{displaymath} a=a_{\bar{z}}\dif\bar{z}+ a_{\bar{v}}\dif\bar{v}+a_t\dif t \in\widetilde\Omega^{0,1}\bigl(\bbC^2\times\bbR\bigr) , \end{displaymath}
where we write \smash{$\widetilde\Omega^{0,p}\bigl(\bbC^2\times\bbR\bigr)$} for the space of $p$-forms built solely from the antiholomorphic forms~$\dif\bar z$,~$\dif\bar v$ along $\bbC^2$ and the real form $\dif t$ along $\bbR$. Abusing terminology slightly, we will refer to elements of \smash{$\widetilde\Omega^{0,p}\bigl(\bbC^2\times\bbR\bigr)$} as $(0,p)$-forms. The kinetic term of the action is constructed using the operator
\begin{displaymath} \dift = \bar\partial_{\bbC^2} + \dif_\bbR
= \dif\bar{z}\partial_{\bar{z}} + \dif\bar{v}\partial_{\bar{v}}+\dif t\partial_t . \end{displaymath}
This differential gives the space \smash{$\widetilde\Omega^{0,\bullet}\bigl(\bbC^2\times\bbR\bigr)= C^\infty\bigl(\bbC^2\times\bbR\bigr)[\dif\bar z,\dif\bar v,\dif t]$} of arbitrary $(0,\bullet)$-forms the structure of a commutative differential-graded algebra, analogous to the Dolbeault complex of a complex manifold. The $(1,0)$ tangent bundle \smash{$T^{1,0}\bigl(\bbC^2\times\bbR\bigr)$} is spanned by smooth vectors $V=V^z\partial_z + V^v\partial_v$ which annihilate \smash{$\widetilde\Omega^{0,1}\bigl(\bbC^2\times\bbR\bigr)$} by contraction. This bundle may be endowed with a `holomorphic' structure in a similar way.

The next ingredient we need to define the theory is the $(2,0)$-form $\Omega = \dif z\wedge\dif v$, together with a corresponding Poisson bivector
\begin{displaymath} \Pi = \p_z\vee\p_v . \end{displaymath}
This bivector is fixed by the property that it lives in \smash{$\wedge^2T^{1,0}\bigl(\bbC^2\times\bbR\bigr)$} and obeys $\Pi\ip\Omega = -1$. For the non-commutative theory, we will quantize the Poisson bivector associated to $\Pi$ using the Moyal formula. This gives us an associative star product
\begin{displaymath}
F\star G = c\circ\exp\frac{\fq}{2}\bigl(\p_z\otimes\p_v - \p_v\otimes\p_z)(F\otimes G) = \sum_{m,n=0}^\infty\frac{\fq^{m+n}}{m!n!2^{m+n}}(-)^n\p_z^m\p_v^nF\p_z^n\p_v^mG .
 \end{displaymath}
Here \smash{$F,G\in C^\infty\bigl(\bbC^2\times\bbR\bigr)$} and \smash{$c\colon C^\infty\bigl(\bbC^2\times\bbR\bigr)^{\otimes 2}\to C^\infty\bigl(\bbC^2\times\bbR\bigr)$} is the multiplication map.\footnote{In this work, we will take $\fq$ to be a formal parameter and strictly we should tensor our spaces of functions and forms with the ring of formal power series $\bbC\llbracket\fq\rrbracket$. For the sake of brevity, we will omit this tensor product, with the understanding that in the non-commutative setting fields are valued in modules over this ring.} The star product naturally extends to elements \smash{$f,g\in\widetilde\Omega^{0,\bullet}\bigl(\bbC^2\times\bbR\bigr)$} if we understand the vector fields $\p_z$ and $\p_v$ act by Lie derivative.\footnote{In fact, there is really no ambiguity in how $V$ a $(1,0)$ vector field could act on $\omega\in\widetilde\Omega^{0,\bullet}(\bbC\times\bbR)$ since $\cL_V\omega = [V\ip,\dif]\omega = [V\ip,\p]\omega = V\ip\p\omega$.} It is immediate that $\dift$ distributes over $\star$, so that $\bigl(\smash{\widetilde\Omega^{0,\bullet}}\bigl(\bbC^2\times\bbR\bigr),\smash{\dift},\star\bigr)$ has the structure of a \emph{differential graded algebra} (dga). More precisely, it defines a~sheaf of dgas.

With these ingredients, the action of \emph{$5d$ non-commutative Chern--Simons theory} is
\begin{align}
S_\star[a] &{}= \frac{\im}{2\pi}\int_{\bbC^2\times\bbR}\Omega\wedge\biggl(\frac{1}{2}a\star\dift a + \frac{1}{3\fq}a\star a\star a\biggr) \nonumber \\
&{}= \frac{\im}{2\pi}\int_{\bbC^2\times\bbR}\Omega\wedge\biggl(\frac{1}{2}a\wedge\dift a + \frac{1}{6}a\wedge[a,a]_\star\biggr) . \label{eq:ncCS}
\end{align}
Here the (graded) Moyal bracket is defined as \smash{$[f,g]_\star = \fq^{-1}\bigl(f\star g - (-)^{\deg f\deg g}g\star f\bigr)$}.\footnote{Costello defines $5d$ non-commutative Abelian Chern--Simons using the vertex $a\star a\star a$ without dividing by~$\fq$~\cite{Costello:2016nkh}. This is because he considers both the Abelian and non-Abelian theory, and in the latter case $\tr(a\star a\star a)$ is non-vanishing in the $\fq\to0$ limit. We could adopt the same convention, but prefer to retain the order $\fq^0$ part of $a$ which we will see shortly has a geometric interpretation.} This depends on $\fq$ only through its square. In the second equality, we have replaced the $\star$ product in the kinetic term with an ordinary wedge product, and similarly replaced one of the star products in the cubic vertex. This is possible due to the following identity for $f,g\in\widetilde\Omega^{0,\bullet}(\bbC\times\bbR)$ (assuming sufficient rapid falloffs at infinity)
\begin{align}
\frac{\im}{2\pi}\int_{\bbC^2\times\bbR}\Omega\wedge(f\star g) & = \frac{\im}{2\pi}\int_{\bbC^2\times\bbR}\Omega\wedge f\wedge g = \frac{\im}{2\pi}\int_{\bbC^2\times\bbR}\Omega\wedge g\wedge f \nonumber\\
& = \frac{\im}{2\pi}\int_{\bbC^2\times\bbR}\Omega\wedge(g\star f) . \label{eq:pairing}
\end{align}
Notice that in order for this integral to be non-vanishing an odd and even degree form must pair with one another, so no sign is generated when exchanging $f$, $g$. The equations of motion of this theory read
\begin{displaymath} \dift a + \frac{1}{2}[a,a]_\star = \dift a + \frac{1}{\fq}a\star a = 0 . \end{displaymath}
Solutions related by gauge transformations, acting infinitesimally by
\begin{displaymath} \delta a = \dift\eps + [a,\eps]_\star \end{displaymath}
for $\eps\in C^\infty(\bbC^2\times\bbR)$, are taken to be physically equivalent. Gauge invariance of the action \eqref{eq:ncCS} follows from the dga structure on the space of fields, together with its compatibility with the integration pairing \eqref{eq:pairing}. (More precisely, the BV action obeys the classical master equation. We refer the reader to \cite{Costello:2016nkh} for details.)

The equations of motion imply that the poly-differential operator $\dbar + [a,~\,]_\star$ is integrable. Resolving the (sheaf of) dga(s) \smash{$\bigl(\widetilde\Omega^{0,\bullet}\bigl(\bbC^2\times\bbR\bigr),\dift + [a,~\,]_\star,\star\bigr)$} yields a deformation of the sheaf of mixed topological holomorphic functions on $\bbC^2\times\bbR$ equipped with the non-commutative product~$\star$.\footnote{Here the sheaf of mixed topological holomorphic functions is the box product of the sheaf of holomorphic functions $\cO_{\bbC^2}$ on $\bbC^2$ and the constant sheaf $\underline\bbC$ on $\bbR$.} In this way \smash{$\dift + [a,~\,]_\star$} deforms $\bbC^2\times\bbR$ as a non-commutative manifold. Since \smash{$\dift + [a,~\,]_\star$} respects the integration pairing \eqref{eq:pairing}, we can say that this deformation preserves $\Omega$. Gauge transformations act by inner automorphisms of the (sheaf of) dga(s) \smash{$\bigl(\widetilde\Omega^{0,\bullet}\bigl(\bbC^2\times\bbR\bigr),\dift + [a,~\,]_\star,\star\bigr)$}, which can be interpreted as the non-commutative analogue of diffeomorphisms.

Currently the action \eqref{eq:ncCS} should be interpreted as a holomorphic functional on a complex space of fields; this is sufficient to define the theory perturbatively, but physically it is natural to restrict to a real contour in field space. The overall factor of $\im/2\pi$ is adapted to the reality conditions we impose in Section~\ref{sec:PS}.

Non-commutative Chern--Simons theory on $\bbC^2\times\bbR$ is non-renormalizable by power-counting. However, in \cite{Costello:2016nkh} Costello shows that, allowing counterterms accompanied by powers of $\hbar/\fq^3$, a translationally invariant perturbative quantization exists to all loop orders. Allowing $\hbar/\fq^3$ to appear in counterterms can be interpreted heuristically as treating $\hbar\ll\fq^3$. (Under scale transformations of the $(z,v,t)$ co-ordinates the action transforms homogeneously with mass dimension $[S_\star] = -6$ if we assign $[\fq] = -2$ and $[a] = -2$. As a result, $\hbar/\fq^3$ is dimensionless.)

Even in the Abelian case, the non-commutative theory has a non-trivial interacting limit as~${\fq\to0}$. Taking this limit in the action \eqref{eq:ncCS} gives
\begin{equation} \label{eq:PCS} S_{\rm PCS}[a] = \frac{\im}{2\pi}\int_{\bbC^2\times\bbR}\Omega\wedge\biggl(\frac{1}{2}a\wedge\dift a + \frac{1}{6}a\wedge\{a,a\}\biggr) , \end{equation}
where
\begin{displaymath} \{f,g\} = \cL_{\p_z}f\wedge\cL_{\p_v}g - (-)^{\deg f\deg g}\cL_{\p_v}f\wedge\cL_{\p_z}g \end{displaymath}
extends the Poisson bracket associated to $\Pi$ to forms $f,g\in\widetilde\Omega^{0,\bullet}\bigl(\bbC^2\times\bbR\bigr)$. We will refer to this as \emph{Poisson--Chern--Simons theory}. Its equations of motion imply the integrability of the Beltrami differential $\dbar + \{a,~\,\}$ of $\bbC^2\times\bbR$. Gauge transformations are $(1,0)$ Poisson diffeomorphisms, and invariance of the action \eqref{eq:PCS} is ensured by the fact that $\bigl(\widetilde\Omega^{0,\bullet}(\bbC^2\times\bbR),\dift,\{~ ,~\,\}\bigr)$ forms a~differential graded Poisson algebra, together with its compatibility with the integration pairing~\eqref{eq:pairing}. Therefore, Poisson--Chern--Simons describes Hamiltonian deformations of $\bbC^2\times\bbR$ as a~transversally holomorphically foliated manifold.

At one-loop, a translationally invariant quantization of Poisson--Chern--Simons still exists \big(as~does a one-loop quantization of non-commutative Chern--Simons without allowing counterterms involving powers of $\hbar/\fq^3$\big), essentially because in a holomorphic theory the one-loop diagrams are finite and one-loop anomalies vanish in odd dimensions \cite{Gwilliam:2021zkv}. We note that although the results of \cite{Costello:2016nkh} do not show the existence of a higher loop quantization of Poisson--Chern--Simons theory, they do not preclude one either.

\subsection[The theory on a more general five-manifold X]{The theory on a more general five-manifold $\boldsymbol{X}$} \label{subsec:ambiguity}

We have seen how to define non-commutative Chern--Simons theory on $\bbC^2\times\bbR$, now we define non-commutative Chern--Simons on a more general $5d$ real manifold $X$.

To achieve this, we require that $X$ possesses a nowhere vanishing 2-form $\Omega\in\Omega^2_\bbC(X)$ which is~closed, $\dif\Omega \!=\! 0$, and simple, $\Omega\wedge\Omega \!=\! 0$. Given this basic ingredient, we can define ${\ell \! =\! \ker\Omega\!\subset\! T_\bbC X}$, the complex distribution consisting of vectors which contract to zero against $\Omega$. Closure of $\Omega$ implies that $\ell$ is integrable, and simplicity ensures that it has complex rank three. We can view~$\ell$ as an avatar of the $(0,1)$ tangent bundle of a complex manifold, and we will sometimes refer to its sections as $(0,1)$ vector fields accordingly.

We write $\ell_\perp\subset T_\bbC^*X$ for the annihilator of $\ell$, the subbundle of the complex cotangent bundle consisting of covectors which wedge to zero against $\Omega$. The simplicity of $\Omega$ ensures that $\ell_\perp$ has complex rank two. $\ell_\perp$ generates a subsheaf $\cS$ of $\Omega^\bullet_\bbC(X)$ consisting of all complex differential forms which wedge to zero against $\Omega$. We can then further form the quotient \smash{$\widetilde\Omega^{0,\bullet}(X) = \Omega^\bullet_\bbC(X)/\cS$}. The de Rham differential $\dif$ descends to the quotient \smash{$\widetilde\Omega^{0,\bullet}(X)$}. To see why, fix a representative $\eta\in\Omega^\bullet(X)$ for some equivalence class \smash{$[\eta]\in\widetilde\Omega^{0,\bullet}(X)$}. Then the ambiguity in the representative is $\eta\sim\eta + \zeta$ where $\Omega\wedge\zeta = 0$. The ambiguity in $\dif\eta$ is $\dif\eta\sim\dif\eta + \dif\zeta$; however,
\begin{displaymath} \Omega\wedge\dif\zeta = \dif(\Omega\wedge\zeta) = 0 \end{displaymath}
by the closure of $\Omega$. Therefore, \smash{$[\dif\eta]\in\widetilde\Omega^{0,\bullet}(X)$} is well defined. We write $\dift$ for the resulting differential on \smash{$\widetilde\Omega^{0,\bullet}(X)$}.

In general, the fibres of $\ell$ (resp.\ $\ell_\perp$) may intersect those of the real tangent bundle $TX$ (resp.\ real cotangent bundle $T^*X$) in real subspaces of dimension one, two or three (resp.\ zero, one or two). If the intersection has dimension one everywhere, then $\ell$ contains a real non-vanishing vector field whose integral curves are the leaves of a \emph{transverse holomorphic foliation} of $X$. In~this~work, we will be concerned with 2-forms $\Omega$ for which this property fails, but at worst on some submanifold of at least co-dimension one.

In a sufficiently small open neighbourhood $U\subset X$ of a point $p\in X$ for which $\ell_p\cap T_pX$ has dimension one, we can always find holomorphic co-ordinates $(v,z,t)\in\bbC^2\times\bbR$ such that $\Omega = \dif z\wedge\dif v$. As before, we can then represent each equivalence class in $\widetilde\Omega^{0,\bullet}(U)$ by an element of $C^\infty(U)[\dif\bar z,\dif\bar v,\dif t]$, and for such representatives the differential has the expected form $\dift = \dif\bar z\p_{\bar z} + \dif\bar v\p_{\bar v} + \dif t\p_t$. Therefore, at least locally we are in the situation described in Section~\ref{subsec:C2xR}; however, on $\PS$ it will be more natural to view $(0,\bullet)$-forms as living in the quotient.

Next, let us attempt to define the Poisson bracket, which is obtained by inverting $\Omega$. On $X$, the avatar of the $(1,0)$ tangent bundle of a complex manifold is the quotient $T^{1,0}X = T_\bbC X/\ell$, and $\Omega$ can be inverted to give an element $\Pi\in\wedge^2T^{1,0}X$. However, if we try to use $\Pi$ to define a~Poisson bracket on \smash{$\widetilde\Omega^{0,\bullet}(X)$}, we run into a problem: The action of a $(1,0)$ vector field \big(defined to be an equivalence class in $T^{1,0}X$\big) on a section of \smash{$\widetilde\Omega^{0,\bullet}(X)$} (also an equivalence class) is certainly ambiguous.

In the open neighbourhood $U$ of a point $p\in X$ as above, we can give a definition by representing sections of $T^{1,0}X|_U$ as vector fields $V = V^z\p_z + V^v\p_v$ and taking $\Pi = \p_z\vee\p_v$. But globally the best we can do is pick a section \smash{$\cD\colon T^{1,0}X\to T_\bbC X$} (often we will confuse $\cD$ and its image) and define \smash{$\widetilde\Omega^{0,\bullet}_\cD(X)$} to be the subsheaf of differential forms generated by $\cD_\perp$. Each equivalence class in \smash{$\widetilde\Omega^{0,\bullet}(X)$} has a unique representative in \smash{$\widetilde\Omega^{0,\bullet}_\cD(X)$} and on these representatives \smash{$\dift$} is represented by the differential \smash{$\dift_\cD$}. The action of Lie derivatives by sections of $\cD$ on elements of \smash{$\widetilde\Omega^{0,\bullet}_\cD(X)$} is unambiguous, and so the Poisson bracket $\{~,~\}_\cD$ is well-defined.

We now come to an important point: although we need to pick a choice of section $\cD$ to define Poisson--Chern--Simons theory, \emph{the dependence on this choice can always be removed by a field redefinition}. The reason for this is that, given two choices $\cD$ and $\cD^\prime$, the difference between the corresponding Poisson brackets $\{~ ,~\,\}_\cD$ and $\{~ ,~\,\}_{\cD^\prime}$ involves derivatives along vector fields in $\ell$, which are trivial in BRST cohomology. This implies that there always exists a field redefinition $a \to a^\prime(a)$ such that $S_{\cD}[a] = S_{\cD^\prime}[a^\prime]$. Working to lowest non-trivial order, we will demonstrate this explicitly in Appendix~\ref{app:Dindependence} for the choice of $\cD$ used below and another choice $\cD^\prime$ that arises naturally from minitwistor space. That the field redefinition exists to all orders follows from a generalization of \cite[Lemma~9.2.1]{Costello:2021jvx}: In brief, it is clear that using the equations of motion we can trade derivatives along $\ell$ for higher-order terms; less clear is why the resulting higher-order vertices will ultimately vanish. In order to write down vertices beyond cubic order, we~must contract $a$ against $(0,1)$ vector fields, but it is precisely such contractions which cancel against Lie derivatives in the $(0,1)$ directions in cohomology. This ensures that there's a field redefinition eliminating both contractions and derivatives along $\ell$. Mathematically, this is the statement that \smash{$\bigl(\widetilde\Omega^{0,\bullet}_\cD(X),\dift_\cD,\{~,~\}_\cD\bigr)$} is independent of $\cD$ up to quasi-isomorphism. The utility of this fact is that we are free to make a convenient choice for $\cD$ which need not adapted to a~holomorphic co-ordinate frame. On $\PS$, this simplifies life dramatically.

In order to define non-commutative Chern--Simons theory on $X$, we would further need to turn on non-commutativity for the Poisson bracket $\Pi$. One could appeal to general results regarding the existence of deformation quantizations, but in the case of interest where $X=\PS$ we will be able to employ the Moyal formula to get an explicit non-commutative product.

\section{Correspondence space theories} \label{sec:PS}

In this section, we introduce Poisson--Chern--Simons theory and non-commutative Chern--Simons theory on minitwistor correspondence space. As we will show later, these are equivalent to the dKP and KP field theories \eqref{eq:dKPaction} and \eqref{eq:KPaction}.

\subsection[Poisson--Chern--Simons on PS]{Poisson--Chern--Simons on $\boldsymbol{\PS}$} \label{subsec:correspondence-PCS}

To define $5d$ Poisson--Chern--Simons theory on $\PS$ we first need to choose a simple, non-vanishing, closed 2-form $\Omega\in\Omega^2(\PS)$. Following \cite{Bittleston:2020hfv,Costello:2021bah}, we will find a suitable $\Omega$ by pulling back a~meromorphic form from $\mT$. Fixing a real reference spinor $\iota_\al$ (and extending this by a second real reference spinor $o_\al$ to a dyad $\la o\iota\ra=1$ for later convenience), let
\begin{displaymath} \Omega = \frac{\la\lambda\dif\lambda\ra\wedge\dif x^{\al\beta}\lambda_\al\lambda_\beta}{\la\lambda\iota\ra^4} . \end{displaymath}
Closure fails on the locus of the quartic pole, and the price we pay is that our fields will have to obey compensating `boundary conditions' there.\footnote{Strictly speaking, the locus $\lambda\sim\iota$ has real co-dimension two. Nevertheless, we shall refer to the constraints we impose there as boundary conditions.} Notice also that whilst generically the fibres of $\ell_\perp$ intersect the real cotangent bundle trivially, on the locus $\lambda\sim\bar\lambda$ the intersection is one-dimensional. This reflects the fact that in Lorentzian signature correspondence space is not transversally holomorphically fibred over $\mT$.

As explained in Section~\ref{subsec:ambiguity}, to write the theory down we will also need to fix a rank two integrable subbundle $\cD\subset T_\bbC\PS$. A particularly convenient choice is
\begin{displaymath} \cD = \bigg\la\iota^\al\frac{\p}{\p\lambda_\al}\bigg|_x ,\iota^\al\iota^\beta\frac{\p}{\p x^{\al\beta}}\bigg\ra_\bbC , \end{displaymath}
with frame
\begin{equation} \label{eq:const-frame-vect} \p_0 = \la\lambda\iota\ra\iota^\al\frac{\p}{\p\lambda^\al}\bigg|_x ,\qquad\p_1 = \iota^\al\iota^\beta\frac{\p}{\p x^{\al\beta}} . \end{equation}
Here the derivative in $\lambda$ is taken at fixed $x\in\bbR^{1,2}$. Since $\rho_*\p_1 = \p_+$ and $\p_1$ is constant up the fibres of $\rho$, it is harmless to confuse it with $\p_+$. We will refer to \eqref{eq:const-frame-vect} and its associated distribution as the `constant frame'.\footnote{On the locus $\lambda\sim\iota$, the distribution $\cD$ intersects $T^{0,1}\PS$ and so strictly is not a good section of $T^{1,0}\PS$. Since this is the locus on which $\Omega$ is also singular, it can be accounted for by suitably restricting the behaviour of the component fields there.}

This choice of $\cD$ also fixes representatives for $\til\Omega^{0,\bullet}(\PS)$ which lie in its annihilator. (From this section, we will drop the subscript $\cD$ when it is unambiguous.) In this case,
\begin{displaymath} \cD_\perp = \big\la\overline{\la\lambda\dif\lambda\ra},\dif x^{\al\beta}\iota_\al\big\ra_\bbC \end{displaymath}
with frame
\begin{equation} \label{eq:const-frame-form} \tilde\sigma^0 = \frac{\overline{\la\lambda\dif\lambda\ra}}{\overline{\la o\lambda\ra}^2} ,\qquad \tilde\sigma^\al = \dif x^{\al\beta}\iota_\beta . \end{equation}
Integrability of $\cD$ follows from the trivial structure equations $\dif\tilde\sigma^0 = \dif\tilde\sigma^\al = 0$. (The specific choice of $\tilde\sigma^0$ here is largely irrelevant -- for convenience, we have placed the anti-holomorphic double pole at $\lambda\sim o$ so it does not complicate the behaviour of the component fields at $\lambda\sim\iota$.)

Away from $\lambda\sim\iota$, we can extend the frame \eqref{eq:const-frame-form} to all of $\Omega^1_\bbC(\PS)$ by adjoining the $(1,0)$-forms
\begin{displaymath} {\rm e}^0 = \la\lambda\dif\lambda\ra ,\qquad {\rm e}^1 = \dif x^{\al\beta}\lambda_\al\lambda_\beta . \end{displaymath}
Note that these are globally defined in $\PS$, at the cost of transforming as sections of the line bundle $\cO(2)$. We could trivialise using the divisor $\la\lambda\iota\ra$, but the $(1,0)$-cotangent bundle is independent of $\iota$ and we prefer to preserve this feature in the frame. The structure equations are non-trivial
\begin{gather*}
\dif {\rm e}^0 = \la\dif\lambda\wedge\dif\lambda\ra = 2\frac{\la\dif\lambda\iota\ra}{\la\lambda\iota\ra}\wedge {\rm e}^0 ,\\
\dif {\rm e}^1 = - 2\dif x^{\al\beta}\lambda_\al\wedge\dif\lambda_\beta = 2\frac{\la\dif\lambda\iota\ra}{\la\lambda\iota\ra}\wedge {\rm e}^1 + 2\frac{\lambda_\al}{\la\lambda\iota\ra} {\rm e}^0\wedge\tilde\sigma^\al .
\end{gather*}
The terms involving $\la\dif\lambda\iota\ra$ signal that ${\rm e}^0$, ${\rm e}^1$ transform as sections of $\cO(2)$; however, the final term has an important consequence
\begin{displaymath}
\bigl(\dift\p_0\bigr)\ip {\rm e}^1 = \bigl[\dift,\p_0\ip\bigr]{\rm e}^1 = 2\la\lambda\iota\ra\tilde\sigma^\al\lambda_\al .
\end{displaymath}
Here $[~ ,~]$ denotes the graded commutator, and since $\dift$ and $\p_0\ip$ are odd operations on forms, it~behaves as an anti-commutator in the above equation. We infer that
\begin{equation} \label{eq:dift-p0}
\dift\p_0 = 2\frac{\tilde\sigma^\al\lambda_\al}{\la\lambda\iota\ra}\otimes\p_1 .
\end{equation}
Our frame \eqref{eq:const-frame-vect} is therefore not `holomorphic'. We will see that this is essential in understanding the origin of the dispersive term in the KP equation.

Having fixed the constant frame, the Poisson bivector is $\Pi = \p_0\vee\p_1$ which clearly obeys $\Pi\ip\Omega = -1$. Although the vector field $\p_1$ is not `holomorphic', the bivector itself is since
\begin{displaymath} \dift\Pi = 2\frac{\tilde\sigma^\al\lambda_\al}{\la\lambda\iota\ra}\otimes(\p_1\vee\p_1) = 0 . \end{displaymath}

Finally, for future reference, it will be useful to specify a frame for the $(0,1)$-tangent bundle~$\ell$ which is dual to \eqref{eq:const-frame-form}. It is spanned by the vector fields $\tilde\p_0$, $\tilde\p_\al$ obeying
\begin{displaymath} \tilde\p_0\ip \tilde\sigma^0 = 0 ,\qquad \tilde\p_0\ip\tilde\sigma^\al = 0 ,\qquad \tilde\p_\al\ip\tilde\sigma^0 = 0 ,\qquad \tilde\p_\al\ip\tilde\sigma^\beta = \delta_\al^{~\,\beta} . \end{displaymath}
These equations are solved by
\begin{displaymath} \tilde\p_0 = - \overline{\la o\lambda\ra}o^\al\frac{\p}{\p\bar\lambda^\al}\bigg|_x ,\qquad \tilde\p_\al = P_\al^{~\,\beta}\frac{\lambda^\gamma}{\la\lambda\iota\ra}\frac{\p}{\p x^{\beta\gamma}}, \end{displaymath}
where
\begin{displaymath} P_\al^{~\,\beta} = \frac{\iota_\al\lambda^\beta - 2\lambda_\al\iota^\beta}{\la\lambda\iota\ra} = \iota_\alpha o^\beta - 2o_\alpha\iota^\beta - \frac{\la o\lambda\ra}{\la\lambda\iota\ra}\iota_\alpha\iota^\beta . \end{displaymath}
Note that \smash{$\eps^{\beta\al}P_\al^{~\,\gamma}P_\beta^{~\,\delta} = 2\epsilon^{\delta\gamma}$} so that $P$ is invertible. The derivative with respect to $\bar\lambda$ is taken at fixed $x\in\bbR^{1,2}$. Given dual frames of~$\ell$ and~$\cD^\vee$, we can write down an expression for the differential
\begin{equation} \label{eq:dift-const} \tilde\dif = \tilde\sigma^0\tilde\p_0 + \tilde\sigma^\al\tilde\p_\al = \dbar_{\CP^1_x} + \dif x^{\al\beta}\iota_\al P_\beta^{~\,\gamma}\frac{\lambda^\delta}{\la\lambda\iota\ra}\frac{\p}{\p x^{\gamma\delta}} . \end{equation}

Having set up our frames correctly, we are now in position to define Poisson--Chern--Simons theory on $\PS$. The dynamical field is the partial connection $a \in \til\Omega^{0,1}(\PS)$, whose physical part has components $a_0\tilde\sigma^0 + a_\al\tilde\sigma^\al$. The action is
\begin{align}
S_{\rm PCS}[a] &{}= \frac{\im}{2\pi}\int_\PS\Omega\wedge\biggl(\frac{1}{2}a\wedge\dift a + \frac{1}{6}a\wedge\{a,a\}\biggr) \nonumber\\
&{}= \frac{\im}{2\pi}\int_\PS\Omega\wedge\biggl(\frac{1}{2}a\wedge\dift a + \frac{1}{3}a\wedge\cL_{\p_0}a\wedge\cL_{\p_1}a\biggr), \label{eq:PCS-PS}
\end{align}
where $\cL_{\p_0} a = \p_0\ip\p a$ and $\cL_{\p_1} a = \p_1\ip\p a$.\footnote{When we use the same distribution $\cD$ to define the $(1,0)$ tangent bundle and space of forms $\til\Omega^{0,\bullet}_\cD(\PS)$, the Lie derivative along $(1,0)$ vector fields is unambiguous.} The equations of motion state that
\begin{displaymath} \dift a + \frac{1}{2}\{a,a\} = 0 , \end{displaymath}
ensuring nilpotency of the deformed differential $\dift + \{a,~\,\}$. Gauge transformations are $(1,0)$ Poisson diffeomorphisms generated infinitesimally by Hamiltonians $\eps\in\til\Omega^0(\PS)$ and acting by $\delta a = \dift\eps + \{a,\eps\}$. At least away from the locus $\lambda\sim\bar\lambda$, correspondence space is fibred over $\mT$ with real one-dimensional fibres. Entering a gauge in which the pullback of $a$ to these fibres vanishes, the equations of motion tell us that the remaining components of $a$ are constant. Therefore, we can understand $a$ as being pulled back from a Hamiltonian complex structure deformation of $\mT$. The surviving gauge transformations are Poisson diffeomorphims on $\mT$ (and we will see shortly that these must vanish to second order at $\lambda\sim\iota$). In this way, at least qualitatively, we recover the minitwistor description of the dKP equation reviewed in Section~\ref{subsec:Hitchin}. This makes Poisson--Chern--Simons on $\PS$ a natural candidate for the twistorial uplift of dKP theory.

Our reality conditions state that $a$ should be invariant under pullback by $S\colon(x,\lambda)\mapsto\bigl(x,\bar\lambda\bigr)$ ensuring that the resulting Hamiltonian deformations describe real Einstein--Weyl geometries. The action $S_\mrm{PCS}$ is real: the sign generated by conjugating $\im/2\pi$ is compensated by $S$ reversing the orientation of $\PS$.

The price of introducing a pole in the measure is that we must impose boundary conditions to eliminate potential contributions when varying the action. Let us determine these now.

\subsection{Boundary conditions} \label{subsec:bcs}

Varying the kinetic term of \eqref{eq:PCS-PS} gives
\begin{displaymath}
\delta S_\mrm{kin.}[a] = - \frac{\im}{4\pi}\int_\PS\Omega\wedge\dift(a\wedge\delta a) + \frac{\im}{2\pi}\int_\PS\delta a\wedge(\text{linearised equations of motion}) .
\end{displaymath}
On-shell this reduces to the boundary contribution, which is easiest to isolate by excising a small disc $D^2_\eps = \{|\tilde z|\leq\eps\}$ around $\tilde z = - \la\lambda\iota\ra/\la o\lambda\ra = 0$ along the fibres of $\PS\to\bbR^{1,2}$, yielding
\begin{displaymath}
\delta S_\mrm{kin.}[a] \simeq \frac{\im}{4\pi}\int_{\bbR^{1,2}\times S^1_\eps} \Omega\wedge a\wedge\delta a .
\end{displaymath}
Since \smash{$\Omega = \la\lambda\dif\lambda\ra\wedge\dif x^{\gamma\delta}\lambda_\gamma\lambda_\delta/\la\lambda\iota\ra^4$} saturates the integral over $S^1_\eps$, only the components of $a$ in~the~${\tilde\sigma^\al = \dif x^{\al\beta}\iota_\beta}$ directions contribute. Expanding $a = a_0\tilde\sigma^0 + a_\al\tilde\sigma^\al$, we have
\begin{displaymath}
\delta S_{\rm kin.}[a] \simeq \frac{\im}{8\pi}\int_{\bbR^{1,2}\times S^1_\eps} \Omega\wedge\tilde\sigma^\al\wedge\tilde\sigma_\al\,a^\beta\delta a_\beta .
\end{displaymath}
Certainly, $\dif x^{\gamma\delta}\lambda_\gamma\lambda_\delta\wedge\tilde\sigma^\al\wedge\tilde\sigma_\al = - \sqrt{2} \dif^3x \la\lambda\iota\ra^2$, so the boundary term is
\begin{displaymath}
\delta S_{\rm kin.}[a] \simeq - \frac{\im}{4\sqrt{2}\pi}\int_{\bbR^{1,2}}\dif^3x\oint_{S^1_\eps}\frac{\la\lambda\dif\lambda\ra}{\la\lambda\iota\ra^2}a^\beta\delta a_\beta .
\end{displaymath}
From this expression, it is clear that requiring $a_\al = \cO(\la\lambda\iota\ra)$ is sufficient to kill the boundary term. (That is, $a_\al$ should vanish holomorphically to first order on the locus $\lambda\sim\iota$.) But this is needlessly strong, in fact there is a weaker condition that will also do the trick. We require that~$a_\al$ is regular on the locus $\lambda\sim\iota$, and that the combination
\begin{equation} \label{eq:constant-bc}
2\la\lambda\iota\ra o^\gamma a_\gamma + \la o\lambda\ra\iota^\gamma a_\gamma = \lambda^\gamma a_\gamma + \la\lambda\iota\ra o^\gamma a_\gamma = \cO\bigl(\la\lambda\iota\ra^2\bigr) .
\end{equation}
This implies $\iota^\gamma a_\gamma = \cO(\la\lambda\iota\ra)$ as for the na\"{i}ve boundary condition, but $o^\gamma a_\gamma$ can be non-vanishing. Although this appears to depend on the choice of $o$, under a shift $o\to o + m\iota$ the left-hand side changes by $m\la\lambda\iota\ra\iota^\gamma a_\gamma = \cO\bigl(\la\lambda\iota\ra^2\bigr)$ which is consistent with the boundary condition \eqref{eq:constant-bc}.

For the sake of clarity, we also write out the boundary condition in inhomogeneous co-ordinates. Writing $\tilde z = - \la\iota\lambda\ra/\la o\lambda\ra$ and identifying $a_- = o^\gamma a_\gamma$, \smash{$a_y = - \iota^\gamma a_\gamma/\sqrt{2}$} the boundary condition \eqref{eq:constant-bc} tell us that $a_y$ is divisible by $\tilde z$ and $\sqrt{2}\tilde z a_- + a_y$ is divisible by $\tilde z^2$.

To see that this really does eliminate the boundary term, we first exploit the Schouten identity $\eps^{\beta\al} = \iota^\beta o^\al - o^\beta\iota^\al$ to write
\begin{displaymath} \delta S_{\rm kin.}[a] \simeq - \frac{\im}{4\sqrt{2}\pi}\int_{\bbR^3}\dif^3x\oint_{S^1_\eps}\frac{\la\lambda\dif\lambda\ra}{\la\lambda\iota\ra^2} (o^\gamma a_\gamma\iota^\eps\delta a_\eps - \iota^\gamma a_\gamma o^\eps\delta a_\eps ) . \end{displaymath}
We can use the condition \eqref{eq:constant-bc} to eliminate $\iota^\gamma a_\gamma$, $\iota^\eps\delta a_\eps$ since shifts of order $\la\lambda\iota\ra^2$ are harmless. This leaves
\begin{displaymath}
\delta S_{\rm kin.}[a] \simeq \frac{\im}{2\sqrt{2}\pi}\int_{\bbR^3}\dif^3x\oint_{S^1_\eps}\frac{\la\lambda\dif\lambda\ra}{\la o\lambda\ra\la\lambda\iota\ra}o^\gamma o^\eps (a_\gamma\delta a_\eps - a_\gamma\delta a_\eps ) = 0 .
\end{displaymath}
Under a gauge transformation with parameter $\eps$, the left-hand side of equation \eqref{eq:constant-bc} transforms~as
\begin{displaymath}
\bigl(o^\al\iota^\beta - 2\iota^\al o^\beta\bigr)\lambda_\al P_\beta^{~\,\gamma}\lambda^\delta\p_{\gamma\delta}\eps = 2o^\al\lambda^\beta\p_{\al\beta}\eps,
\end{displaymath}
where we recall \smash{$P_\al^{~\,\beta} = \bigl(\iota_\al\lambda^\beta - 2\lambda_\al\iota^\beta\bigr)/\la\lambda\iota\ra$}. Assuming no restrictions on the $x$ dependence of~$\eps$, this can only preserve the boundary condition if \smash{$\eps = \cO\bigl(\la\lambda\iota\ra^2\bigr)$}. This motivates using the boundary condition \smash{$a_0 = \cO\bigl(\la\lambda\iota\ra^2\bigr)$} for the components of $a$ along the fibres of $\PS\to\bbR^3$. \big(The combination $\tilde\sigma^0a_0$ should be non-singular at $\lambda\sim o$, since nothing interesting happens there. This requires a second order anti-holomorphic zero in $a_0$, but this is merely an artefact of the frame.\big) There are no boundary contributions generated when varying the interaction vertex of Poisson--Chern--Simons. This is because both $\p_0$, $4\p_1$ commute with $\la\lambda\iota\ra$ and so are insensitive~to~$\Omega$.

We have seen that the boundary condition \eqref{eq:constant-bc} is valid, but it is still not clear why we should make this particular choice above any other. We will see in Section~\ref{subsec:dKP-Lax} that it correctly reproduces the Lax formulation of the dKP equation and in Section~\ref{subsec:compactify} that the action \eqref{eq:PCS-PS} is equivalent to dKP theory on $\bbR^{1,2}$. If we instead try to impose the na\"{i}ve boundary condition $a_\al = \cO(\la\lambda\iota\ra)$ we find that the linearised equations of motion are not solvable.

\subsection[Non-commutative Chern--Simons on PS]{Non-commutative Chern--Simons on $\boldsymbol{\PS}$} \label{subsec:correspondence-ncCS}

In order to write down the non-commutative theory, we need to quantize the Poisson bracket
\begin{displaymath} \Pi = \p_0\vee\p_1 . \end{displaymath}
This is achieved by applying the Moyal formula, which requires splitting the bivector into the skew product of two commuting derivations of $C^\infty(\PS)$ to get a map
\begin{displaymath} \Pi_{\rm split}\colon\ C^\infty(\PS)\otimes C^\infty(\PS)\to C^\infty(\PS)\otimes C^\infty(\PS) \end{displaymath}
obeying $c\circ\Pi_{\rm split} = \{~ ,~\,\}$. Fortunately $[\p_0,\p_1] = 0$, so we can simply apply the Moyal formula with the na\"{i}ve splitting $\Pi_{\rm split} = \p_0\otimes\p_1 - \p_1\otimes\p_0$ to get
\begin{gather} \label{eq:star}
F\star G = c\circ\exp\frac{\fq}{2}\bigl(\p_0\otimes\p_1 - \p_1\otimes\p_0\bigr)(F\otimes G) = \sum_{m,n\geq 0}\frac{\fq^{m+n}}{m!n!2^{m+n}}(-)^n\p_0^m\p_1^nF\p_1^n\p_0^mG\!
\end{gather}
for $F,G\in C^\infty(\PS)$. This naturally extends to elements of \smash{$\til\Omega^{0,\bullet}(\PS)$} if we understand $\p_0$, $\p_1$ to act by Lie derivative. But there is a catch: the vector field $\p_0$ is not holomorphic (equation~\eqref{eq:dift-p0}). This means that $\dift$ does not distribute over $\star$. Indeed, we show in Appendix~\ref{app:distributor} that
\begin{equation} \label{eq:distributor} \dift(f\star g) = \dif f \star g + (-)^{\mrm{deg}\,f}f\star\dift g + \frac{\fq^2}{4}\dif x^-\wedge\cL_{\p_1}(\cL_{\p_1}f\star \cL_{\p_1}g) . \end{equation}
Mathematically, this means that \smash{$\bigl(\til\Omega^{0,\bullet}(\PS),\dift,\star\bigr)$} is not a dga, and so it cannot be used to define a consistent non-commutative gauge theory.

One method of resolving this issue is to instead split the bivector into commuting holomorphic derivations. We will see in Section~\ref{subsec:holomorphic-split} that this is indeed possible; however, there is an alternative approach. Rather than modifying the star product, we can change the differential. Consider
\begin{equation} \label{eq:Dift} \Dift = \dift - \frac{\fq^2}{12}\dif x^-\wedge\cL_{\p_1}^3 , \end{equation}
Certainly, this is nilpotent
\begin{displaymath} \Dift^2 = \dift^2 - \frac{\fq^2}{12}\bigl[\dift,\dif x^-\wedge\cL_{\p_1}^3\bigr] = 0 , \end{displaymath}
and so defines a differential on \smash{$\til\Omega^{0,\bullet}(\PS)$}. (As above, $[~ ,~]$ denotes the graded commutator.) It also distributes correctly over $\star$, as can be seen by evaluating
\begin{displaymath} \begin{aligned}
\Dift(f\star g)={}& \dift(f\star g) - \frac{\fq^2}{12}\dif x^-\wedge\cL_{\p_1}^3(f \star g) \\
={}&\dift(f\star g) - \frac{\fq^2}{12}\bigl(\dif x^-\wedge\cL_{\p_1}^3f\bigr)\star g- (-)^{\mrm{deg}\,f}\frac{\fq^2}{12}f\star\bigl(\dif x^-\wedge\cL_{\p_1}^3g\bigr)\\
&{-}\, \frac{\fq^2}{4}\dif x^-\wedge\cL_{\p_1}(\cL_{\p_1}f\star \cL_{\p_1}g) .
\end{aligned} \end{displaymath}
Applying the identity \eqref{eq:distributor}, we see the final term cancels leaving
\begin{displaymath} \Dift(f\star g) = \Dift f\star g+ (-)^{\mrm{deg}\,f}f\star\Dift g . \end{displaymath}
This shows that \smash{$\bigl(\til\Omega^{0,\bullet}(\PS),\Dift,\star\bigr)$} is a dga, and therefore that
\begin{align}
S_\star[a] &{}= \frac{\im}{2\pi}\int_\PS\Omega\wedge\biggl(\frac{1}{2}a\star \Dift a + \frac{1}{3\fq}a\star a\star a\biggr)\nonumber\\
&{} = \frac{\im}{2\pi}\int_\PS\Omega\wedge\biggl(\frac{1}{2}a\wedge\Dift a + \frac{1}{6}a\wedge[a,a]_\star\biggr) \label{eq:ncCS-PS}
\end{align}
is a consistent non-commutative gauge theory on $\PS$, i.e., the BV action satisfies the classical master equation. As in Section~\ref{subsec:C2xR}, in the second equality we have used the analogue of the identity~\eqref{eq:pairing} on correspondence space to remove a star product from each term. We continue to use the reality condition $a=S^*a$ and the boundary conditions \eqref{eq:constant-bc} since the new vertices involve only $(1,0)$ derivatives $\p_0$, $\p_1$ which cannot generate any extra boundary contributions.

The equations of motion of the theory \eqref{eq:ncCS-PS} read
\begin{displaymath} \Dift a + \frac{1}{2}[a,a]_\star = 0 , \end{displaymath}
and infinitesimal gauge transformations act by $\delta a = \Dift\eps + [a,\eps]_\star$. As in the case of Poisson--Chern--Simons, away from the locus $\lambda\sim\bar\lambda$ correspondence space fibres over $\mT$ with one-dimensional fibres. We can exploit this to enter a gauge in which the pullback of $a$ to the fibres vanishes and on-shell the remaining components are constant. In this gauge, $a$~is obtained by pullback from $\mT$. Residual gauge transformations are also constant in the fibre direction, and so are pulled back from $\mT$. Hence, at least qualitatively, solutions to the equations of motion of~\eqref{eq:ncCS-PS} describe non-commutative deformations of minitwistor space. Following the proposal of Strachan~\cite{Strachan:1994kp}, it is natural to expect that the theory \eqref{eq:ncCS} describes the KP equation on space-time.

We will verify this explicitly in Section~\ref{sec:off-shell-reduction} by showing that the action \eqref{eq:ncCS-PS} with the boundary conditions \eqref{eq:constant-bc} is equivalent to full KP theory on $\bbR^{1,2}$, where the constant $1/\sigma^2$ controlling the dispersive term is identified with the formal parameter $\fq^2$. Remarkably, we find that the~$3d$ interaction vertex is unmodified, essentially because the higher derivative corrections to the interaction vertex on $\PS$ involve many $\iota$s which contract to zero with one another upon reduction. The modification to the kinetic term through $\Dift$ results in the dispersive term in KP.

\subsection{Splitting the bivector holomorphically} \label{subsec:holomorphic-split}

In the previous subsection,
we quantized the bivector $\Pi$ using the Moyal formula adapted to the na\"{i}ve splitting into $\p_0$, $\p_1$. This gave a star product $\star$ over which $\dift$ failed to distribute, but fortunately we were able to identify a modified differential $\Dift$ which did distribute properly.

Perhaps a more natural approach would have been to split the bivector holomorphically in the first instance. Here we show that doing so results in an equivalent theory on correspondence space. To this end, we seek a modified splitting $\Pi^\prime_{\rm split} = v_0\otimes\p_1 - \p_1\otimes v_0$ where $v_0 = \p_0 + s\p_1$ such that $[v_0,\p_1] = - (\p_1s)\p_1 = 0$ and
\begin{displaymath} \dift v_0 = \dift\p_0 + \dift s\otimes\p_1 = \biggl(2\frac{\tilde\sigma^\al\lambda_\al}{\la\lambda\iota\ra} + \dift s\biggr)\otimes\p_1 = 0 . \end{displaymath}
A candidate is
\begin{displaymath} s = - 2\frac{x^{\al\beta}\iota_\al\lambda_\beta}{\la\lambda\iota\ra} . \end{displaymath}
Then the resulting holomorphic $\ast$ product is
\begin{equation} \label{eq:ast} F*G = c\circ\exp\frac{\fq}{2}\bigl(v_0\otimes\p_1 - \p_1\otimes v_0\bigr)(F\otimes G) . \end{equation}
This extends to $(0,\bullet)$-forms in the usual way, and $\bigl(\til\Omega^{0,\bullet}(\PS),\dift,*\bigr)$ is a dga. We can use this to define a non-commutative Chern--Simons using the now familiar action \eqref{eq:ncCS} and boundary conditions \eqref{eq:constant-bc}.

The cost of splitting the bivector holomorphically is that the associative product $\ast$ manifestly breaks translational invariance on $\bbR^{1,2}$. This seems incompatible with the translational invariance of the KP equation. Furthermore, since the kinetic term of the non-commutative gauge theory coincides with that of Poisson Chern--Simons it is hard to see how a dispersive term could be generated in $3d$. Nevertheless, the dgas $\bigl(\til\Omega^{0,\bullet}(\PS),\Dift,\star\bigr)$ and $\bigl(\til\Omega^{0,\bullet}(\PS),\dift,\ast\bigr)$ are isomorphic and so define equivalent theories on correspondence space.\footnote{They are isomorphic at the chain level, indicating that at least classically the corresponding non-commutative Chern--Simons theories are equivalent off-shell.}

To understand why, let
\begin{displaymath} R = \frac{1}{2}\bigl(\Pi_{\rm split} - \Pi_{\rm split}^\prime\bigr) =
\frac{1}{2} (\p_1\otimes(s\p_1) - (s\p_1)\otimes\p_1 ) . \end{displaymath}
It is crucial here that the tensor products are taken over $\bbC$ rather then $C^\infty(\PS)$, so that $s$ cannot pass from one factor to the other. Composing with the multiplication map $c$ allows us to swap~$s$ between the factors, so that $c\circ R = 0$.

By a quick computation,
\begin{displaymath} [\Pi_{\rm split},R] = (x^-\p_1)\otimes\p_1^2 + \p_1^2\otimes(x^-\p_1) . \end{displaymath}
The object appearing on the right-hand side is built solely from $x^-$, $\p_1$ which commute with~$s$,~$\p_0$ and therefore with both $\Pi_{\rm split}$ and $R$. It can be treated as central.

We can now directly compare the star products $\star$ \eqref{eq:star} and $\ast$ \eqref{eq:ast}. From the Baker--Campbell--Hausdorff formula,
\begin{displaymath} \begin{aligned}
\exp\Bigl(\frac{\fq}{2}\Pi^\prime_{\rm split}\Bigr) &= \exp\Bigl(\frac{\fq}{2}\Pi_{\rm split} - \fq R\Bigr) \\
&= \exp\biggl(- \frac{\fq^2}{4}[\Pi_{\rm split},R]\biggr)\circ\exp(-\fq R)\circ\exp\Bigl(\frac{\fq}{2}\Pi_{\rm split}\Bigr) .
\end{aligned} \end{displaymath}
Now that all instances of $R$ now appear to the left of $\Pi_{\rm split}$ (and recalling that $[\Pi_{\rm split},R]$ is effectively central), we can act with the multiplication map to get
\begin{displaymath} \ast = c\circ\exp\Bigl(\frac{\fq}{2}\Pi^\prime_{\rm split}\Bigr) = c\circ\exp\biggl(-\frac{\fq^2}{4}\bigl((x^-\p_1)\otimes\p_1^2 + \p_1^2\otimes(x^-\p_1)\bigr)\biggr)\circ\exp\Bigl(\frac{\fq}{2}\Pi_{\rm split}\Bigr) . \end{displaymath}
Since $\p_1x^- = 0$, we can commute any instance of $x^-$ in, say, the first factor of the tensor product to the left until it reaches the multiplication symbol. At this stage, it can be safely swapped to the second factor, and commuted to the right past any number of $\p_1$s; therefore, in the above formula any $x^-$ can swap between the two factors of the tensor product. In this situation, we have the identity
\begin{gather*} (x^-\p_1)\otimes\p_1^2 + \p_1^2\otimes(x^-\p_1) = \frac{1}{3}(x^-\otimes 1)(\p_1\otimes 1 + 1\otimes\p_1)^3 - \frac{1}{3}\bigl(x^-\p_1^3\bigr)\!\otimes 1 - \frac{1}{3}1\otimes\!\bigl(x^-\p_1^3\bigr) .
\end{gather*}
All objects built from $x^-4$, $\p_1$ commute with $\Pi_{\rm split}$, so that
\begin{displaymath} \begin{aligned}
\ast ={}& c\circ\exp\biggl(-\frac{\fq^2}{12}(x^-\otimes1)(\p_1\otimes 1 + 1\otimes\p_1)^3\biggr)\circ\exp\Bigl(\frac{\fq}{2}\Pi_{\rm split}\Bigr) \\
&{\circ}\,\biggl(\exp\biggl(\frac{\fq^2}{12}x^-\p_1^3\biggr)\otimes\exp\biggl(\frac{\fq^2}{12}x^-\p_1^3\biggr)\biggr) \\
={}& \exp\biggl(-\frac{\fq^2}{12}x^-\p_1^3\biggr) c\circ\exp\Bigl(\frac{\fq}{2}\Pi_{\rm split}\Bigr)\circ\biggl(\exp\biggl(\frac{\fq^2}{12}x^-\p_1^3\biggr)\otimes\exp\biggl(\frac{\fq^2}{12}x^-\p_1^3\biggr)\biggr).
\end{aligned} \end{displaymath}
Defining $\psi = \exp(\fq^2x^-\p_1^3/12)$ this is $\psi(F\ast G) = (\psi F) \star (\psi G)$. Extending $\psi$ to act on $(0,\bullet)$-forms in the usual way, it is immediate that
\begin{displaymath} \psi\,\dift = \Dift\,\psi , \end{displaymath}
so \smash{$\psi\colon \bigl(\til\Omega^{0,\bullet}(\PS),\dift,\ast\bigr)\to\bigl(\til\Omega^{0,\bullet}(\PS),\Dift,\star\bigr)$} is an isomorphism of dgas.

It is gratifying to find that quantizing the Poisson bivector in a manifestly holomorphic way is consistent with the approach taken in Section~\ref{subsec:correspondence-ncCS}. Nevertheless, it does not explain how a theory on correspondence space which manifestly breaks translation invariance on $\bbR^{1,2}$ can describe the~KP equation. We resolve this issue in Section~\ref{subsec:holomorphic-split-action}, where we observe that the dispersive term in~KP can be eliminated by a field redefinition at the cost of breaking translation invariance. We expect that this is the form of KP theory obtained by directly compactifying non-commutative Chern--Simons using the holomorphic star product \eqref{eq:ast}.

\section{From correspondence space to space-time} \label{sec:off-shell-reduction}

In this section, we show that Poisson--Chern--Simons theory on minitwistor correspondence space describes dKP, and furthermore that non-commutative Chern--Simons on minitwistor correspondence space describes KP.

More precisely, it is long been known that solutions to the dKP equation are in correspondence with minitwistor spaces dressed by suitable holomorphic data and admitting appropriate involutive automorphisms. Furthermore, it has more recently been proposed that solutions to the KP equation are equivalent to certain non-commutative minitwistor spaces. Solutions to the equations of motion of Poisson--Chern--Simons theory and non-commutative Chern--Simons theory engineer minitwistor spaces corresponding to solutions of dKP and KP, and so the on-shell equivalence of the theories is not surprising.

In this section, we show that by partially imposing the equations of motion of Poisson--Chern--Simons theory and non-commutative Chern--Simons it is possible to recover an off-shell scalar field $\phi\in C^\infty\bigl(\bbR^{1,2}\bigr)$. Evaluating the correspondence space actions on this partially on-shell solution gives dKP \eqref{eq:dKPaction} and KP \eqref{eq:KPaction}, respectively. This demonstrates the off-shell equivalence of the theories.

\subsection{Recovering the dKP Lax} \label{subsec:dKP-Lax}

Here we show that gauge fixing the correspondence space field $a$ of Poisson--Chern--Simons and partially imposing the equations of motion recovers a Lax formulation of the dKP equation. Repeating the exercise for non-commutative Chern--Simons yields a Lax for the full KP equation.

To recover the Lax pair, we will mimic the approach adopted in \cite{Bittleston:2020hfv, Costello:2019tri}. The first step is to trivialise $a$ on the real minitwistor lines $\CP^1_x$ for $x\in\bbR^{1,2}$. `Gauge transformations' in Poisson--Chern--Simons are Poisson diffeomorphisms whose infinitesimal counterparts are Hamiltonian vector fields. We write $\eps\in C^\infty(\PS)$ for an infinitesimal gauge parameter, $V_\eps = \{\eps,~\,\}$ for the corresponding Hamiltonian vector field and $f_\eps = \exp V_\eps$ for the Poisson diffeomorphism it generates.

Pulling back $a$ to the minitwistor line $\CP^1_x$ yields $\tilde\sigma^0a_0\in\Omega^{0,1}\bigl(\CP^1_x\bigr)$ with $a_0 = \cO\bigl(\la\lambda\iota\ra^2\bigr)$ at $\lambda\sim\iota$. To trivialise, we seek a Poison diffeomorphism $f_\eps$ such that
\begin{equation} \label{eq:trivialise} V_{a_0} = \{a_0,~\,\} = f_\eps^{-1}\tilde\p_0f_\eps . \end{equation}
We can get an explicit formula for $a_x$ in terms of $\eps$ using the derivative of the exponential
\begin{displaymath} a_0 = \frac{1-{\rm e}^{-V_\eps}}{V_\eps}\tilde\p_0\eps . \end{displaymath}
The Poisson diffeomorphism $f_\eps$ is determined by its action on the co-ordinate functions $\lambda^\al$ and~$x^{\al\beta}$. Defining
\begin{displaymath} \Lambda^\al(x,\lambda) = f_\eps \lambda^\al ,\qquad X^{\al\beta}(x,\lambda) = f_\eps x^{\al\beta} , \end{displaymath}
we can recast equation \eqref{eq:trivialise} as
\begin{displaymath} \tilde\p_0\Lambda^\al = - \iota^\al\p_1a_0(X,\Lambda) ,\qquad \tilde\p_0X^{\al\beta} = \iota^\al\iota^\beta\p_0a_0(X,\Lambda) . \end{displaymath}
All fields and gauge parameters are required to be invariant under simultaneous conjugation and pullback by $S\colon (x,\lambda)\mapsto \bigl(x,\bar\lambda\bigr)$, ensuring that $(X,\Lambda)$ are similarly equivariant. These equations can be solved order by order in $\tilde\sigma^0a_0$, subject to suitable boundary conditions; we will assume the resulting perturbative formula converges. When \smash{$\dbar_{\CP^1_x}$} acts on sections of $\cO(1)$, such as $\Lambda^\al$, it has two-dimensional kernel. We can fix this ambiguity by requiring that \smash{$\Lambda^\al - \lambda^\al = \cO\bigl(\la\lambda\iota\ra^2\bigr)$}. Acting on functions the $\dbar_{\CP^1_x}$ operator has one-dimensional kernel, so it is enough to fix $X^{\al\beta} - x^{\al\beta} = \cO(\la\lambda\iota\ra)$. Together these conditions amount to taking $\eps = \cO(\la\lambda\iota\ra)$. Generically, this does not describe a true gauge transformation, which instead must obey $\eps = \cO\bigl(\la\lambda\iota\ra^2\bigr)$. In fact, the gauge invariant data contained in $a_0$, a scalar field $\phi\in C^\infty\bigl(\bbR^{1,2}\bigr)$, is precisely the leading order term in $\eps$ as we approach $\lambda\sim\iota$:
\begin{equation} \label{eq:gauge-invariant-data}
\phi(x) = \la o\lambda\ra o^\al\p_{\lambda_\al}\eps\big|_{\lambda\sim\iota} .
\end{equation}
Indeed, we are free to make any choice for $\eps$ so long as it vanishes holomorphically to first order on the locus $\lambda\sim\iota$ with derivative \eqref{eq:gauge-invariant-data} but for the moment we will leave it unspecified.

Having fixed the gauge, we next impose the equations of motion in the fibre directions. These are easiest to interpret by writing
\begin{equation} \label{eq:holomorphic-gauge}
\dift + \{a,~\,\} = f_\eps^{-1}\bigl(\dift + \bigl\{a^\prime,~\,\bigr\}\bigr)f_\eps
\end{equation}
for $f_\eps$ as above, $a^\prime$ is the gauge field in the `illegal' gauge $a_0^\prime = 0$. The equations of motion tell us that the components of~$a^\prime$ are meromorphic in $\lambda$ at fixed $x$
\begin{displaymath} \dbar_{\CP^1_x} a^\prime_\al = 0 . \end{displaymath}
But since the gauge $a_0^\prime = 0$ is not attainable with a true gauge transformation, the~$a^\prime_\al$ do not obey our original boundary conditions \eqref{eq:constant-bc} at $\lambda\sim\iota$. Instead they must be chosen so that the~$a_\al$ obey these boundary conditions, which we recall are given by
\begin{displaymath}
2\la\lambda\iota\ra o^\gamma a_\gamma + \la o\lambda\ra\iota^\gamma a_\gamma = \cO\bigl(\la\lambda\iota\ra^2\bigr)
\end{displaymath}
with $a$ finite. Certainly, this requires that $\iota^\gamma a_\gamma = \cO(\la\lambda\iota\ra)$. Inverting equation \eqref{eq:holomorphic-gauge} yields
\begin{displaymath}
a^\prime_\al = a_\al + \frac{1-{\rm e}^{V_\eps}}{V_\eps}\bigl(\tilde\p_\al\eps + \{a_\al,\eps\}\bigr) .
\end{displaymath}
In order to impose the boundary condition, it will be useful to keep track of not merely the piece of order $\la\lambda\iota\ra$ in $\eps$, but also the quadratic term. We write
\begin{displaymath} \eps = \frac{\la\lambda\iota\ra}{\la o\lambda\ra}\phi + \frac{\la\lambda\iota\ra^2}{\la o\lambda\ra^2}\chi + \cO\bigl(\la\lambda\iota\ra^3\bigr) . \end{displaymath}
Then
\begin{displaymath} \begin{aligned}
\iota^\al a^\prime_\al ={}& \iota^\al a_\al - \frac{2\iota^\al\lambda^\beta}{\la\lambda\iota\ra}\p_{\al\beta}\eps - \frac{\iota^\al\lambda^\beta}{\la\lambda\iota\ra}\{\eps,\p_{\al\beta}\eps\} + \cO\bigl(\la\lambda\iota\ra^2\bigr) \\
={}& \iota^\al a_\al - 2\iota^\al\iota^\beta\p_{\al\beta}\phi - \frac{2\la\lambda\iota\ra}{\la o\lambda\ra} o^\al\iota^\beta\p_{\al\beta}\phi - \frac{2\la\lambda\iota\ra}{\la o\lambda\ra}\iota^\al\iota^\beta\p_{\al\beta}\chi\\
&{-}\, \frac{\la\lambda\iota\ra}{\la o\lambda\ra}\bigl\{\phi,\iota^\al\iota^\beta\p_{\al\beta}\phi\bigr\} + \cO\bigl(\la\lambda\iota\ra^2\bigr),
\end{aligned} \end{displaymath}
so that $\iota^\al a_\al^\prime = - 2\iota^\al\iota^\beta\p_{\al\beta}\phi$. Similarly,
\begin{displaymath} \begin{aligned}
o^\al a^\prime_\al &= o^\al a_\al - \frac{o^\al\lambda^\beta}{\la\lambda\iota\ra}\p_{\al\beta}\eps + \frac{\la o\lambda\ra\iota^\al\lambda^\beta}{\la\lambda\iota\ra^2}\p_{\al\beta}\eps
+ \frac{\la o\lambda\ra\iota^\al\lambda^\beta}{2\la\lambda\iota\ra^2}\{\eps,\p_{\al\beta}\eps\} + \cO(\la\lambda\iota\ra) \\
&= o^\al a_\al
+ \frac{\la o\lambda\ra}{\la\lambda\iota\ra}\iota^\al\iota^\beta\p_{\al\beta}\phi
+ \iota^\al\iota^\beta\p_{\al\beta}\chi + \frac{1}{2}\bigl\{\phi,\iota^\al\iota^\beta\p_{\al\beta}\phi\bigr\} + \cO(\la\lambda\iota\ra) .
\end{aligned} \end{displaymath}
Hence
\begin{displaymath} \begin{aligned}
2\la\lambda\iota\ra o^\gamma a_\gamma^\prime
&{}= 2\la\lambda\iota\ra o^\gamma a_\gamma + \la o\lambda\ra\iota^\gamma a_\gamma + 2\la o\lambda\ra\iota^\al\iota^\beta\p_{\al\beta}\phi - 2\la\lambda\iota\ra o^\al\iota^\beta\p_{\al\beta}\phi + \cO\bigl(\la\lambda\iota\ra^2\bigr) \\
&{}= \cO\bigl(\la\lambda\iota\ra^2\bigr) ,
\end{aligned} \end{displaymath}
so we should take
\begin{displaymath} o^\al a^\prime_\al = \frac{\la o\lambda\ra}{\la\lambda\iota\ra}\iota^\al\iota^\beta\p_{\al\beta}\phi - o^\al\iota^\beta\p_{\al\beta}\phi . \end{displaymath}
Putting this together,
\begin{displaymath} a^\prime_\al = - \biggl(\iota_\al o^\beta - 2o_\al\iota^\beta - \frac{\la o\lambda\ra}{\la\lambda\iota\ra}\iota_\al\iota^\beta\biggr)\iota^\gamma\p_{\beta\gamma}\phi = - P_\al^{~\,\beta}\iota^\gamma\p_{\beta\delta}\phi . \end{displaymath}
The Lax can be recovered as the space-time components of the connection $\dift + \{a,~\,\}$ in the gauge $a_0 = 0$. It therefore takes the form
\begin{equation} \label{eq:dKP-Lax}
L^\prime_\al = \tilde\p_\al + \bigl\{a_\al^\prime,~\,\bigr\} = P_\al^{~\,\beta}\frac{\lambda^\gamma}{\la\lambda\iota\ra}\p_{\beta\gamma} + \iota_\al u\,\p_1 + P_\al^{~\,\beta}\iota^\gamma\p_{\beta\gamma}u\,\p_0,
\end{equation}
where we have introduced $u=\p_1\phi$. This is not the standard Lax distribution introduced in equation \eqref{eq:dKP-Lax-dist}; however, the two are related by $L^\prime_\al \sim P_\al^{~\,\beta}L_\beta$ and so generate the same distribution. The Lax operators obtained in equation \eqref{eq:dKP-Lax} have a couple of attractive features. First, by construction they preserve the volume form $\Omega$. Second, they are a true Lax pair, rather than a Lax distribution, for the dKP equation. Indeed, expressed in terms of inhomogeneous co-ordinates they take the form
\begin{gather}
L^\prime_0 = \p_- - z^2\p_+ + u\p_+ - \frac{1}{\sqrt{2}}\p_yu\p_z - z\p_+u\p_z ,\nonumber\\
 L^\prime_1 = 2\biggl( - \frac{1}{\sqrt{2}}\p_y + z\p_+ + \p_+u\p_z\biggr) . \label{eq:dKP-Lax-inhomogeneous}
\end{gather}
The commutator of these two operators is
\begin{displaymath} [L^\prime_0,L^\prime_1] \sim 2\biggl(\p_+\p_-u - \frac{1}{2}\p_y^2u + \p_+(u\p_+u)\biggr)\p_z , \end{displaymath}
vanishing precisely on the support of the dKP equation. This justifies our choice of boundary conditions in Section~\ref{subsec:bcs}.

\subsection{Recovering the KP Lax} \label{subsec:KP-Lax}

The argument of the previous subsection carries over the case of the KP equation with surprisingly little difficulty.

In this case, we have non-commutative 5d Chern--Simons theory on correspondence space. An~infinitesimal gauge transformation with parameter $\eps\in C^\infty(\PS)$ (where as usual we suppress the dependence on the formal variable $\fq$) acts by $a\mapsto \dift\eps + [a,\eps]_\star$. This exponentiates to the finite transformation
\begin{equation} \label{eq:finite-nc-gauge}
a\mapsto a + \frac{1 - {\rm e}^{-[\eps,~\,]_\star}}{[\eps,~\,]_\star}\bigl(\Dift\eps + [a,\eps]_\star\bigr) .
\end{equation}
We can understand this in a more intuitive way by forming the finite gauge parameter
\begin{displaymath} f_\eps = \exp_\star\biggl(\frac{\eps}{\fq}\biggr) = \sum_{m=0}^\infty\frac{1}{m!\fq^m}\underbrace{\eps\star\dots\star\eps}_{m~\text{times}} . \end{displaymath}
Note that at any given order in $\fq$ (which can now be negative) infinitely many terms can contribute; however at any fixed power of $\eps$ and given order in $\fq$ there are only finitely many terms. We will assume the sums over increasing powers of $\eps$ converge. Then non-commutative gauge transformations take the more familiar form
\begin{equation} \label{eq:Abelian-nc-gauge} \Dift + \frac{1}{\fq}a\mapsto f_{-\eps}\star\Dift f_\eps + \frac{1}{\fq}f_{-\eps}\star a\star f_\eps . \end{equation}
Here $f_{-\eps} = \exp_\star(-\eps/\fq)$ should be interpreted as the `inverse' of $f_\eps$, obeying $f_{-\eps}\star f_\eps = f_\eps\star f_{-\eps} = 1$. Although the right-hand side of \eqref{eq:Abelian-nc-gauge} seems to be terribly singular in $\fq$, applying the formula for the derivative of the exponential recovers the transformation \eqref{eq:finite-nc-gauge}.\footnote{This perspective can be made robust if normalize the cubic vertex of non-commutative $5d$ Chern--Simons to be $a\star a\star a$ as in \cite{Costello:2016nkh}. We choose not to do this to get a cubic vertex which is not formal.}

Pulling back $a$ to the minitwistor line $\CP^1_x$ as before, we trivialise with $f_\eps = \exp_\star\eps$ obeying%
\begin{displaymath}
a_0 = \fq f_{-\eps}\star\tilde\p_0f_\eps .
\end{displaymath}
Without loss of generality, we may take $\eps = \cO(\la\lambda\iota\ra)$, but true gauge transformations obey $\eps = \cO\bigl(\la\lambda\iota\ra^2\bigr)$. The gauge invariant data contained in $a_0$ is the leading order term in $\eps$ as we approach $\lambda\sim\iota$
\begin{displaymath}
\phi(x) = \la o\iota\ra o^\al\p_{\lambda_\al}\eps\big|_{\lambda\sim\iota}\in C^\infty\bigl(\bbR^{1,2}\bigr) .
\end{displaymath}
We could now make any choice for $\eps$ as long as it vanishes to first order on the locus $\lambda\sim\iota$ with the above derivative, but we will leave it unspecified.

Having fixed the gauge, we next impose the equations of motion in the fibre directions. This is achieved by writing
\begin{displaymath}
\Dift + \frac{1}{\fq}a = f_{-\eps}\star\Dift f_\eps + \frac{1}{\fq}f_{-\eps}\star a^\prime\star f_\eps .
\end{displaymath}
The equations of motion tell us that $a^\prime$ is a meromorphic function of $\lambda$, which we can determine using the boundary conditions on $a$. Fortunately $\Dift$ and $[~ ,~\,]_\star$ differ from $\dift$ and $\{~ ,~\,\}$ respectively by terms of order $\la\lambda\iota\ra^3$. Since our boundary conditions only constrain the components of $a$ at order $\la\lambda\iota\ra$, the analysis of Section~\ref{subsec:dKP-Lax} goes through without modification and we learn that
\begin{displaymath} a^\prime_\al = - P_\al^{~\,\beta}\iota^\gamma\p_{\beta\gamma}\phi . \end{displaymath}
Expressed in terms of inhomogeneous co-ordinates,
\begin{gather}
\mathscr{L}_0^\prime = o^\al\biggl(\tilde D_\al + \frac{1}{\fq}a^\prime_\al\biggr) = \p_- - z^2\p_+ - \frac{\fq^2}{12}\p_+^3 + \frac{1}{\sqrt{2}\fq}\p_y\phi + \frac{z}{\fq}\p_+\phi , \nonumber\\
\mathscr{L}_1^\prime = \iota^\al\biggl(\tilde D_\al + \frac{1}{\fq}a^\prime_\al\biggr) = 2\biggl( - \frac{1}{\sqrt{2}}\p_y + z\p_+ - \frac{1}{\fq}\p_+\phi\biggr) .\label{eq:KP-Lax}
\end{gather}
The Lax equation takes the form
\begin{displaymath} \begin{aligned}
\bigl[\scrL_0^\prime,\scrL_1^\prime\bigr]_\star &{}= o^\al\tilde D_\al\bigl(\iota^\al a^\prime_\al\bigr) - \iota^\al\tilde D_\al\bigl(o^\al a^\prime_\al\bigr) + \bigl[o^\al a^\prime_\al,\iota^\al a^\prime_\al\bigr]_\star \\
&{}= - 2\biggl(\p_+\p_-\phi - \frac{1}{2}\p_y^2\phi - \frac{\fq^2}{12}\p_+^4\phi + \frac{1}{2}\p_+(\p_+\phi)^2\biggr) = 0 .
\end{aligned} \end{displaymath}
Differentiating with respect to $x^+$ and identifying $u = \p_+\phi$ yields the full KP equation \big(albeit with formal dispersion parameter $\fq^2$ playing the role of $1/\sigma^2$\big). This is not a Lax pair in the usual sense, since it takes the form of an Abelian connection on non-commutative correspondence space rather than a pair of poly-differential operators. We can construct a pair of poly-differential operators commuting on the support of the KP equation from \eqref{eq:KP-Lax} by forming the combinations
\begin{gather*}
o^\al\bigl(\tilde D_\al + \bigl[a^\prime_\al,~\,\bigr]_\star\bigr) = \p_- - z^2\p_+ - \frac{\fq^2}{12}\p_+^3 + \sum_{m\geq0}\frac{\fq^{2m}}{(2m)!2^{2m}}\p_+^{2m}u\p_z^{2m}\p_+ \\
\hphantom{o^\al\bigl(\tilde D_\al + \bigl[a^\prime_\al,~\,\bigr]_\star\bigr) =}{}
- \frac{1}{\sqrt{2}}\sum_{m\geq0}\frac{\fq^{2m}}{(2m+1)!2^{2m}}\p_+^{2m}\bigl(\p_yu + \sqrt{2}z\p_+u\bigr)\p_z^{2m+1} , \\
\iota^\al\bigl(\tilde D_\al + \bigl[a^\prime_\al,~\,\bigr]_\star\bigr) = 2\biggl( - \frac{1}{\sqrt{2}}\p_y + z\p_+ + \sum_{m\geq0}\frac{\fq^{2m}}{(2m+1)!2^{2m}}\p_+^{2m+1}u\p_z^{2m+1}\biggr) .
\end{gather*}
Note that these depend on $\phi$ only through $u$, and in the limit $\fq\to0$ they reproduce the Lax operators for dKP appearing in equation \eqref{eq:dKP-Lax-inhomogeneous}. Although these poly-differential operators are of arbitrary degree in $\p_z$, when acting on holomorphic sections of $\cO(n)$ they will truncate. A~similar non-commutative Lax equation was identified by Strachan in \cite{Strachan:1994tx}, though it does not appear to coincide with the expression we find here.

\subsection{Reducing the action} \label{subsec:compactify}

In this subsection, we will compactify the non-commutative Chern--Simons theory engineered in Section~\ref{subsec:correspondence-ncCS} from $\PS$ to $\bbR^{1,2}$. Recall that its action is
\begin{equation} \label{eq:noncomm*}
S_\star[a] = \frac{\im}{2\pi}\int_\PS\Omega\wedge\biggl(\frac{1}{2}a \star \tilde{\Dif} a + \frac{1}{3\fq}a \star a \star a\biggr) ,
\end{equation}
where
\begin{displaymath}
\Omega = \frac{\la\lambda\dif\lambda\ra\wedge\dif x^{\al\beta}\lambda_\al\lambda_\beta}{\langle\lambda\iota\rangle^4}
\end{displaymath}
and the star product is defined by splitting the bivector using the constant frame
\begin{displaymath}
\p_0 = \la\lambda\iota\ra\iota_\al\frac{\p}{\p\lambda_\al} ,\qquad\p_1 = \iota^\al\iota^\beta \frac{\p}{\p x^{\al\beta}} = \p_+ .
\end{displaymath}
The equations of motion for \eqref{eq:noncomm*} read
\begin{displaymath}
\Omega\wedge\biggl(\tilde{\Dif} a + \frac{1}{2}[a,a]_\star\biggr) = 0 ,
\end{displaymath}
where $\tilde{\Dif} a = \tilde{\dif} a - \fq^2\dif x^-\wedge\cL_{\p_1}^3a/12$ and
\begin{displaymath}
\tilde\dif = \dbar_{\CP^1}\big|_x + \tilde\sigma^\al\tilde\p_\al = \dbar_{\CP^1}\big|_x + \dif x^{\al\beta}\iota_\al\biggl(\iota_\beta o^\gamma - 2o_\beta\iota^\gamma - \frac{\la o\lambda\ra}{\la\lambda\iota\ra}\iota_\beta\iota^\gamma\biggr)\frac{\lambda^\delta}{\la\lambda\iota\ra}\frac{\p}{\p x^{\gamma\delta}} .
\end{displaymath}
It will be useful to note that
\begin{displaymath}
P_\al^{~\,\beta} = \iota_\alpha o^\beta - 2o_\alpha\iota^\beta - \frac{\la o\lambda\ra}{\la\lambda\iota\ra}\iota_\alpha\iota^\beta
\end{displaymath}
obeys \smash{$P^{\al\beta}P_\al^{~\,\gamma} = 2\epsilon^{\gamma\beta}$} as well as \smash{$\iota^\al P_\al^{~\,\beta} = 2\iota^\beta$} and \smash{$o^\al P_\al^{~\,\beta} = o^\beta - \la o\lambda\ra\iota^\beta/\la\lambda\iota\ra = 2o^\beta - \lambda^\beta/\la\lambda\iota\ra$}.

To reduce to space-time, we will solve the mixed $\tilde\sigma^0\,\tilde\sigma^\alpha$-components of the equations of motion:
\begin{displaymath}
\Omega\wedge\tilde\sigma^\al\wedge\biggl(\tilde{\Dif} a + \frac{1}{2}[a,a]_\star\biggr) = 0 .
\end{displaymath}
Our strategy is to work perturbatively. We introduce
a parameter $\delta$ and write
\begin{displaymath} a(\delta) = \sum_{k=1}^\infty a^{(k)}\delta^k , \end{displaymath}
where our original field is $a = a(1)$. The coefficient field $a^{(k)}$ solves an equation of order $k$ whose mixed $0\alpha$-component is given by
\begin{equation} \begin{aligned} \label{eq:0alpha-eom}
&\Omega\wedge\tilde\sigma^\alpha\wedge\Dift a^{(1)} = 0 ,\qquad \Omega\wedge\tilde\sigma^\al\wedge\biggl(\tilde{\Dif} a^{(2)} + \frac{1}{2}\bigl[a^{(1)},a^{(1)}\bigr]_\star\biggr) = 0 , \\
&\Omega\wedge\tilde\sigma^\al\wedge\bigl(\Dift a^{(3)} + \bigl[a^{(1)},a^{(2)}\bigr]_\star\bigr) = 0
\end{aligned} \end{equation}
and so on.

As usual for reductions to space-time, we gauge fix the pullback of $a$ to the $\CP^1_x$ fibres of correspondence space over $\bbR^{1,2}$. We cannot set it to zero, since gauge transformations must vanish to second order on the divisor, but we can instead fix
\begin{displaymath} a_{\CP^1_x}(\delta) = \la\lambda\iota\ra^2\phi(x)\tilde\rho^0\delta \end{displaymath}
for some function $\phi(x)$ depending only on space-time. Here we have chosen a Fubini--Study metric on $\CP^1$ inducing an anti-holomorphic antipodal map $\lambda\mapsto\hat\lambda$ with $\hat o = \iota$. \smash{$\tilde\rho^0 = \bigl\la\hat\lambda\dif\hat\lambda\bigr\ra/\bigl\la\lambda\hat\lambda\bigr\ra^2$} is the harmonic representative of $H^1\bigl(\CP^1,\cO(-2)\bigr)$ for this metric.\footnote{In the notation of Section~\ref{subsec:KP-Lax}, this amounts to writing $a_{\CP^1_x}(\delta)$ in `pure gauge' for an illegal finite gauge transformation $f_\eps = \exp_\star\eps$ with \smash{$\eps = \phi(x)\bigl\la\lambda\iota\bigr\ra\bigl\la\iota\hat\lambda\bigr\ra \delta/\bigl\la\lambda\hat\lambda\bigr\ra + \cO\bigl(\delta^2\bigr)$}. The higher-order terms $\delta$ are at least \smash{$\cO\bigl(\la\lambda\iota\ra^2\bigr)$}, so that \smash{$\la o\lambda\ra o^\al\p_{\lambda_\al}\eps|_{\lambda\sim\iota} = \phi(x)\delta$} as required.} We emphasize that we have fixed the \emph{exact} form of \smash{$a_{\CP^1_x}(\delta)$}; this component is purely linear in $\delta$. This leads to many simplifications that are key to the success of this perturbative approach.

Firstly, when we insert the solutions to the mixed $0\alpha$-components of the equations of motion back into the action, all contributions to the kinetic terms of the form
\begin{displaymath} \frac{\im}{2\pi}\int_\PS\Omega\wedge\biggl(\frac{1}{2}a^{(1)}\wedge\tilde{\Dif}a^{(k)} + \frac{1}{2}a^{(k)}\wedge\tilde{\Dif}a^{(1)}\biggr) = \frac{\im}{2\pi}\int_\PS\Omega\wedge a^{(k)}\wedge\tilde{\Dif}a^{(1)} \end{displaymath}
must vanish when $k>1$. This is because $a^{(k)}$ has no $\tilde\rho^0$ component when $k>1$, so this term is proportional to the leading $0\alpha$ equation of motion. In fact, remarkably the only terms in the full non-commutative theory $S_\star[a]$ which can contribute to the space-time action are the lowest-order kinetic and Poisson terms
\begin{equation} \label{eq:contributing-action} \frac{\im}{2\pi}\int_\PS\Omega\wedge\biggl(\frac{1}{2}a^{(1)}\wedge\Dift a^{(1)} + \frac{1}{6}a^{(1)}\wedge\bigl\{a^{(1)},a^{(1)}\bigr\}\biggr) . \end{equation}
Higher terms drop out because they contain too many powers of $\iota_\al$ which must contract with each other when we integrate over the $\CP^1$ fibres of $\PS\to\bbR^{1,2}$.

To understand this, let us use the notation $\cO\sim [p,q]$ to indicate that an object $\cO$ involves~$p$ instances of $\iota$ and $q$ space-time derivatives. $q$ is \emph{not} the mass dimension under scaling, it simply counts derivatives. For example, our gauge fixing gives $a_{\CP_x^1} = \la\lambda\iota\ra^2\phi(x)\tilde\rho^0\delta\sim[2,0]$ while \smash{$\dbar_{\CP^1}\big|_x\sim [0,0]$}. We declare that the non-commutativity parameter $\fq\sim[-4,-1]$ so as to ensure that $\tilde\sigma^\al\tilde D_\alpha$ scales homogeneously as $\sim [0,1]$. Then equation \eqref{eq:0alpha-eom} shows that for $k>1$ we have \smash{$a^{(k)} = \tilde\sigma^\al a^{(k)}_\al\sim[6k-4,k-1]$}. Now consider the perturbative expansion \smash{$S_\star[a(\delta)] = \sum_{m=2}^\infty S^{(m)}\delta^m$} of the action evaluated on the solutions to \eqref{eq:0alpha-eom}. Recalling that the holomorphic $2$-form $\Omega\sim[-4,0]$, we have $S^{(m)}\sim[6(m-2),m]$. Finally, expanding in powers of the non-commutativity parameter as $S^{(m)}= \sum_{n\geq0}S^{(m,n)}\fq^{2n}$ the coefficients $S^{(m,n)}\sim[6m+8n-12,m+2n]$. Because $\la\iota\iota\ra=0$, when we integrate out the $\CP^1$ each power of $\iota$ must end up being contracted with a spatial derivative $\p_{\beta\gamma}$, and each such derivative can carry at most two $\iota$s. Hence to get a non-vanishing contribution to the space-time action, we need $m+n\leq3$ so that the only possible contributing terms have $(m,n) = (2,0), (3,0), (2,1)$. Of these cases, $(2,0)$ and $(2,1)$ correspond to the kinetic term in~\eqref{eq:contributing-action}, while $(3,0)$ corresponds to the vertex. Let us now evaluate these terms.

To leading order, the partial equations of motion $\Omega\,\tilde\sigma^\al\,\Dift a^{(1)}=0$ give
\begin{displaymath} \tilde\p_0 a_\al^{(1)} = \la\lambda\iota\ra^2\tilde D_\alpha\phi
= \la\lambda\iota\ra^2\biggl(\frac{P_\al^{\ \beta}\lambda^\gamma}{\la\lambda\iota\ra}\p_{\beta\gamma}\phi - \iota_\al\frac{\fq^2}{12}\p_+^3\phi\biggr) , \end{displaymath}
where we have used the definitions \eqref{eq:Dift} and \eqref{eq:dift-const} of $\tilde{\Dif}$ and $\tilde{\p}_\al$. Solving this, we find that the leading term
\begin{displaymath}
a^{(1)} = \la\lambda\iota\ra^2\phi\,\tilde\rho^0 - \frac{\la\lambda\iota\ra}{\bigl\la\lambda\hat\lambda\bigr\ra}\biggl(P_\al^{~\,\beta}\hat\lambda^\gamma\p_{\beta\gamma}\phi - \iota_\al\frac{\fq^2\bigl\la\hat\lambda\iota\bigr\ra}{12}\p_+^3\phi\biggr)\tilde\sigma^\al .
\end{displaymath}
It is easy to check that this solution obeys the boundary conditions required in the $\bigl\{\tilde\sigma^0,\tilde\sigma^\al\bigr\}$ frame introduced in Section~\ref{subsec:bcs}; furthermore, had we instead tried to impose the stronger boundary condition $a_\al = \cO(\la\lambda\iota\ra)$ then we would have been unable to solve the linearised equation of motion.

The quadratic term in the space-time action comes from inserting this linearised solution into the quadratic term in~\eqref{eq:contributing-action}. This gives
\begin{align*}
&\frac{\im}{4\pi}\int_\PS\Omega\wedge a^{(1)}\wedge\Dift a^{(1)} = \frac{\im}{2\pi}\int_\PS\Omega\wedge\tilde\rho^0\wedge\tilde\sigma^\al\wedge\tilde\sigma^\beta\,\la\lambda\iota\ra^2\phi\tilde D_\al a^{(1)}_\beta \\
&= \frac{1}{\sqrt{2}}\biggl(\frac{\im}{2\pi}\biggr)\int_\PS\omega\wedge\dif^3x\,\frac{\la\lambda\iota\ra}{\bigl\la\lambda\hat\lambda\bigr\ra}\biggl(\frac{P^{\al\beta}\lambda^\gamma}{\la\lambda\iota\ra}\p_{\beta\gamma}\phi - \iota^\al\frac{\fq^2}{12}\p_+^3\phi\biggr)\biggl(P_\al^{~\,\delta}\hat\lambda^\epsilon\p_{\delta\epsilon}\phi - \iota_\al\frac{\fq^2\bigl\la\hat\lambda\iota\bigr\ra}{12}\p_+^3\phi\biggr) \\
&= \frac{1}{\sqrt{2}}\biggl(\frac{\im}{2\pi}\biggr)\int_\PS\omega\wedge\dif^3x \biggl(P^{\al\beta}P_\al^{~\,\delta}\frac{\lambda^\gamma\hat\lambda^\epsilon}{\bigl\la\lambda\hat\lambda\bigr\ra}\p_{\beta\gamma}\phi\p_{\delta\epsilon}\phi - \frac{\fq^2}{12}\p_+^3\phi P^{\al\beta}\iota_\al\frac{\bigl\la\hat\lambda\iota\bigr\ra\lambda^\gamma -\la\lambda\iota\ra\hat\lambda^\gamma}{\bigl\la\lambda\hat\lambda\bigr\ra}\p_{\beta\gamma}\phi\biggr) \\
&= \frac{1}{\sqrt{2}}\biggl(\frac{\im}{2\pi}\biggr)\int_\PS\omega\wedge\dif^3x \biggl(\frac{1}{2}\p^{\al\beta}\phi\p_{\al\beta}\phi - \frac{\fq^2}{12}\p_+^3\phi\p_+\phi\biggr) .
\end{align*}
In going to the second line, we used the fact that the top form
\begin{displaymath} \Omega\wedge\tilde\rho^0\wedge\tilde\sigma^\beta\wedge\tilde\sigma_\beta = \frac{\la\lambda\dif\lambda\ra\wedge\dif x^{\al\beta}\lambda_\al\lambda_\beta}{\la\lambda\iota\ra^4}\wedge\tilde\rho^0\wedge\tilde\sigma^\beta\wedge\tilde\sigma_\beta = \frac{\sqrt{2}\,\omega\wedge\dif^3 x}{\la\lambda\iota\ra^2} , \end{displaymath}
where \smash{$\omega = \la\lambda\dif\lambda\ra\wedge\bigl\la\hat\lambda\dif\hat\lambda\bigr\ra/\bigl\la\lambda\hat\lambda\bigr\ra^2$} is the usual K\"{a}hler form on $\CP^1$. We can now perform the integral over $\CP^1$ using
\begin{displaymath} \frac{\im}{2\pi}\int_{\CP^1}\omega = 1 \end{displaymath}
to get
\begin{equation} \label{eq:kineticsol1} \frac{1}{\sqrt{2}}\int_{\bbR^{1,2}}\dif^3x \biggl(\p_+\phi\p_-\phi - \frac{1}{2}\phi_y^2 + \frac{\fq^2}{12}\bigl(\p_+^2\phi\bigr)^2\biggr) . \end{equation}
Equation \eqref{eq:kineticsol1} is the kinetic term of the KP action, including the dispersive term.

For the vertex, we also need to compute
\begin{displaymath} \begin{aligned}
&\cL_{\p_0}a^{(1)} = 2\frac{\la\lambda\iota\ra^3\bigl\la\hat\lambda\iota\bigr\ra}{\bigl\la\lambda\hat\lambda\bigr\ra}\phi\,\tilde\rho^0 - \frac{\la\lambda\iota\ra}{\bigl\la\lambda\hat\lambda\bigr\ra}\biggl(\frac{\la\lambda\iota\ra\bigl\la\hat\lambda\iota\bigr\ra}{\bigl\la\lambda\hat\lambda\bigr\ra}P_\al^{~\,\beta} - \iota_\al\iota^\beta\biggr)\hat\lambda^\gamma\p_{\beta\gamma}\phi\,\tilde\sigma^\al + \cO\bigl(\fq^2\bigr) ,\\
&\cL_{\p_1}a^{(1)} = \la\lambda\iota\ra^2\p_+\phi\,\tilde\rho^0 - \frac{\la\lambda\iota\ra}{\bigl\la\lambda\hat\lambda\bigr\ra}P_\al^{~\,\beta}\hat\lambda^\gamma\p_{\beta\gamma}\p_+\phi\,\tilde\sigma^\al + \cO\bigl(\fq^2\bigr) ,
\end{aligned} \end{displaymath}
where we can discard terms of order $\fq^2$. In fact, we can also discard the terms involving $\bigl\la\hat\lambda\iota\bigr\ra$ as these will also lead to factors of $\la\iota\iota\ra = 0$ upon integration over $\lambda$. Thus the vertex comes from
\begin{displaymath} \begin{aligned}
&\frac{1}{3}\biggl(\frac{\im}{2\pi}\biggr)\int_\PS\Omega\wedge a^{(1)}\wedge\cL_{\p_0} a^{(1)}\wedge\cL_{\p_1}a^{(1)} = \frac{1}{\sqrt{2}}\biggl(\frac{\im}{2\pi}\biggr)\int_\PS\omega\wedge\dif^3x\,\phi\bigl(\cL_{\p_0}a^{(1)}\bigr)^\al\bigl(\cL_{\p_1}a^{(1)}\bigr)_\al \\
&\qquad{}= - \frac{1}{\sqrt{2}}\biggl(\frac{\im}{2\pi}\biggr)\int_\PS\omega\wedge\dif^3x \,
\frac{\la\lambda\iota\ra^2}{\bigl\la\lambda\hat\lambda\bigr\ra^2}\phi \bigl(\iota^\al\hat\lambda^\beta\p_{\al\beta}\phi\bigr)\bigl(\iota^\gamma P_\gamma^{~\,\delta}\hat\lambda^\eps\p_{\delta\epsilon}\p_+\phi\bigr) \\
&\qquad{}= - \sqrt{2}\biggl(\frac{\im}{2\pi}\biggr)\int_\PS\omega\wedge\dif^3x\,\frac{\la\lambda\iota\ra^2}{\bigl\la\lambda\hat\lambda\bigr\ra^2}\,\phi \bigl(\iota^\al\hat\lambda^\beta\p_{\beta\gamma}\phi\bigr)\bigl(\iota^\gamma\hat\lambda^\delta\p_{\gamma\delta}\p_+\phi\bigr) .
\end{aligned} \end{displaymath}
This can be readily evaluated using
\begin{displaymath}\frac{\im}{2\pi}\int_{\CP^1}\omega\,\frac{\lambda^\al\lambda^\beta\hat\lambda^\gamma\hat\lambda^\delta}{\bigl\la\lambda\hat\lambda\bigr\ra^2} = - \frac{1}{3}\eps^{\al(\gamma}\eps^{\delta)\beta} \end{displaymath}
to give
\begin{displaymath} \frac{1}{\sqrt{2}}\int_{\bbR^{1,2}}\dif^3x\,\frac{1}{3}(\p_+\phi)^3 . \end{displaymath}
This is the vertex from the KP action \eqref{eq:KPaction} on $\bbR^{1,2}$, so altogether non-commutative Chern--Simons descends to
\begin{equation} \label{eq:KPaction-again} S_{\rm KP}[\phi] = \int_{\bbR^{1,2}}\frac{\dif^3x}{\sqrt{2}}\biggl(\p_+\phi\p_-\phi - \frac{1}{2}\p_y\phi^2 + \frac{\fq^2}{12}\bigl(\p_+^2\phi\bigr)^2 + \frac{1}{3}(\p_+\phi)^3\biggr) . \end{equation}
Comparing with equation \eqref{eq:KPaction}, we recognise this as the action of KP theory with formal dispersion parameter $\fq^2$ playing the role of $1/\sigma^2$. \big(At least, up to an irrelevant overall factor of~$1/\sqrt{2}$.\big) We emphasise that this is the \emph{complete} reduction of the full, non-commutative action~$S_\star[a]$ on~$\PS$ at $\delta=1$, to all orders in $\fq$. Any potential higher-order terms vanish on reduction. Thus we have shown that, when the mixed $0\alpha$-components of the equations of motion are imposed,~${S_\star[a]=S_{\rm KP}[\phi]}$.

\subsection{Holomorphically split bivector on space-time} \label{subsec:holomorphic-split-action}

In Section~\ref{subsec:holomorphic-split}, we saw that instead of naively quantizing the Poisson bivector on correspondence space at the cost of deforming the differential, we could instead quantize using a manifestly holomorphic splitting of the bivector. The cost is that translation invariance is explicitly broken. We will not directly compactify this theory to space-time, but will identify a natural candidate for the resulting $3d$ action.

To achieve this, we will perform a linear field redefinition eliminating the quartic vertex in the action \eqref{eq:KPaction-again}. The transformation
\begin{equation} \label{eq:eliminate-dispersion} \tilde\phi = \exp\biggl( - \frac{\fq^2}{12}x^-\p_+^3\biggr)\phi \end{equation}
does the trick, since
\begin{displaymath} \biggl(\p_+\p_- - \frac{1}{2}\p_y^2\biggr)\tilde\phi = \exp\biggl( - \frac{\fq^2}{12}x^-\p_+^3\biggr)\biggl(\p_+\p_- - \frac{1}{2}\p_y^2 - \frac{\fq^2}{12}\p_+^4\biggr)\phi . \end{displaymath}
Notice that the field redefinition \eqref{eq:eliminate-dispersion} is takes an almost identical form to the isomorphism~$\psi^{-1}$ from Section~\ref{subsec:holomorphic-split}. Discarding boundary terms, we also have the crucial identity
\begin{displaymath} \int_{\bbR^{1,2}}\dif^3x\,\tilde f\tilde g = \int_{\bbR^{1,2}}\dif^3x\,f g . \end{displaymath}
Together these can be used to rewrite the action as
\begin{align*}
&S_{\rm KP}[\phi] \propto \int_{\bbR^{1,2}}\dif^3x \biggl(\p_+\tilde\phi\p_-\tilde\phi - \frac{1}{2}\p_y\tilde\phi^2 + \frac{1}{3}\biggl(\exp\biggl(\frac{\fq^2}{12}x^-\p_+^3\biggr)\p_+\tilde\phi\biggr)^3\biggr) \\
&\quad{}= \int_{\bbR^{1,2}}\dif^3x\biggl(\p_+\tilde\phi\p_-\tilde\phi - \frac{1}{2}\p_y\tilde\phi^2 + \frac{1}{3}\sum_{p,q,r\geq0}\biggl(\frac{\fq^2}{12}\biggr)^{p+q+r}\frac{(x^-)^{p+q+r}}{p!q!r!}\p_+^{1+3p}\tilde\phi\p_+^{1+3q}\tilde\phi\p_+^{1+3r}\tilde\phi\biggr) .
\end{align*}
Viewed as a Lagrangian for $\tilde\phi$, we expect that this is the result of compactifying non-commutative Chern--Simons with the holomorphic splitting: It has canonical kinetic term, as would be expected if the differential on correspondence space is unmodified. The price is an infinite tower of cubic vertices and explicit violation of translational invariance in $x^-$ (which plays the role of time in the KP equation). The cubic vertex can be somewhat simplified by integrating by parts to remove all derivatives from one of the $\tilde\phi$s. The result is
\begin{displaymath}
\int_{\bbR^{1,2}}\dif^3x \biggl(\p_+\tilde\phi\p_-\tilde\phi - \frac{1}{2}\p_y\tilde\phi^2 - \frac{1}{3}\tilde\phi\sum_{p,q\geq0}\biggl(-\frac{\fq^2}{4}\biggr)^{p+q}\frac{(x^-)^{p+q}}{p!q!}\p_+\bigl(\p_+^{1+2p+q}\tilde\phi\p_+^{1+p+2q}\tilde\phi\bigr)\biggr) . \end{displaymath}

\section{Amplitudes in KP theory} \label{sec:amplitudes}

A general argument of Costello \cite{Costello:2022wso} shows that any local holomorphic theory on twistor space should correspond to an integrable theory on $\bbR^4$. One consequence of this is that all amplitudes should vanish (at least for general kinematics). This is guaranteed by the Coleman--Mandula theorem: the infinitely many conserved charges of an integrable theory necessitate a perturbatively trivial S-matrix.

The arguments of \cite{Costello:2022wso} can also be applied to mixed topological-holomorphic theory on $\PS$. In particular, scalar scattering states $\phi_k(x) = \exp(\im k\cdot x)$ in $\bbR^{1,2}$ can be realised via the Penrose transform by cohomology representatives
\begin{displaymath} {\rm e}^{\im\omega\mu/\la\lambda\iota\ra^2}\bar\delta_{\lambda\sim\kappa} \in H^1(\mT;\cO(-2)) , \end{displaymath}
where $k_{\al\beta} = \omega\kappa_\al\kappa_\beta$. Pulling back these states to $\PS$ and dressing them with the divisor generates solutions to the linearised equations of motion of non-commutative and Poisson--Chern--Simons~theory
\begin{displaymath} a_k = \la\lambda\iota\ra^2 {\rm e}^{\im\omega x^{\al\beta}\lambda_\al\lambda_\beta/\la\lambda\iota\ra^2}\bar\delta_{\lambda\sim\kappa} . \end{displaymath}
Crucially, such states are localised at points on the $\CP^1$ base of $\mT$. When computing amplitudes in a mixed topological-holomorphic theory it is natural to fix the gauge using a metric. In our case we use a linear combination of the flat metric $\eta$ on $\bbR^{1,2}$ and the Fubini--Study metric on the $\CP^1$ fibres. Any gauge invariant quantity will be independent of the metric, which includes the S-matrix. We can therefore choose to scale up the Fubini--Study metric on $\CP^1$, suppressing any interactions between states supported at different values of $\lambda$. For generic kinematics, all external particles in a scattering process will be supported at different values of $\lambda\sim\kappa_i$, and so in this limit any interactions between them can be made arbitrarily weak. In this way, we see that the S-matrix is trivial.

At the classical level, integrability of the dKP and KP equations is well established, and so it is natural to expect vanishing of all tree level amplitudes. We verify explicitly that this expectation is borne out by evaluating the four-point tree amplitude in both theories. In Section~\ref{subsec:n-pt}, we extend this result to $n$-points by on-shell recursion.

However, the minitwistor description buys us more than the vanishing of the trees: Since $\text{dim}_\bbR\PS=5$, our non-commutative Chern--Simons theory $S_\star[a]$ cannot have any perturbative anomaly at one-loop. Thus, at one-loop, the corresponding space-time theory $S_{\rm KP}[\phi]$ should remain integrable and still have no amplitudes. We leave an explicit verification of this at higher loop order to future work.

Let us denote the $n$-particle tree amplitude in momentum space as
\begin{displaymath} \cA^n_{\rm (d)KP}(k_i) = (2\pi)^3\delta^{(3)}\Biggl(\sum_{i=1}^n k_i\Biggr)A^n_{\rm (d)KP}(k_i) . \end{displaymath}
In this section, we will show explicitly that \smash{$\cA^4_{\rm (d)KP}(k_i)=0$} and then in Section~\ref{sec:3d/2d-correspondence} argue that the \smash{$A^n_{\rm (d)KP}(k_i)$} obey an on-shell recursion relation implying \smash{$\cA^n_{\rm (d)KP}(k_i)=0$} inductively for all $n\geq 4$.

\subsection{Vanishing of the four-point tree amplitude} \label{subsec:tree-amps}

We begin with the dKP theory. Recall that this theory has action
\begin{displaymath} \int_{\bbR^{1,2}}\dif^3x \biggl(\frac{1}{2}\p^\mu\phi\p_\mu\phi + \frac{1}{3}(\p_+\phi)^3\biggr) . \end{displaymath}
It will be helpful to normalize our scattering states as
\begin{equation} \label{eq:scattering-states}
\hat\phi_k(x) = \frac{1}{\im k_+}{\rm e}^{\im k\cdot x}
= \frac{1}{\im\la\kappa\iota\ra^2}{\rm e}^{\im x^{\al\beta}\kappa_\al\kappa_\beta} ,
\end{equation}
where $k_{\al\beta} = \kappa_\al\kappa_\beta$ is null.\footnote{In this section, we absorb the energy $\omega$ into the scale of $\kappa$.} The normalization of these states has been chosen to simplify the amplitudes by eliminating irrelevant dressing by derivatives of the external states. With this normalization, the three-point amplitude is simply $A^3_{\rm dKP} = 2$.\footnote{Unlike three-point amplitudes in YM or gravity in $d=4$, this amplitude is non-zero even in Lorentzian signature $\bbR^{1,2}$. This is not in conflict with integrability, because momentum conservation forces these three null particles to all be collinear.}

Let us now check that $A_{\rm dKP}^4$ indeed vanishes. This amplitude receives contributions from three Feynman diagrams commonly referred to as the $s$-channel, $t$-channel and $u$-channel. Eliminating~$k_4$ using momentum conservation, these three diagrams give
\begin{displaymath} A_{\rm dKP}^4 = \frac{2\bigl(\la1\iota\ra^2 + \la2\iota\ra^2\bigr)^2}{\la12\ra^2} +\frac{2\bigl(\la2\iota\ra^2 + \la3\iota\ra^2\bigr)^2}{\la23\ra^2} +\frac{2\bigl(\la3\iota\ra^2 + \la1\iota\ra^2\bigr)^2}{\la31\ra^2},
\end{displaymath}
where we recall that $k_+ = \la\kappa\iota\ra^2$. Combining all the terms and using the fact that
\begin{displaymath} 0 = k_4^2 = (k_1+k_2+k_3)^2 = 2\bigl(\la12\ra^2+\la23\ra^2+\la31\ra^2\bigr), \end{displaymath}
one finds after a modest amount of algebra that
\begin{displaymath} A_{\rm dKP}^4 = - \frac{4\la1\iota\ra\la2\iota\ra\la3\iota\ra}{\la12\ra\la23\ra\la31\ra}\bigl(\la1\iota\ra\la23\ra + \la2\iota\ra\la31\ra + \la 3\iota\ra\la12\ra\bigr), \end{displaymath}
which vanishes by the usual Schouten identity. Thus the four-point tree amplitude vanishes. This is exactly as expected since the dKP equation is integrable.

Let us now consider full KP theory. Compared to the dKP case, we must now account for the dispersion term in the action. This means we can still use the same scattering states \eqref{eq:scattering-states} as~above, but now the external momenta obey the modified dispersion relation
\begin{equation} \label{eq:KP-dispersion}
k^2 + \frac{\fq^2}{6}k_+^4 = 0 .
\end{equation}
This is solved by
\begin{displaymath} k_{\alpha\beta} = \kappa_\alpha\kappa_\beta - \frac{\fq^2}{12}\iota_\alpha\iota_\beta\langle\kappa\iota\rangle^6 . \end{displaymath}
Accounting for the dispersive term, the KP propagator is
\begin{displaymath}\frac{1}{k^2 + \frac{\fq^2}{6}k_+^4} . \end{displaymath}
It simplifies for $k = k_1 + k_2$ with $k_1$, $k_2$ obeying the dispersion relation \eqref{eq:KP-dispersion} since
\begin{gather}
(k_1+k_2)^2 + \frac{\fq^2}{6}(k_1+k_2)_+^4 \nonumber\\
\qquad{}= 2\langle12\rangle^2 - \frac{\fq^2}{6}\bigl(\la1\iota\ra^2\la2\iota\ra^2\bigl(\la1\iota\ra^4+\la2\iota\ra^4\bigr) - 2\la1\iota\ra^2\la2\iota\ra^2\bigl(2\la1\iota\ra^4 + 3\la1\iota\ra^2\la2\iota\ra^2 + 2\la2\iota\ra^4\bigr)\bigr) \nonumber\\
\qquad{}= 2\langle12\rangle^2 + \frac{\fq^2}{2}\la1\iota\ra^2\la2\iota\ra^2\bigl(\la1\iota\ra^2+\la2\iota\ra^2\bigr)^2 .\label{eq:KP-prop-denominator}
\end{gather}
The four-point tree amplitude is therefore
\begin{displaymath} \cA_{\rm KP}^4 = \frac{2\bigl(\la1\iota\ra^2 + \la2\iota\ra^2\bigr)\bigl(\la3\iota\ra^2 + \la4\iota\ra^2\bigr)}{\la12\ra^2 + \frac{\fq^2}{4}\la1\iota\ra^2\la2\iota\ra^2\bigl(\la1\iota\ra^2+\la2\iota\ra^2\bigr)^2} + \text{cyclic permutations of \{2,3,4\}} . \end{displaymath}
We evaluate this in Appendix~\ref{app:four-point-KP} and verify that it vanishes to all orders in $\fq^2$.

\subsection[Vanishing of n-point tree amplitudes]{Vanishing of $\boldsymbol{n}$-point tree amplitudes} \label{subsec:n-pt}

In Section~\ref{sec:3d/2d-correspondence}, we will see that the holomorphic collinear singularities of tree amplitudes in dKP are described by $w_{1+\infty}$, so in particular they obey an on-shell recursion. That is, the singularity in the amplitude as the spinor helicity variables $\kappa_i$, $\kappa_j$ become collinear takes the form
\begin{displaymath} A^{n+1}_{\rm dKP}(k_i) \sim \cS_{\rm dKP}^{\omega_i,\omega_j}(\kappa_i,\kappa_j)A^{n}_{\rm dKP}\bigl(k_1,\dots,\widehat{k_i},\dots,k_{n+1}\bigr) . \end{displaymath}
Here \smash{$\cS_{\rm dKP}^{\omega_i,\omega_j}(\kappa_i,\kappa_j)$} is a splitting operator which crucially includes derivatives in $\kappa_j$.

Therefore, $\cA^5_{\rm dKP}$ is free from all holomorphic collinear singularities, since $\cA^4_{\rm dKP}=0$. But it is clear that, for our chosen normalization, $\cA^5_{\rm dKP}$ has vanishing weight under simultaneous scaling of the helicity variables. Since there can be no $\iota$ dependence in denominators, the ambiguity at five-points is a constant independent of all helicity variables. This constant must be proportional to $\iota^8$. Since these spinors can only contract among themselves, we infer that the constant is zero and the five-point amplitude vanishes. This argument applies inductively at $n$-points: all holomorphic singularities vanish as for the five-point and since all lower-point trees vanish there can be no multiple-particle singularities. The constant ambiguity must be proportional to $\iota^{4(n-3)}$ and so vanishes. We infer that all tree amplitudes vanish in the dKP theory.

To see that this inductive argument also applies to the full KP theory, let us expand all propagators as
\begin{equation} \label{eq:2-valent-vertex}
\frac{1}{\langle12\rangle^2 + \frac{\fq^2}{4}\la1\iota\ra^2\la2\iota\ra^2\bigl(\la1\iota\ra^2 + \la2\iota\ra^2\bigr)^2} = \frac{1}{\langle12\rangle^2}\sum_{\ell=0}^\infty\frac{\fq^{2\ell}\la1\iota\ra^{2\ell}\la2\iota\ra^{2\ell}\bigl(\la1\iota\ra^2 + \la2\iota\ra^2\bigr)^{2\ell}}{2^{2\ell}\langle12\rangle^{2\ell}},
\end{equation}
which amounts to viewing \smash{$\fq^2\bigl(\p_+^2\phi\bigr)^2/12$} as a $2$-point vertex deforming dKP. At order $\fq^0$ we do not call on this vertex at all, and so by the above argument in dKP the $n$-point tree level amplitude vanishes. If we use the $2$-point vertex $k$ times, then by basic graph theory all diagrams with $n$ external legs will still have $n-2$ trivalent vertices but $k+n-3$ propagators. This means that such a diagram scales like
\begin{displaymath}
\fq^{2k}\kappa^{-2n - 4(k+n-3) + 6(n-2) + 8k}\iota^{-2n + 6(n-2) + 8k} = \fq^{2k}\kappa^{4k}\iota^{4(n-3)+8k} .
\end{displaymath}
Since we still cannot have any collinear singularities and now there also cannot be any $\iota$ dependence in the denominator, some of the $8k$ $\iota$s must pair up among themselves as there are only~$4k$ new~$\kappa$s. This dependence on $k$ is reflected in \eqref{eq:2-valent-vertex}. We learn that the five-point amplitude vanishes at all orders in $\fq^2$ and hence inductively all tree level amplitudes vanish in KP (again the vanishing of the lower-point trees ensures there are no multi-particle poles).

\section[The 3d minitwistorial theory/2d vertex algebra correspondence]{The $\boldsymbol{3d}$ minitwistorial theory/$\boldsymbol{2d}$ vertex algebra\\ correspondence} \label{sec:3d/2d-correspondence}

In this section, we show that surface defects wrapping the fibres of correspondence space in Poisson- and non-commutative Chern--Simons theory support $w_{1+\infty}$ and $W_{1+\infty}$ vertex algebras respectively. We then recover these directly on space-time from the collinear splitting of form factors in dKP and KP theory respectively.

\subsection{Koszul duality on correspondence space} \label{subsec:Koszul-Duality}

Here we outline how Koszul duality on correspondence space yields an equivalence between data of (d)KP theory (or more generally any minitwistorial theory) and that of a $2d$ vertex algebra~$\cC$ living on minitwistor lines. This is a $3d$ counterpart of the Costello--Paquette correspondence in $4d$ \cite{Costello:2022wso}. For an introduction to the role of Koszul duality in describing defects in mixed topological-holomorphic quantum field theory, we refer the reader to \cite{Costello:2020jbh,Paquette:2021cij}.

To summarise: Linearised solutions to the classical equations of motion of the $3d$ theory are identified with states in the vertex algebra $\cC$. Of particular interest are complexified momentum eigenstates, which correspond to so-called `hard' generators. The operator products of these hard generators are captured by collinear splitting in space-time. The conformal blocks of $\cC$, i.e., the possible ways of defining correlators compatible with operator products, are in bijection with local operators of the minitwistorial theory. Finally, amplitudes of the $3d$ theory with a~local operator inserted at a point (form factors) equate to correlators of hard generators in the corresponding conformal block. That collinear singularities of form factors are universal, i.e., independent of the inserted local operator, is a consequence of the vanishing of all space-time amplitudes. These identifications are listed in Table \ref{tab:3d/2d-correspondence} for convenience. In this work, we have only engineered the correspondence space uplift of (d)KP theory at tree level, and so we can only trust a classical version of the correspondence; however, we expect that at the very least a~one-loop version persists.

\begin{table}[!ht]
 \centering
	\caption{Summary of the correspondence between $3d$ minitwistorial theories and $2d$ vertex algebras.}
	\label{tab:3d/2d-correspondence}
\vspace{1mm}

 \begin{tabular}{c c}
	\toprule
	$3d$ minitwistorial theory & $2d$ vertex algebra \\ \midrule
 momentum eigenstates $\phi_k(x)$ & hard generators $v(\kappa,\omega)$ \\
 local operators $\cO(0)$ & conformal blocks $\la\dots\ra_\cO$ \\
 form factors $F_\cO(k_i)$ & correlators $\la v(\kappa_1,\omega_1,)\cdots v(\kappa_n,\omega_n)\ra_\cO$ \\
 collinear splitting $\cS^{\omega_1,\omega_2}(\kappa_1,\kappa_2)$ & OPEs $v(\kappa_1,\omega_1)v(\kappa_2,\omega_2)\sim\cdots$ \\
	\bottomrule
	\end{tabular}
\end{table}

Let us briefly sketch where this correspondence comes from, concentrating on the case of dKP for simplicity. The linearised equations of motion read
\begin{displaymath}
\bigl(2\p_+\p_- - \p_y^2\bigr)\phi = 0
\end{displaymath}
with solution $\phi_k(x) = \exp(\im k\cdot x)$ for $k^2 = 2k_+k_- - k_y^2 = 0$. (We choose not to normalise as we did in Section~\ref{sec:amplitudes}.) The massless condition is solved by $k_{\al\beta} = \omega\kappa_\al\kappa_\beta$ with the redundancy~${(\omega,\kappa_\al)\sim\bigl(t^{-2}\omega,t\kappa_\al\bigr)}$ for $t\in\bbC^\times$. \big(At $\omega = 0$, the $\CP^1$ factor shrinks to a point, so complexified null momenta belong to $\bbC^2/\bbZ_2$.\big) We will sometimes write $\phi_k(x) = \phi^\omega(x;\kappa)$ to make this parametrisation explicit. This differs from our convention in the previous section, where we absorbed the scale $\omega$ into $\kappa$ which was treated as an inhomogeneous variable. We should view~$\omega\la\kappa\iota\ra^2$ as the energy of the wave, and $\kappa$ as a point on the `celestial' sphere. (For~$\phi_k(x)$ to be real, we must have $\kappa\sim\bar\kappa$ so that this celestial sphere is an $S^1$ as expected for~$\bbR^{1,2}$. The sign of~$\omega$ determines whether the momentum is past- or future-pointing.) Expanding in powers of $\omega$ is a soft expansion
\begin{equation} \label{eq:soft-expansion} \phi_k(x) = \sum_{s=1}^\infty\frac{(\im\omega)^{s-1}}{(s-1)!}\phi_s(x;\kappa) , \end{equation}
where \smash{$\phi_s(x;\kappa) = \bigl(x^{\al\beta}\kappa_\al\kappa_\beta\bigr)^{s-1}$} is a `soft' state.

The momentum eigenstates $\phi^\omega(x;\kappa)$ lift via the Penrose transform to states on correspondence space which pull-back from Dolbeault cohomology representatives on $\mT$. Explicitly, they take the form
\begin{displaymath} a^\omega(x,\lambda;\kappa) = \la\lambda\iota\ra^2 {\rm e}^{\im\omega x^{\al\beta}\lambda_\al\lambda_\beta\la\kappa\iota\ra^2/\la\lambda\iota\ra^2}\bar\delta_{\lambda\sim\kappa} . \end{displaymath}
The precise correspondence between the space-time and twistor data is
\begin{displaymath} \phi^\omega(x;\kappa) = \int_{\CP^1_x}\frac{\la\lambda\dif\lambda\ra}{\la\lambda\iota\ra^2}\wedge a^\omega(x,\lambda;\kappa) . \end{displaymath}
Amplitudes in the $3d$ theory vanish, since on $\PS$ these states cannot talk to one another: Feynman diagrams can be evaluated in a gauge where they are arbitrarily far apart.

However, suppose we insert a local operator at the origin $x=0$, denoted by $\cO(0)$. The results of \cite{Costello:2022wso} show that it lifts to a compact surface defect wrapping the minitwistor line $\CP^1_{x=0}$. This defect is the correlation function of a universal vertex algebra $\cC$: the Koszul dual to the algebra of bulk local operators restricted to the defect. The states of $\cC$ are denoted $v_s(\lambda)$ for $s\geq1$ and have conformal dimension $1-s$; they couple to the bulk gauge theory via
\begin{displaymath} I[a,v_s] = \sum_{s\geq1}\int_{\CP^1_{x=0}}\frac{\la\lambda\dif\lambda\ra}{\la\lambda\iota\ra^{2s}}\wedge\frac{v_s(\lambda)\cL_{\p_1}^{s-1}a}{(s-1)!} . \end{displaymath}
With these conventions, the state $v_s(\lambda)$ must vanish to order $\la\lambda\iota\ra^{2(s-1)}$ at $\lambda\sim\iota$. The operator~$\cO$ is identified with conformal block of $\cC$, denoted $\la\dots\ra_\cO$, and we have
\begin{displaymath} \cO(0) = \la\exp{I[a,v_s]}\ra_\cO . \end{displaymath}
The form factor of $\cO(0)$ can be computed on correspondence space using the states $a^\omega(x,\lambda;\kappa)$. These only talk to one another through the defect, where they pick out `hard generators'
\begin{equation} \label{eq:hard-generators} I[a^\omega(x,\lambda;\kappa),v_s] = v(\kappa,\omega) = \sum_{s\geq1}\frac{(\im\omega)^{s-1}}{(s-1)!}v_s(\kappa) . \end{equation}
Comparing equations \eqref{eq:soft-expansion} and \eqref{eq:hard-generators}, we can see that the operators $v_s(\kappa)$ can be identified with soft states $\phi_s(x;\kappa)$. (In fact, this is the motivation for putting the factor of $1/\la\lambda\iota\ra^{2s}$ in the measure.)

The form factor $F_\cO(k_i)$ equates to a correlation function of hard generators in the conformal block $\la\dots\ra_\cO$.
\begin{displaymath} F_\cO(k_i) = \la v(\kappa_1,\omega_1)\cdots v(\kappa_n,\omega_n)\ra_\cO . \end{displaymath}
In order for this equality to hold, the collinear singularities of the form factor as $\kappa_i$, $\kappa_j$ coincide must agree with the operator product of hard generators. But the operator products of the~$v_s(\lambda)$ are also fixed by gauge invariance of the bulk-defect coupling on correspondence space. In the next subsection, we will see how these can be determined straightforwardly.

\subsection{Koszul duality for Poisson--Chern--Simons} \label{subsec:w(1+infty)-PS}

Here we determine the operator products of the vertex algebra $\cC$ corresponding to dKP theory directly on correspondence space. This is achieved by demanding gauge invariance of the bulk-defect coupling $\exp{I[a,v_s]}$.

At tree level the variation of the bi-local term \smash{$\tfrac{1}{2}I[a,v_s]^2$} under the linear part of an infinitesimal gauge transformation cancels against the non-linear variation of the local term $I[a,v_s]$. The former picks out the singular part of the operator product of states $v_s(\lambda)$, whereas the latter is determined by the cubic vertex in Poisson--Chern--Simons theory.

At tree level rather than computing the operator products it is easier to determine the Lie brackets of the $v_s(\lambda)$ modes. These can be extracted from the bulk-defect coupling using states
\begin{equation} \label{eq:soft-modes}
a_{s,m} = \frac{\im}{2\pi}\frac{\la o\lambda\ra^{s-1+m}}{\la \lambda\iota\ra^{s-1+m}}\biggl(\frac{x^{\al\beta}\lambda_\al\lambda_\beta}{\la\lambda\iota\ra^2}\biggr)^{s-1}\dift\Theta
\end{equation}
for $\Theta$ a step function taking the value $1$ for $|z| = |\la o\lambda\ra/\la\lambda\iota\ra| \leq 1$ and $0$ for $|z| = |\la o\lambda\ra/\la\lambda\iota\ra| > 1$. In the \v{C}ech language, these are represented by meromorphic ($\dift$-closed) functions
\begin{equation} \label{eq:Cech-reps}
\Check{a}_{s,m} = \frac{\la o\lambda\ra^{s-1+m}}{\la \lambda\iota\ra^{s-1+m}}\biggl(\frac{x^{\al\beta}\lambda_\al\lambda_\beta}{\la\lambda\iota\ra^2}\biggr)^{s-1}
\end{equation}
on the overlap of the north and south patches of the $\lambda$-sphere ($\lambda\not\sim\iota$ and $\lambda\not\sim o$, respectively). Plugging the states \eqref{eq:soft-modes} into the bulk-defect coupling yields
\begin{equation} \label{eq:extract-modes} \tilde v_{s,m} = I[a_{s,m},v_{s^\prime}] = \frac{1}{2\pi\im}\oint_{S^1}\frac{\la\lambda\dif\lambda\ra}{\la \lambda\iota\ra^{2s}}\,\frac{\la o\lambda\ra^{s-1+m}}{\la \lambda\iota\ra^{s-1+m}}v_s(\lambda) . \end{equation}
Since the cubic vertex of Poisson--Chern--Simons theory is determined by the Poisson bracket, the Lie algebra obeyed by the modes $\tilde v_{s,m}$ is the Poisson algebra of \v{C}ech representatives \eqref{eq:Cech-reps}. We immediately find that
\begin{align}
\{\Check{a}_{s_1,m},\Check{a}_{s_2,n}\} &{}= \p_0\Check{a}_{s_1,m}\p_1\Check{a}_{s_2,n} - \p_1\Check{a}_{s_1,n}\p_0\Check{a}_{s_2,m}\nonumber\\
&{}= (m(s_2-1) - n(s_1-1))\Check{a}_{s_1+s_2-2,m+n} ,\label{eq:w(1+infinity)-mode-algebra} \end{align}
so \smash{$[\tilde v_{s_1,m},\tilde v_{s_2,n}] = (m(s_2-1) - n(s_1-1))\tilde v_{s_1+s_2-2,m+n}$}. This is the mode algebra of $w_{1+\infty}$ (without central extension).\footnote{Often mathematicians refer to the universal enveloping algebra of this Lie algebra as the `mode algebra'.} Note that the $\{\tilde v_{2,m}\}_{m\in\bbZ}$ generate a Witt subalgebra.

We have performed a sleight of hand here -- the mode expansion \eqref{eq:extract-modes} does not take the standard form for a field of conformal dimension $1-s$; instead, this is the natural expansion for ${\tilde v_s(\lambda) = \la\lambda\iota\ra^{2-4s}v_s(\lambda)}$ of conformal dimension $s$. It is really these shifted fields $\tilde v_s(\lambda)$ whose operator products obey the $w_{1+\infty}$ vertex algebra. The price for performing this shift is that the vacuum module at $\lambda\sim\iota$ has been modified.

The vacuum module $|0\ra_{\lambda\sim o}$ at $\lambda\sim o$, or equivalently $z=0$, is the standard one for $w_{1+\infty}$. It~can be read off from \eqref{eq:extract-modes} by noting that $\tilde v_{s,m}|0\ra_{\lambda\sim o} = 0$ for $m\geq 1-s$, since in this case the contour integral can be contracted to a point in the northern hemisphere. Let us remind ourselves why this defines a module. For $m\geq1-s_1$, $n\geq1-s_2$, we require that
\begin{displaymath} [\tilde v_{s_1,m},\tilde v_{s_2,n}]|0\ra_{\lambda\sim o} = (m(s_2-1) - n(s_1-1))\tilde v_{s_1+s_2-2,m+n}|0\ra_{\lambda\sim o} = 0 . \end{displaymath}
Now $\tilde v_{s_1+s_2-2,m+n}|0\ra_{\lambda\sim o} = 0$ for $m+n\geq 1-(s_1+s_2-2)$. This leaves the case $m+n = 2-s_1-s_2$, for which $m=1-s_1$ and $n=1-s_2$ so that the structure constant $m(s_2-1) - n(s_1-1)=0$. The standard $w_{1+\infty}$ vacuum module $|0\ra_{\lambda\sim\iota}$ at $\lambda\sim\iota$ similarly obeys $\tilde v_{s,m}|0\ra_{\lambda\sim\iota}$ for $m\leq s-1$. The global symmetries of $w_{1+\infty}$, i.e., the Lie subalgebra of modes preserving both vacua, is the wedge subalgebra generated by $\{\tilde v_{s,m}\}_{|m|\leq s-1}$.\footnote{This is not the wedge subalgebra appearing in the context of celestial holography, which has extra generators corresponding to spin representations of the global $\fsl_2(\bbC)$ conformal symmetry.} In particular, we find a global $\bbC^\times$ symmetry generated by $j_0 = \tilde v_{1,0}$ and a $\mathrm{PSL}_2(\bbC)$ global conformal symmetry generated by $L_{-1} = \tilde v_{2,-1}$, $L_0 = \tilde v_{2,0}$, $L_1 = \tilde v_{2,1}$.

Let us write $|\text{dKP}\ra_{\lambda\sim\iota}$ for the modified vacuum at $\lambda\sim\iota$ in the case of dKP. Recalling that the boundary condition on $v_s(\lambda)$ requires that it should vanish to order $2(s-1)$ at $\lambda\sim\iota$, we can see from equation \eqref{eq:extract-modes} that $\tilde v_{s,m}|\text{dKP}\ra_{\lambda\sim\iota} = 0$ for $m\leq-1-s$ by contracting the contour in the southern hemisphere.
This defines a consistent vertex algebra module, since for any~$s_1$,~$s_2$ and $m\leq-1-s_1$, $n\leq-1-s_2$ we have $m+n\leq-2-s_1-s_2< - 1 - (s_1+s_2-2)$. The module~$|\text{dKP}\ra_{\lambda\sim\iota}$ breaks all of the global symmetry which was present in the $|0\ra_{\lambda\sim\iota}$ vacuum. Indeed, any modes $\tilde v_{s,m}$ preserving both vacua would need to have $1-s\leq m\leq-1-s$.

This shift of the vacuum module is crucial for the validity of the $3d/2d$ correspondence outlined in Section~\ref{subsec:Koszul-Duality}. Indeed, the simplest local operator of dKP theory is $\phi$ itself. This should correspond to a conformal block of the vertex algebra $\cC$, or equivalently a family of~$w_{1+\infty}$ correlators compatible with the operator product but in which the states $v_s(\lambda)$ may have poles of order $\la\lambda\iota\ra^{-2s}$. This family of correlators is seeded by
\begin{displaymath} \la\tilde v_1(\lambda)\ra_{\phi(0)} = \frac{1}{\la\lambda\iota\ra^2} \end{displaymath}
with all other one-point functions set to zero. (Here by seeded we mean that all other correlators are determined by recursively applying the OPE until one lands on this one-point function.) In~terms of the hard generators $v(\kappa;\omega)$, this is $\la v(\kappa,\omega)\ra_{\phi(0)} = 1 = \phi_k(0)$. Similarly, the operator~${u = \p_+\phi}$ is identified with the block
\begin{displaymath}\la\tilde v_2(\lambda)\ra_{u(0)} = \frac{\im}{\la\lambda\iota\ra^2} \end{displaymath}
with all other one-point functions set to zero. This yields
\begin{displaymath} \la v(\kappa,\omega)\ra_{u(0)} = \im\omega\la\kappa\iota\ra^2 = \p_+\phi_k(0) . \end{displaymath}
Translation invariance ensures that the form factor of $\cO(x)$ can be obtained from $\cO(0)$ by dressing with ${\rm e}^{{\rm i}K\cdot x}$ for $K$ the total momentum of all hard scattering states. Hence
\begin{displaymath}
\la v(\kappa,\omega)\ra_{u(x)} = \im\omega\la\kappa\iota\ra^2{\rm e}^{\im k\cdot x} = \p_+\phi_k(x) .
\end{displaymath}
Notice that $w_{1+\infty}$ with its usual vacua has no non-trivial conformal blocks, so the correspondence would fail immediately.

\subsection{Collinear splitting in dKP} 

\begin{figure}[t]
 \centering
 \includegraphics[scale=0.3]{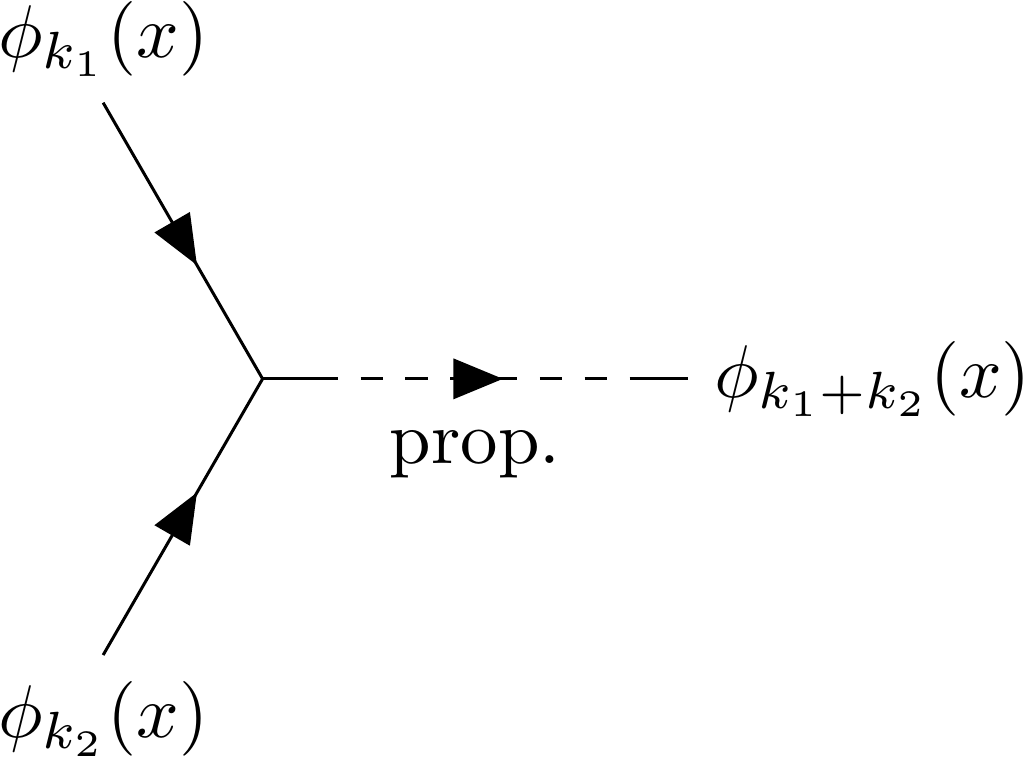}
 \caption{The tree splitting function in dKP theory is determined by the singular part of the above diagram as $\la12\ra\to0$.}
 \label{fig:perturbiner}
\end{figure}

In this subsection, we compute the tree splitting function, or equivalently the singular part of the perturbiner, in dKP. It is computed as the partially off-shell Feynman diagram illustrated in Figure~\ref{fig:perturbiner}. From the arguments of Section~\ref{subsec:Koszul-Duality}, we expect to recover the mode algebra directly recovered from correspondence space.
Together with the states $\phi_k(x)$ we recall the propagator~$- \im/k^2$ and vertex $k_+^3/3$. The perturbiner is
\begin{displaymath} \cS_{\rm dKP}^{\omega_1,\omega_2}(\kappa_1,\kappa_2) = - \frac{2\im k_{1+}k_{2+}(k_1+k_2)_+}{(k_1+k_2)^2}{\rm e}^{\im(k_1+k_2)\cdot x} . \end{displaymath}
Rewriting in terms of energy and spinor-helicity variables, this is
\begin{equation} \label{eq:dKP-perturbiner}
\cS_{\rm dKP}^{\omega_1,\omega_2}(\kappa_1,\kappa_2) = - \frac{\im\la1\iota\ra^2\la2\iota\ra^2\bigl(\omega_1\la1\iota\ra^2 + \omega_2\la2\iota\ra^2\bigr)}{\la12\ra^2}{\rm e}^{\im(k_1+k_2)\cdot x} .
\end{equation}
This is an off-shell solution, but the singular part of this expression in the collinear limit $\la12\ra\to0$ is expressible in terms of on-shell states. To see why, we exploit the Schouten identity to write
\begin{displaymath} \kappa_1 = \frac{\la1\iota\ra}{\la2\iota\ra}\kappa_2 - \frac{\la12\ra}{\la2\iota\ra}\iota . \end{displaymath}
This allows us to expand the perturbiner \eqref{eq:dKP-perturbiner} around $\kappa_1\sim\kappa_2$ to get
\begin{displaymath}
{\rm e}^{\im(k_1+k_2)\cdot x} = \biggl(1 - 2\im\frac{\la12\ra\la1\iota\ra}{\la2\iota\ra^2}\omega_1x^{\al\beta}\kappa_{2\al}\iota_\beta + \cO\bigl(\la12\ra^2\bigr)\biggr){\rm e}^{\im K_+ x^{\al\beta}\kappa_{2\al}\kappa_{2\beta}/\la2\iota\ra^2} ,
\end{displaymath}
where $K_+ = (k_1+k_2)_+ = \omega_1\la1\iota\ra^2 + \omega_2\la2\iota\ra^2$. Whilst the leading term in $\la12\ra$ is clearly an on-shell state, this is not immediate for the subleading piece linear in $x$. However, it can be written in the form
\begin{displaymath}
 \biggl(1 - \la12\ra\frac{\omega_1\la1\iota\ra}{K_+}\iota^\al\p_{\kappa_2^\al} + \cO\bigl(\la12\ra^2\bigr)\biggr)\phi^{K_+/\la2\iota\ra^2}(x;\kappa_2) .
 \end{displaymath}
Substituting into equation \eqref{eq:dKP-perturbiner} yields
\begin{displaymath} \begin{aligned}
\cS_{\rm dKP}^{\omega_1,\omega_2}(\kappa_1,\kappa_2)\sim{}& {-}\im\frac{\la1\iota\ra^2\la2\iota\ra^2\bigl(\omega_1\la1\iota\ra^2 + \omega_2\la2\iota\ra^2\bigr)}{\la12\ra^2}\phi^{K_+/\la2\iota\ra^2}(x;\kappa_2) \\
&{+}\, \im\frac{\la1\iota\ra^3\la2\iota\ra^2\omega_1}{\la12\ra}\iota^\al\p_{\kappa_2^\al}\phi^{K_+/\la2\iota\ra^2}(x;\kappa_2) .
\end{aligned} \end{displaymath}
Here the symbol $\sim$ indicates that we work modulo regular terms in $\la12\ra$. This is currently written in terms of hard scattering states - expanding in powers of $\omega_1$, $\omega_2$ we can extract the OPE of the soft states $v_m(\kappa)$.
\begin{displaymath} \begin{aligned}
v_{s_1}(\kappa_1)v_{s_2}(\kappa_2)\sim{}& \frac{(s_1+s_2-2)}{\la12\ra^2}\frac{\la1\iota\ra^{2s_1}}{\la2\iota\ra^{2s_1-6}}v_{s_1+s_2-2}(\kappa_2) \\
&{}{+}\, \frac{(s_1-1)}{\la12\ra}\frac{\la1\iota\ra^{2s_1-1}}{\la2\iota\ra^{2s_1-6}}\iota^\al\p_{\kappa_2^\al}v_{s_1+s_2-2}(\kappa_2) .
\end{aligned} \end{displaymath}
This OPE depends explicitly on $\iota$, though this dependence can be removed through an appropriate redefinition of the states. If we let $\tilde v_s(\kappa) = \la\kappa\iota\ra^{2-4s}v_s(\kappa)$, so that $\tilde v_s(\kappa)$ has conformal spin $s$, then
\begin{displaymath}
\begin{aligned}
\tilde v_{s_1}(\kappa_1)\tilde v_{s_2}(\kappa_2)\sim{}&\frac{(s_1+s_2-2)}{\la12\ra^2}\frac{\la1\iota\ra^{2-2s_1}}{\la2\iota\ra^{2-2s_1}}\tilde v_{s_1+s_2-2}(\kappa_2)\\
&{}{-}\, \frac{(s_1-1)}{\la12\ra}\frac{\la1\iota\ra^{1-2s_1}}{\la2\iota\ra^{2-2s_1}}\iota^\al\p_{\kappa_2^\al}\tilde v_{s_1+s_2-2}(\kappa_2) .
\end{aligned}
\end{displaymath}
Although this appears to depend on $\iota$, in fact this dependence is only through the regular part. Expressed in terms of the inhomogeneous co-ordinate $w = \la o\kappa\ra/\la\kappa\iota\ra$, we have
\begin{equation} \label{eq:w(1+infinity)} \tilde v_{s_1}(w_1)\tilde v_{s_2}(w_2)\sim\frac{(s_1+s_2-2)}{(w_1-w_2)^2}\tilde v_{s_1+s_2-2}(w_2) + \frac{(s_1-1)}{w_1-w_2}(\p_w\tilde v_{s_1+s_2-2})(w_2) . \end{equation}
These are the defining relations of the $w_{1+\infty}$ vertex algebra.\footnote{We emphasise that this is \emph{not} the wedge subalgebra of $w_{1+\infty}$, nor the loop algebra of some Lie algebra. These are the defining OPEs of $w_{1+\infty}$ on the nose.}\! (The inhomogeneous co-ordinate~$w$ is adapted to $\iota$, but since the singular part of the OPE is independent of this choice we'd get the same result in any holomorphic co-ordinates.) The $w_{1+\infty}$ algebra in particular contains a~Witt subalgebra generated by $\tilde v_2(w)$ under which the $\tilde v_s(w)$ transform as conformal primaries.

To connect with Section~\ref{subsec:w(1+infty)-PS}, it is instructive to perform a mode expansion
\begin{displaymath}
\tilde v_s(w) = \sum_{m\in\bbZ}\frac{\tilde v_{s,m}}{w^{m+s}} ,
\end{displaymath}
or equivalently in terms of the homogeneous co-ordinate $\kappa$
\begin{displaymath} \tilde v_s(\kappa) = \sum_{m\in\bbZ}\frac{\tilde v_{s,m}}{\la o\kappa\ra^{s+m}\la\kappa\iota\ra^{s-m}} . \end{displaymath}
In terms of the modes $\tilde v_{s,m}$ the OPE \eqref{eq:w(1+infinity)} becomes the commutator
\begin{displaymath}
[\tilde v_{s_1,m},\tilde v_{s_2,n}] = (m(s_2-1) - n(s_1-1))\tilde v_{s_1+s_2-2,m+n} .
\end{displaymath}
We recognise this as the mode algebra of $w_{1+\infty}$ \eqref{eq:w(1+infinity)-mode-algebra}.

\subsection{Koszul duality for non-commutative Chern--Simons} \label{subsec:W(1+infty)-PS}

Now let us understand the consequences of switching on the dispersive term in KP. The linearised equation of motion for the potential $\phi$ is more complicated, it reads
\begin{displaymath}
\bigl(2\p_+\p_- - \p_y^2\bigr)\phi = \frac{\fq^2}{6}\p_+^4\phi
\end{displaymath}
with plane wave solutions $\Phi_k(x) = \exp(\im k\cdot x)$ for $2k_+k_- - k_y^2 + \fq^2k_+^4/6 = 0$. We can parametrise solutions to this dispersion relation by the same data as in the dispersionless case: by $(\omega,\kappa_\al)\sim\bigl(t^{-2}\omega,t\kappa_\al\bigr)$,
\begin{displaymath}
k_{\al\beta} = \omega\kappa_\al\kappa_\beta - \frac{\fq^2}{12}\omega^3\la\kappa\iota\ra^6\iota_\al\iota_\beta .
\end{displaymath}
We can then perform a soft expansion as above
\begin{equation} \label{eq:KP-soft-modes}
\Phi^\omega(x;\kappa) = \sum_{s=1}^\infty\frac{(\im\omega)^{s-1}}{(s-1)!}\Phi_s(x;\kappa) ,
\end{equation}
where the soft states $\Phi_s(x;\kappa)$ are no longer homogeneous polynomials in $x$.

We can lift these states to correspondence space, although from the discussion in Sections~\ref{subsec:correspondence-ncCS} and \ref{subsec:holomorphic-split} there are different isomorphic presentations of the defining dga on $\PS$. Splitting the bivector holomorphically to obtain the presentation \smash{$\bigl(\til\Omega_\PS,\dift,\ast\bigr)$}, we can continue to use the correspondence space representatives
\begin{displaymath}
a^\omega(x,\lambda;\kappa) = \la\lambda\iota\ra^2 {\rm e}^{\im\omega x^{\al\beta}\lambda_\al\lambda_\beta\la\kappa\iota\ra^2/\la\lambda\iota\ra^2}\bar\delta_{\lambda\sim\kappa} .
\end{displaymath}
To get their counterparts in the presentation \smash{$\bigl(\til\Omega_\PS,\Dift,\star\bigr)$}, we simply apply the isomorphism ${\psi = \exp\bigl(\fq^2x^-\cL_{\p_1}^3/12\bigr)}$ to get
\begin{displaymath}
\psi(a^\omega(x,\lambda;\kappa)) = \la\lambda\iota\ra^2 \exp\im\omega\la\kappa\iota\ra^2x^{\al\beta}\biggl(\frac{\lambda_\al\lambda_\beta}{\la\lambda\iota\ra^2} - \frac{\fq^2}{12}\omega^2\la\kappa\iota\ra^4\iota_\al\iota_\beta\biggr)\bar\delta_{\lambda\sim\kappa} .
\end{displaymath}
The linear Penrose transform is then
\begin{displaymath}
\Phi^\omega(x;\kappa) = \int_{\CP^1_x}\frac{\la\lambda\dif\lambda\ra}{\la\lambda\iota\ra^2}\wedge\psi(a^\omega(x,\lambda;\kappa)) .
\end{displaymath}
We can introduce a universal vertex algebra $\cC$ living on $\CP^1_{x=0}$ as in Section~\ref{subsec:Koszul-Duality}. This is simplest using the presentation \smash{$\bigl(\til\Omega_\PS,\dift,\ast\bigr)$}. Plugging the correspondence space avatars of the soft modes $\Phi_s(x;\kappa)$ into the bulk-defect coupling yields states $\cV_s(\kappa)$ of conformal dimension $1-s$ for $s\in\bbZ_{\geq1}$. Again, these vanish to order $\la\lambda\iota\ra^{2(s-1)}$ at $\lambda\sim\iota$.

It is easiest to compute the mode algebra of $\cC$ directly on correspondence space. Since $\dift$ is undeformed we can continue to use the \v{C}ech representatives \eqref{eq:Cech-reps} for the modes $a_{s,m}$. Recall that the star product is
\begin{displaymath}
F\ast G = c\circ\exp\frac{\fq}{2} (v_0\otimes\p_1 - \p_1\otimes v_0 )(F\otimes G),
 \end{displaymath}
where $v_0 \!=\! \p_0 + s\p_1$ for \smash{$s \!=\! -2x^{\al\beta}\iota_\al\lambda_\beta/\la\lambda\iota\ra$}. The vector field $v_0$ has the property ${v_0\bigl(x^{\al\beta}\lambda_\al\lambda_\beta\bigr) \!=\! 0}$ and $v_0\la o\lambda\ra = \la\lambda\iota\ra$ making it straightforward to compute
\begin{gather}
[\Check{a}_{s_1,m},\Check{a}_{s_2,n}]_\ast = \frac{1}{\fq}(\Check{a}_{s_1,m}\ast\Check{a}_{s_2,n} - \Check{a}_{s_2,n}\ast\Check{a}_{s_1,m}) \nonumber\\
\hphantom{[\Check{a}_{s_1,m},\Check{a}_{s_2,n}]_\ast }{}
= \sum_{\ell\geq 0}\frac{\fq^{2\ell}}{(2\ell+1)!2^{2\ell}} \sum_{k=0}^{2\ell+1}(-)^k\binom{2\ell+1}{k}[s_1-1+m]_{2\ell+1-k}[s_2-1]_{2\ell+1-k}\nonumber\\
\hphantom{[\Check{a}_{s_1,m},\Check{a}_{s_2,n}]_\ast = \sum_{\ell\geq 0}\frac{\fq^{2\ell}}{(2\ell+1)!2^{2\ell}} \sum_{k=0}^{2\ell+1}}{}
\times[s_2-1+n]_k[s_1-1]_k\Check{a}_{s_1+s_2-2-2\ell,m+n}\nonumber \\
\hphantom{[\Check{a}_{s_1,m},\Check{a}_{s_2,n}]_\ast }{}
= \sum_{\ell\geq0}\frac{\fq^{2\ell}}{(2\ell+1)!2^{2\ell}}N^{s_1,s_2}_{2\ell}(m,n)\Check{a}_{s_1+s_2-2-2\ell,m+n},\label{eq:W(1+infinity)-mode-algebra}
\end{gather}
where $[a]_b = a(a-1)\cdots (a-(b-1))$
denotes the descending factorial and we have introduced the structure constants
\begin{displaymath} N^{s_1,s_2}_\ell(m,n) = \sum_{k=0}^{\ell+1}(-)^k\binom{\ell+1}{k}[s_1-1]_k[s_2-1]_{\ell+1-k}[s_1-1+m]_{\ell+1-k}[s_2-1+n]_k . \end{displaymath}
Notice that if $\ell+1>s_1+s_2-2$ in the above then for every term in the sum either $k>s_1-1$ or $\ell+1-k>s_2-1$, so that either $[s_1-1]_k=0$ or $[s_2-1]_{\ell+1-k}=0$. As a result the sum over~$\ell$ in equation truncates and the bracket closes. The Lie algebra \eqref{eq:W(1+infinity)-mode-algebra} is in fact isomorphic to the mode algebra of $W_{1+\infty}$.

To see why, notice that in effect we are quantizing the Poisson algebra of regular functions on~${\bbC^\times\times\bbC}$ equipped with the bivector $\p_z\vee\p_v$. The non-commutative product is defined by applying the Moyal formula to this bivector. Let us write $\mathscr{O}_\fq(\bbC^\times\times\bbC)$ for this associative algebra. The commutator induces a Lie algebra structure on $\mathscr{O}_\fq(\bbC^\times\times\bbC)$ which is isomorphic to \eqref{eq:W(1+infinity)-mode-algebra}. The Moyal product on $\mathscr{O}_\fq(\bbC^\times\times\bbC)$ is equivariant under the $\bbC^\times$ action simultaneously scaling $\fq$ and $v$. Absorbing the parameter $\fq$ into $v$ yields $\mathscr{O}_1(\bbC^\times\times\bbC)$ which is isomorphic to $\mrm{Diff}(\bbC^\times)$, the algebra of holomorphic differential operators on $\bbC^\times$. Here we are identifying $v\leftrightarrow - \dif/\dif z$. The commutator of differential operators endows $\mrm{Diff}(\bbC^\times)$ with a Lie algebra structure isomorphic to the mode algebra of $W_{1+\infty}$ \cite{Pope:1990be}.\footnote{The bivector $\p_z\vee\p_v$ differs from the standard bivector $z\p_z\vee\p_v$ on $\bbC^\times\times\bbC$, hence the structure constants appearing in equation \eqref{eq:W(1+infinity)-mode-algebra} differ from those of \cite{Pope:1990be}. But the Moyal formula applied to the two bivectors yields isomorphic associative algebras.}

Now that we have determined the Lie algebra of the modes
\begin{displaymath} \til\cV_{s,m} = \frac{1}{2\pi\im}\oint_{S^1}\frac{\la\lambda\dif\lambda\ra}{\la \lambda\iota\ra^{2s}}\,\frac{\la o\lambda\ra^{s-1+m}}{\la \lambda\iota\ra^{s-1+m}}\cV_s(\lambda), \end{displaymath}
we should take care to keep track of the vacuum modules. The \smash{$\til\cV_{s,m}$} are the modes of the states \smash{$\til\cV_s(\lambda) = \la\lambda\iota\ra^{2-4s}\cV_s(\lambda)$}, and these have vacuum module \smash{$\til\cV_{s,m}|\text{KP}\ra_{\lambda\sim\iota} = 0$} for $m\leq-1-s$ at $\lambda\sim\iota$. (It is straightforward to verify that this is compatible with the operator product.) However, the vacuum module at $\lambda\sim o$ is more complicated. We expect this to be the familiar vacuum module of the $W_{1+\infty}$ vertex algebra, but our generators \smash{$\tilde\cV_s(\lambda)$} are not standard. (See \cite{Pope:1990be} for details of the isomorphism.) Indeed, they are not even quasi-primaries under the Witt subalgebra. It is not clear to us how to recover the usual $W_{1+\infty}$ vacuum module on correspondence space.

\subsection{Collinear splitting in KP} 

Finally, we rederive the $W_{1+\infty}$ vertex algebra \eqref{eq:W(1+infinity)-mode-algebra} from collinear splitting in KP with dispersion. Recall that scattering states $\Phi_k(x) = \exp(\im k\cdot x)$ are labelled by $(\omega,\kappa)\sim\bigl(t^{-2}\omega,t\kappa\bigr)$ in terms of which the momentum is
\begin{displaymath} k_{\al\beta}(\omega,\kappa) = \omega\kappa_\al\kappa_\beta - \frac{\fq^2}{12}\omega^3\la\kappa\iota\ra^6\iota_\al\iota_\beta . \end{displaymath}
The vertex $k_+^3/3$ is unmodified, but the propagator is changed to \smash{$-\im/\bigl(k^2+\tfrac{\fq^2}{6}k_+^4\bigr)$}. The perturbiner evaluates to
\begin{displaymath} \cS_{\rm KP}^{\omega_1,\omega_2}(\kappa_1,\kappa_2) = - \frac{2\im k_{1+}k_{2+}(k_1+k_2)_+}{(k_1+k_2)^2 + \frac{\fq^2}{6}(k_1+k_2)_+^4}{\rm e}^{\im(k_1+k_2)\cdot x} . \end{displaymath}
Exploiting the identity \eqref{eq:KP-prop-denominator}, we can rewrite this in terms of spinor helicity variables to obtain
\begin{equation} \label{eq:KP-perturbiner}
\cS_{\rm KP}^{\omega_1,\omega_2}(\kappa_1,\kappa_2) = - \frac{\im \la1\iota\ra^2\la2\iota\ra^2\bigl(\omega_1\la1\iota\ra^2+\omega_2\la2\iota\ra^2\bigr)}{\la12\ra^2 + \frac{\fq^2}{4}\la1\iota\ra^2\la2\iota\ra^2\bigl(\omega_1\la1\iota\ra^2+\omega_2\la2\iota\ra^2\bigr)^2}{\rm e}^{\im(k_1+k_2)\cdot x} .
\end{equation}
We need to rewrite the singular part of the exponential $\exp(\im(k_1+k_2)\cdot x)$ in terms of on-shell states. In order to do so, we exploit the identity
\begin{displaymath} \begin{aligned}
(k_1+k_2)_{\al\beta} ={}& \omega_1\kappa_{1\al}\kappa_{1\beta} + \omega_2\kappa_{2\al}\kappa_{2\beta} - \frac{\fq^2}{12}\bigl(\omega_1^3\la1\iota\ra^6 + \omega_2^3\la2\iota\ra^6\bigr)\iota_\al\iota_\beta \\
={}& k_{\al\beta}\biggl(\frac{\omega_1\la1\iota\ra^2+\omega_2\la2\iota\ra^2}{\la2\iota\ra^2},\kappa_2\biggr) - 2\omega_1\frac{\la12\ra\la1\iota\ra}{\la2\iota\ra}\iota_{(\al}\kappa_{2\beta)} + \omega_1\frac{\la12\ra^2}{\la2\iota\ra^2}\iota_\al\iota_\beta \\
&{+}\, \frac{\fq^2}{4}\omega_1\omega_2\la1\iota\ra^2\la2\iota\ra^2\bigl(\omega_1\la1\iota\ra^2+\omega_2\la2\iota\ra^2\bigr)\iota_\al\iota_\beta \\
={}& k_{\al\beta}\biggl(\frac{K_+}{\la2\iota\ra^2},\frac{\omega_2\la2\iota\ra^2\kappa_2 + \omega_1\la1\iota\ra\la2\iota\ra\kappa_1}{K_+}\biggr)\! +\! \frac{\omega_1\omega_2}{K_+}\biggl(\la12\ra^2 + \frac{\fq^2}{4}\la1\iota\ra^2\la2\iota\ra^2K_+^2\biggr)\iota_\al\iota_\beta,
\end{aligned} \end{displaymath}
where as above $K_+ = (k_1+k_2)_+$. It is no surprise that the exponent can be written this way as the denominator \eqref{eq:KP-perturbiner} measures the failure of $k_1+k_2$ to be on-shell. Since
\begin{displaymath} \exp\biggl(\frac{\omega_1\omega_2}{K_+}\biggl(\la12\ra^2 + \frac{\fq^2}{4}\la1\iota\ra^2\la2\iota\ra^2K_+^2\biggr)x^-\biggr) = 1 + \cO\biggl(\la12\ra^2 + \frac{\fq^2}{4}\la1\iota\ra^2\la2\iota\ra^2K_+^2\biggr), \end{displaymath}
this term does not affect the singular part of the splitting function; therefore, modulo regular~terms
\begin{gather} \label{eq:KP-splitting} \cS_{\rm KP}^{\omega_1,\omega_2}(\kappa_1,\kappa_2) \sim - \frac{\im\la1\iota\ra^2\la2\iota\ra^2K_+}{\la12\ra^2 + \frac{\fq^2}{4}\la1\iota\ra^2\la2\iota\ra^2K_+^2}\Phi^{K_+/\la2\iota\ra^2}\biggl(x;\frac{\omega_2\la2\iota\ra^2\kappa_2 + \omega_1\la1\iota\ra\la2\iota\ra\kappa_1}{K_+}\biggr) . \end{gather}
We can somewhat simplify the spinor argument appearing on the right-hand side by noticing~that
\begin{displaymath} \frac{\omega_2\la2\iota\ra^2\kappa_2 + \omega_1\la1\iota\ra\la2\iota\ra\kappa_1}{K_+} = \kappa_2 - \frac{\omega_1\la12\ra\la1\iota\ra}{K_+} . \end{displaymath}
Let us try to unpack the formula \eqref{eq:KP-splitting} in terms of soft modes as defined in equation \eqref{eq:KP-soft-modes}. This requires expanding the propagator perturbatively in $\fq^2$
\begin{displaymath} \frac{K_+}{\la12\ra^2 + \frac{\fq^2}{4}\la1\iota\ra^2\la2\iota\ra^2K_+^2} = \sum_{\ell\geq0}\frac{(-)^\ell\fq^{2\ell}}{2^{2\ell}}\frac{\la1\iota\ra^{2\ell}\la2\iota\ra^{2\ell}K_+^{2\ell+1}}{\la12\ra^{2\ell+2}} . \end{displaymath}
We have
\begin{gather*}
\frac{\la1\iota\ra^2\la2\iota\ra^2K_+}{\la12\ra^2 + \frac{\fq^2}{4}\la1\iota\ra^2\la2\iota\ra^2K_+^2}\Phi^{K_+/\la2\iota\ra^2}\biggl(x;\kappa_2 - \frac{\omega_1\la12\ra\la1\iota\ra}{K_+}\iota\biggr) \\
\qquad{}= \sum_{\ell\geq0,s\geq1}\frac{\im^{2\ell+s}\fq^{2\ell}}{(s-1)!2^{2\ell}}\frac{\la1\iota\ra^{2(\ell+1)}\la2\iota\ra^{2(\ell+2-s)}K_+^{2\ell+s}}{\la12\ra^{2\ell+2}}\Phi_s\biggl(x;\kappa_2 - \frac{\omega_1\la12\ra\la1\iota\ra}{K_+}\iota\biggr) \\
\qquad{}\sim \sum_{\ell\geq0,s\geq1}\frac{\im^{2\ell+s}\fq^{2\ell}}{(s-1)!2^{2\ell}}\sum_{k=0}^{2\ell+1}(-)^k\frac{\la1\iota\ra^{2\ell+2+k}\la2\iota\ra^{3-2s+k}\omega_1^kK_+^{2\ell+s-k}}{k!(2\ell+1-k)!} \\
\qquad\hphantom{\sim \sum_{\ell\geq0,s\geq1}\frac{\im^{2\ell+s}\fq^{2\ell}}{(s-1)!2^{2\ell}}\sum_{k=0}^{2\ell+1}}{}
\times\bigl(\iota^\al\p_{\kappa_1^\al}\bigr)^{2\ell+1-k}\biggl(\frac{1}{\la12\ra}\biggr)\bigl(\iota^\al\p_{\kappa_2^\al}\bigr)^k\Phi_s(x;\kappa_2) .
\end{gather*}
Now let us eliminate the variable $s$ in favour of $t=2\ell+s$ taking values in the range $t\geq2\ell+1$. The splitting function is
\begin{displaymath} \begin{aligned}
\sum_{\ell\geq0,t\geq2\ell+1}\sum_{k=0}^{2\ell+1}{}&{}\frac{\im^t\fq^{2\ell}}{2^{2\ell}}\frac{(-)^k\la1\iota\ra^{2\ell+2+k}\la2\iota\ra^{3-2(t-2\ell)+k}\omega_1^kK_+^{t-k}}{k!(2\ell+1-k)!(t-2\ell-1)!} \\
&\times\bigl(\iota^\al\p_{\kappa_1^\al}\bigr)^{2\ell+1-k}\biggl(\frac{1}{\la12\ra}\biggr)\bigl(\iota^\al\p_{\kappa_2^\al}\bigr)^k\Phi_{t-2\ell}(x;\kappa_2) .
\end{aligned} \end{displaymath}
Concentrating on the $\omega_i$ dependence, we expand
\begin{displaymath}\omega_1^kK_+^{t-k} = \sum_{q=0}^{t-k}\binom{t-k}{q}\la1\iota\ra^{2(t-q-k)}\la2\iota\ra^{2q}\omega_1^{t-q}\omega_2^q . \end{displaymath}
Identifying the soft modes $\Phi_s(x;\kappa)$ with chiral algebra states $\cV_s(\kappa)$, we can extract the OPE.

To do so, we identify $q=s_2-1$, and $t=s_1+s_2-2$. The restriction $t\geq2\ell+1$ now limits the sum over $0\leq2\ell\leq s_1+s_2-3$. Furthermore, $t-k\geq q$ becomes $k\leq s_1-1$, so that $0\leq k \leq \min\{s_1-1,2\ell+1\}$. In full, the soft OPE is
\begin{gather}
\cV_{s_1}(\kappa_1)\cV_{s_2}(\kappa_2)\label{eq:KP-softOPE}\\
\qquad{}\sim \sum_{\ell=0}^{\lfloor(s_1+s_2-3)/2\rfloor}\frac{\fq^{2\ell}}{2^{2\ell}}\frac{(s_1-1)!}{(s_1+s_2-3-2\ell)!}\sum_{k=0}^{\min\{s_1-1,2\ell+1\}}\frac{(-)^k(s_1+s_2-2-k)!}{k!(s_1-1-k)!(2\ell+1-k)!} \nonumber\\
\quad\qquad{}\times\la1\iota\ra^{2(\ell+s_1)-k}\la2\iota\ra^{2(2\ell+2-s_1)+k+1}\bigl(\iota^\al\p_{\kappa_1^\al}\bigr)^{2\ell+1-k}\biggl(\frac{1}{\la12\ra}\biggr)\bigl(\iota^\al\p_{\kappa_2^\al}\bigr)^k\cV_{s_1+s_2-2-2\ell}(\kappa_2) \nonumber\\
\qquad{}= \sum_{\ell=0}^{\lfloor(s_1+s_2-3)/2\rfloor}\frac{\fq^{2\ell}}{(2\ell+1)!2^{2\ell}}\sum_{k=0}^{2\ell+1}(-)^k\binom{2\ell+1}{k}[s_1+s_2-2-k]_{2\ell+1-k}[s_1-1]_k
\nonumber\\
\quad\qquad{}\times\la1\iota\ra^{2(\ell+s_1)-k}\la2\iota\ra^{2(2\ell+2-s_1)+k+1}\bigl(\iota^\al\p_{\kappa_1^\al}\bigr)^{2\ell+1-k}\biggl(\frac{1}{\la12\ra}\biggr)\bigl(\iota^\al\p_{\kappa_2^\al}\bigr)^k\cV_{s_1+s_2-2-2\ell}(\kappa_2) .\nonumber
\end{gather}
In the final formula, we have eliminated the condition $k\leq s_1-1$ with the understanding that the descending factorial $[a]_b = a(a-1)\cdots (a-(b-1))$ vanishes when $b>a$ (for~$a$,~$b$ positive integers).

As is immediately clear from our formula for the hard OPE, equation \eqref{eq:KP-softOPE} is symmetric under simultaneous exchange of $\kappa_1\leftrightarrow\kappa_2$, $s_1\leftrightarrow s_2$. We can make this symmetry manifest by rewriting the OPE as
\begin{align} \label{eq:sym-KP-OPE}
&\cV_{s_1}(\kappa_1)\cV_{s_2}(\kappa_2) \\
&\qquad{}\sim\sum_{\ell\geq0}\frac{\fq^{2\ell}}{(2\ell+1)!2^{2\ell}}\la1\iota\ra^{2s_1-1}\la2\iota\ra^{4\ell-2s_1+5}M^{s_1,s_2}_{2\ell}(\la1\iota\ra\la\iota\p_{\kappa_1}\ra,\la2\iota\ra\la\iota\p_{\kappa_2}\ra)\frac{\cV_{s_1+s_2-2\ell-2}(\kappa_2)}{\la12\ra}
\nonumber
\end{align}
and
\begin{displaymath} M^{s_1,s_2}_\ell(m,n) = \sum_{k=0}^{\ell+1}(-)^k\binom{\ell+1}{k}[s_1-1]_k[s_2-1]_{\ell+1-k}m^{\ell+1-k}n^k . \end{displaymath}
In equation \eqref{eq:sym-KP-OPE}, we removed the upper restriction on $\ell$ since \smash{$M^{s_1,s_2}_{2\ell}$} vanishes outside this range anyway. To see that this reproduces equation \eqref{eq:KP-softOPE}, we expand
\begin{displaymath} \la2\kappa\ra^k\la\iota\p_{\kappa_2}\ra^k\biggl(\frac{\cV_s(\kappa_2)}{\la12\ra}\biggr) = \sum_{i=0}^k(-)^{k-i}\binom{k}{i}\la1\iota\ra^{k-i}\la\iota\p_{\kappa_1}\ra^{k-i}\biggl(\frac{1}{\la12\ra}\biggr)\la2\kappa\ra^i\la\iota\p_{\kappa_2}\ra^i\cV_s(\kappa_2) . \end{displaymath}
The sum over $i\leq k\leq2\ell+1$ can then be performed directly
\begin{displaymath}
\sum_{k=i}^{2\ell+1}\frac{[s_1-1]_k[s_2-1]_{2\ell+1-k}}{(2\ell+1-k)!(k-i)!} = \frac{[s_1-1]_i[s_1+s_2-2-i]_{2\ell+1-i}}{(2\ell+1-i)!} ,
\end{displaymath}
as desired. Rewriting the OPE in terms of rescaled states \smash{$\til\cV_s(\kappa) = \la1\iota\ra^{2(1-2s)}\cV_s(\kappa)$} of conformal spin~$s$, it reads
\begin{displaymath}
\til\cV_{s_1}(\kappa_1)\til\cV_{s_2}(\kappa_2)\sim \sum_{\ell\geq0}\frac{\fq^{2\ell}}{(2\ell+1)!2^{2\ell}}\frac{\la2\iota\ra^{2s_1-1}}{\la1\iota\ra^{2s_1-1}} M^{s_1,s_2}_{2\ell}\biggl(\frac{\la1\iota\ra\la\iota\p_{\kappa_1}\ra}{\la2\iota\ra^2},\frac{\la\iota\p_{\kappa_2}\ra}{\la2\iota\ra}\biggr)\frac{\til\cV_{s_1+s_2-2\ell-2}(\kappa_2)}{\la12\ra} .
\end{displaymath}
The ratio $\la1\iota\ra/\la2\iota\ra$ is independent of $\iota$ modulo terms of order $\la12\ra$, and $\la\iota\p_\kappa\ra/\la\kappa\iota\ra$ is also independent of $\iota$ modulo the addition of the homogeneity operator. Therefore, the singular part of the OPE is independent of the reference spinor $\iota$. In terms of the inhomogeneous co-ordinate $w = \la o\kappa\ra/\la\kappa\iota\ra$, the OPE reads
\begin{displaymath} \til\cV_{s_1}(w_1)\til\cV_{s_2}(w_2) \sim - \sum_{\ell\geq0}\frac{\fq^{2\ell}}{(2\ell+1)!2^{2\ell}}M^{s_1,s_2}_{2\ell}\bigl(\p_{w_1},\p_{w_2}\bigr)\frac{\til\cV_{s_1+s_2-2\ell-2}(w_2)}{w_{12}} . \end{displaymath}
It is not immediate that this operator product associates. To verify this, it is useful to expand in~modes.

The corresponding modes are
\begin{displaymath} \til\cV_{s,m} = \frac{1}{2\pi\im}\oint_{w=0}\dif w\,w^{s-1+m}\til\cV_s(w) , \end{displaymath}
and their algebra is determined by
\begin{align*}
&\frac{1}{(2\pi\im)^2}\oint_{w_2=0}\oint_{w_{12}=0}w_1^{s_1-1+m}w_2^{s_2-1+n}\p_{w_1}^{2\ell+1-k}\p_{w_2}^k\biggl(\frac{\tilde\cV_{s_1+s_2-2\ell-2}(w_2)}{w_{12}}\biggr) \\
&\quad{}= (-)^{2\ell+1}[s_1-1+m]_{2\ell+1-k}[s_2-1+n]_k\oint_{w_2=0}w_2^{s_1+s_2-2+m+n-(2\ell+1)}\til\cV_{s_1+s_2-2\ell-2}(w_2) \\
&\quad{}= (-)^{2\ell+1}[s_1-1+m]_{2\ell+1-k}[s_2-1+n]_k\til\cV_{s_1+s_2-2\ell-2,m+n} .
\end{align*}
Therefore, the mode algebra is
\begin{displaymath}
\bigl[\til\cV_{s_1,m},\til\cV_{s_2,n}\bigr] = \sum_{\ell\geq0}\frac{\fq^{2\ell}}{(2\ell+1)!2^{2\ell}}N^{s_1,s_2}_{2\ell}(m,n)\til\cV_{s_1+s_2-2\ell-2,m+n} . \end{displaymath}
We recognise this as the mode algebra of $W_{1+\infty}$ obtained directly from correspondence space in Section~\ref{subsec:W(1+infty)-PS} (and therefore it obeys the Jacobi identity).

\section{Discussion} \label{sec:discussion}

There are a number of natural extensions to this work.

{\bf Other examples of minitwistorial theories.} We have shown that Abelian Poisson- and non-commutative $5d$ Chern--Simons theory on $\PS$ with
\begin{displaymath} \Omega = \frac{\la\lambda\dif\lambda\ra\wedge\dif x^{\al\beta}\lambda_\al\lambda_\beta}{\la\lambda\iota\ra^4} \end{displaymath}
describe the dKP and KP equation on space-time respectively. There are a number of natural modifications of this set-up which will engineer other $3d$ minitwistorial theories.

We could instead choose to trivialise the canonical bundle with the divisor $\la o\lambda\ra^2\la\lambda\iota\ra^2$. On-shell arguments suggest that for this measure $5d$ Poisson--Chern--Simons will lead to $3d$ Toda theory (also known as $\gSU_\infty$ Toda) \cite{Ward:1990qt}. We expect that switching on non-commutativity leads to the infinite $2d$ Toda lattice with infinitesimal spacing \cite{Mikhailov:1979tc,Ueno:1984zz}. Trivialising the canonical bundle with $\mu^2$ leads to these same theories, although perturbed around non-trivial backgrounds. One attractive feature of these set-ups is that they make sense in both Lorentzian and Euclidean signature.

Another natural generalisation is to study non-Abelian $5d$ Chern--Simons theory on $\PS$. In~the commutative case, this was introduced in \cite{Bittleston:2020hfv} and depending on the choice of divisor leads to the Manakov--Zakharov--Ward model \cite{Manakov:1981kz,Ward:1988ie} or its pseudo-dual \cite{Mason:1991rf}. It would be interesting to understand the consequences of switching on non-commutativity on correspondence space in these examples. Natural candidates for the resulting space-time theories can be found in the literature \cite{Date:1981opa,Ueno:1984zz} (potentially including the Davey--Stewartson equation \cite{Kajiwara:1990kph}).

Just as for $4d$ space-time, there are BF counterparts of these theories which crucially do not rely on trivialising the canonical bundle. $5d$ mixed topological-holomorphic BF theory on correspondence space descends to the monopole theory; a Lagrangian multiplier type action for the Bogomol'nyi equation \cite{Adamo:2017xaf, Hitchin:1982gh}. It has a gravitational counterpart which on correspondence space describes generic deformations of the transverse holomorphic foliation and descends to a~Lagrange multiplier type action for the Einstein--Weyl equations \cite{Hitchin:1982vry,Jones:1985pla}.

There are also a number of modified twistor constructions appearing in the literature which could be adapted to the correspondence space of $3d$ space-time. These include orbifolds \cite{Cole:2024sje}, higher genus twistor spaces \cite{Jarov:2025hgt} and their singular degenerations \cite{Cole:2023umd}.

{\bf Three-dimensional integrability.} Crucial developments in the study of the KP equation, and indeed integrable partial differential equations more generally, were made by the Japanese school \cite{Date:1981opa,Jimbo:1981tov,Jimbo:1981mpt,Jimbo:1981mpf, Sato:1983se}; the geometric and analytic underpinnings of this approach were subsequently developed by Segal and Wilson \cite{Segal:1985aga}. The central construct is the $\tau$-function $u = \phi_x = -\fq^2(\log\tau)_{xx}$ which is equated with a $2d$ CFT correlator with operator insertions. We expect that the universal vertex algebra supported on minitwistor lines is intimately connected with this $2d$ CFT, and that the inserted operators are generated by defects on correspondence space intersecting the minitwistor line at points. (A link between the $\tau$-function and the minitwistor $\CP^1$ was also suggested in \cite{Mason:2001vj}.) Natural candidates for such defects are topological lines supported at points on $\mT$.

Whilst it seems the most direct connection between this work and more traditional approaches to the KP equation is via Sato theory, it would also be enlightening to interpret the inverse scattering transform \cite{Ablowitz:1991xb, Manakov:1981inv}, Hirota's direct method \cite{Date:1981opa, Hirota:1986uc} and the connection with topological strings \cite{Dijkgraaf:1990rs, Fukuma:1990jw} from the minitwistor perspective.

We also note that there have been alternative proposals for minitwistor formulations of the KP equation through `generalized twistor correspondences' \cite{Mason:1995db,Mason:1991rf}, though whether these have any relation to non-commutative $5d$ Chern--Simons theory is unclear to us.

{\bf Celestial holography.} The Einstein--Weyl geometry relevant to the dKP equation can be lifted to a self-dual metric on a $4d$ space-time of signature $(2,2)$~\cite{Dunajski:2001ea, Dunajski:2000rf}. Part of the original motivation for this work was to seek a model of celestial holography for self-dual Einstein metrics, in the simplified setting of a symmetry reduction to $3d$. From this perspective, $w_{1+\infty}$ and $W_{1+\infty}$ can be viewed as celestial chiral algebras for dKP and KP theory. It is natural to ask whether these embed inside the celestial chiral algebras of $4d$ integrable theories.

{\bf Quantum corrections and twisted holography.} We have worked exclusively at the classical level, but we expect that both Poisson- and non-commutative $5d$ Chern--Simons on correspondence space admit one-loop quantizations. In fact, we expect any $5d$ mixed topological-holomorphic theory on $\PS$ to admit a one-loop quantization, because in odd dimensions one-loop perturbative anomalies vanish. A direct consequence should be the vanishing of one-loop amplitudes in the corresponding $3d$ space-time theories.

In the case of non-commutative $5d$ Chern--Simons, we can hope for more. Costello has constructed the theory perturbatively at all loop orders on $\bbC^2\times\bbR$, albeit at the cost of introducing counterterms accompanied by powers of $\hbar/\fq^3$. We conjecture that a similar quantization exists on $\PS$, leading to a quantum integrable counterpart of KP theory. The $W_{1+\infty}$ vertex algebra has a non-linear deformation $\cW_{1+\infty}$ \cite{Gaberdiel:2012ku,Hornfeck:1994is,Linshaw:2017tvv,Prochazka:2014gqa} depending on an additional parameter $\hbar$; here non-linear refers to the fact that the normal-ordered products of fields can appear on the right-hand side of operator product expansions. The deformation arises from loop corrections to holomorphic surface defects in $5d$ non-commutative Chern--Simons theory \cite{Costello:2016nkh,Gaiotto:2020dsq,Oh:2020hph}. We anticipate that this same deformation arises on correspondence space; a direct check would be to match one-loop corrections to the collinear splitting of form factors in KP theory to the order~$\hbar$ terms in the operator products of $\cW_{1+\infty}$. Similar computations have been performed in~$4d$ self-dual gauge theory \cite{Costello:2022upu} and gravity \cite{Bittleston:2022jeq}.

Finally, $5d$ non-commutative Abelian Chern--Simons theory conjecturally arises as the twist of M-theory in the $\Omega$ background. Holomorphic and topological line defects are the twisted avatars of M2 and M5 branes \cite{Costello:2016nkh,Costello:2017fbo}. They obey twisted and $\Omega$-deformed versions of AdS/CFT. It is natural to ask whether these same dualities can implemented on correspondence space, yielding~$3d$ counterparts of the celestial holograms developed in \cite{Bittleston:2024efo,Costello:2023hmi,Costello:2022jpg}.

\begin{appendix}

\section{Distributor} \label{app:distributor}

In this appendix, we derive the distributor of $\dift$ over the associative product $\star$ appearing in equation \eqref{eq:distributor}. In order to do so we will need to recall equation \eqref{eq:dift-p0}
\begin{displaymath}
\dift\p_0 = 2\frac{\tilde\sigma^\al\lambda_\al}{\la\iota\lambda\ra}\otimes\p_1 ,
\end{displaymath}
and furthermore
\begin{equation} \label{eq:crucial-Lie-derivative}
\cL_{\p_0}\frac{\tilde\sigma^\al\lambda_\al}{\la\lambda\iota\ra} = \dif x^{\al\beta}\iota_\al\iota_\beta = \dif x^- .
\end{equation}
We can write the star product $\star$ of \smash{$f,g\in\til\Omega^{0,\bullet}(\PS)$} as
\begin{displaymath}
f\star g = \sum_{\ell\geq0}\frac{\fq^\ell}{\ell!2^\ell}\eps^{j_1i_1}\cdots\eps^{j_\ell i_\ell}\p_{i_1}\cdots\p_{i_\ell}f\p_{j_1}\cdots\p_{j_\ell}g .
\end{displaymath}
In our chosen conventions $\eps^{10} = 1$, all $(1,0)$-vector fields are implicitly acting by Lie derivatives and we suppress wedge products. Acting with $\dift$ gives
\begin{align} \label{eq:distribute-step1}
&\dift(f\star g) = \dift f\star g + (-)^{\mrm{deg}\,f}f\star\dift g \\
&\qquad{}+ \sum_{\ell\geq1}\frac{\fq^\ell}{\ell!2^\ell}\eps^{j_1i_1}\cdots\eps^{j_\ell i_\ell}\bigl(\bigl[\dift,\p_{i_1}\cdots\p_{i_\ell}\bigr]f\p_{j_1}\cdots\p_{j_\ell}g + (-)^{\mrm{deg}\,f}\p_{i_1}\cdots\p_{i_\ell}f\bigl[\dift,\p_{j_1}\cdots\p_{j_\ell}\bigr]g\bigr) .\nonumber
\end{align}
Now
\begin{displaymath} \begin{aligned}
\bigl[\dift,\p_{i_1}\cdots\p_{i_\ell}\bigr] &= \sum_{k=1}^\ell\p_{i_1}\cdots\p_{i_{k-1}}\bigl[\dift,\p_{i_k}\bigr]\p_{i_{k+1}}\cdots\p_{i_\ell} \\
&= 2\sum_{k=1}^\ell\delta^0_{i_k}\p_{i_1}\cdots\p_{i_{k-1}}\frac{\tilde\sigma^\al\lambda_\al}{\la\iota\lambda\ra}\p_{i_{k+1}}\cdots\p_{i_\ell}\p_1 .
\end{aligned} \end{displaymath}
Here we have migrated $\p_1$ to the right-hand side of the expression harmlessly. We'd like to pull out the $(0,1)$-form $\tilde\sigma^\al\lambda_\al/\la\iota\lambda\ra$ but this must be done with care since $\p_0$ can act on it. Using equation~\eqref{eq:crucial-Lie-derivative}, we have
\begin{displaymath} \begin{aligned}
\bigl[\dift,\p_{i_1}\cdots\p_{i_\ell}\bigr] ={}& 2\frac{\tilde\sigma^\al\iota_\al}{\la\iota\lambda\ra}\sum_{k=1}^\ell\delta^0_{i_k}\p_{i_1}\cdots\p_{i_{k-1}}\p_{i_{k+1}}\cdots\p_{i_\ell}\p_1 \\
&{+}\, 2\dif x^-\sum_{k=2}^\ell\sum_{m=1}^{k-1}\delta^0_{i_m}\delta^0_{i_k}\p_{i_1}\cdots\p_{i_{m-1}}\p_{i_{m+1}}\cdots\p_{i_{k-1}}\p_{i_{k+1}}\cdots\p_{i_\ell}\p_1 .
\end{aligned} \end{displaymath}
When plugging this into equation \eqref{eq:distribute-step1}, we find that precisely which indices $i_k$, $i_m$ have been fixed to zero is immaterial, as they are dummy variables and the vector fields $\p_i$ commute with one another. The sum over $k$, $m$ therefore generates a factor of $\ell(\ell-1)/2$ and the sum over~$k$ alone gives $\ell$. The sign factor $(-)^{\mrm{deg}\,f}$ in the final term of \eqref{eq:distribute-step1} cancels when we bring a $1$-form past~$f$. The only remaining signs are generated when $\epsilon$-symbols have their indices fixed. We~find
\begin{gather*}
\dift(f\star g) = \dift f\star g + (-)^{\mrm{deg} f}f\star\dift g \\
\hphantom{\dift(f\star g) =}{}
+\fq\frac{\tilde\sigma^\al\lambda_\al}{\la\iota\lambda\ra} \sum_{\ell\geq1}\frac{\fq^{\ell-1}}{(\ell-1)!2^{\ell-1}}\eps^{j_1i_1}\cdots\eps^{j_{\ell-1}i_{\ell-1}} \\
\hphantom{\dift(f\star g) =+\fq\frac{\tilde\sigma^\al\lambda_\al}{\la\iota\lambda\ra}\!\sum_{\ell\geq1}}{}
\!\times\!\bigl(\p_{i_1}\cdots\p_{i_{\ell-1}}\p_1f\p_{j_1}\cdots\p_{j_{\ell-1}}\p_1g - \p_{i_1}\cdots\p_{i_{\ell-1}}\p_1f\,\p_{j_1}\cdots\p_{j_{m-1}}\p_1g\bigr) \\
\hphantom{\dift(f\star g) =}{}
+ \frac{\fq^2}{4}\dif x^- \sum_{\ell\geq2}\frac{\fq^{\ell-2}}{(\ell-2)!2^{\ell-2}}\eps^{j_1i_1}\cdots\eps^{j_{\ell-2}i_{\ell-2}} \\
\hphantom{\dift(f\star g) =+ \frac{\fq^2}{4}\dif x^-\!\sum_{\ell\geq2}}{}
\!\times\!\bigl(\p_{i_1}\cdots\p_{i_{\ell-2}}\p_1f\,\p_{j_1}\cdots\p_{j_{\ell-2}}\p_1^2g + \p_{i_1}\cdots\p_{i_{\ell-1}}\p_1f\,\p_{j_1}\cdots\p_{j_{\ell-1}}\p_1^2g\bigr) .
\end{gather*}
Now the first sum vanishes as a result of the relative sign $\eps^{01} = -\eps^{10}$. The second term does not vanish, leading to the succinct formula
\begin{displaymath} \dift(f\star g) = \dift f\star g + (-)^{\mrm{deg}\,f}f\star\dift g + \frac{\fq^2}{4}\dif x^-\p_1(\p_1f\star\p_1g) . \end{displaymath}
This is equation \eqref{eq:distributor} from the manuscript. Notice that the extra term is a total derivative, consistent with the difference between the Moyal and commutative products being exact.

\section[Independence of Poisson--Chern--Simons from the distribution D]{Independence of Poisson--Chern--Simons\\ from the distribution $\boldsymbol{\cD}$}
\label{app:Dindependence}

In Section~\ref{subsec:correspondence-PCS}, we made a choice of rank two integrable subbundle $\cD\subset T_\bbC\PS$ to represent~$T^{1,0}\PS$. We were motivated to make this choice because quantization of the Poisson bivector was particularly simple in this frame. However, this frame had the property that the pushforward
\begin{displaymath} \pi_*\p_1 = \la\lambda\iota\ra^2\p_\mu + \overline{\la\iota\lambda\ra}^2\p_{\bar\mu} \end{displaymath}
does not lie in $T^{1,0}\mT$. As explained in Section~\ref{subsec:ambiguity}, this is harmless because the translation $\p_{\bar\mu}$ is BRST exact.

Nevertheless, it is natural to seek a more geometric definition of the theory. In this appendix we first introduce an alternative choice of frame that arises from a natural holomorphic lift of the holomorphic tangent bundle on minitwistor space. We then consider $5d$ Chern--Simons theory defined using this holomorphic frame. After a field redefinition, we find that this theory is equivalent to the definition of $5d$ Chern--Simons employing the constant frame of Section~\ref{subsec:correspondence-PCS}.

\subsection[5d Chern--Simons theory in the holomorphic frame]{$\boldsymbol{5d}$ Chern--Simons theory in the holomorphic frame} 

Consider the alternative choice of distribution
\begin{displaymath}
\cD^\prime = \biggl\la\bar\lambda^\al\frac{\p}{\p\lambda^\al}\bigg|_x ,\bar\lambda^\al\bar\lambda^\beta\frac{\p}{\p x^{\al\beta}}\biggr\ra_\bbC ,
\end{displaymath}
with frame
\begin{displaymath}
\xi_0 = \frac{\bar\lambda^\al}{\bigl\la\lambda\bar\lambda\bigr\ra}\frac{\p}{\p\lambda^\al}\bigg|_x ,\qquad\xi_1 = \frac{\bar\lambda^\al\bar\lambda^\beta}{\bigl\la\lambda\bar\lambda\bigr\ra^2}\frac{\p}{\p x^{\al\beta}} .
\end{displaymath}
Note that $\xi_0$, $\xi_1$ transform as sections of \smash{$T^{1,0}_{\cD^\prime}\PS\otimes\cO(-2)$}. As for the constant frame, the derivative with respect to $\lambda$ is taken at fixed \smash{$x\in\bbR^{1,2}$. $\cD^\prime$} also fixes representatives for \smash{$\til\Omega^{0,\bullet}(\PS)$} which lies in its annihilator. In this case,
\begin{displaymath} \cD^\prime_\perp = \big\la\overline{\la\lambda\dif\lambda\ra} ,\dif x^{\al\beta}\bar\lambda_\beta\big\ra_\bbC \end{displaymath}
with
\begin{equation} \label{eq:lambda-bar-frame} \tilde {\rm e}^0 = \frac{\la\bar\lambda\dif\bar\lambda\ra}{\bigl\la\lambda\bar\lambda\bigr\ra^2} ,\qquad \tilde {\rm e}^\al = \frac{\dif x^{\al\beta}\bar\lambda_\beta}{\bigl\la\lambda\bar\lambda\bigr\ra} \end{equation}
as the corresponding frame.

Unlike the distribution $\cD$ used in the text, the distribution $\cD^\prime$ has the property ${\pi_*\cD^\prime\subset T^{1,0}\mT}$ whilst retaining the feature $\rho_*\xi_0 = 0$, so that $\xi_0$ is tangent to real minitwistor lines. However,~$\cD^\prime$~suffers from a significant drawback: on the locus $\lambda\sim\bar\lambda$ it non-trivially intersects~$T^{0,1}\PS$ and is therefore not a good section. Whilst the choice of frame \eqref{eq:lambda-bar-frame} is natural, it is also singular~here.

Let us now consider $5d$ Poisson--Chern--Simons theory defined on $\PS$ using $\cD^\prime$. We write \smash{$\cA = \cA_0\tilde {\rm e}^0 + \cA_\al\tilde {\rm e}^\al\in\til\Omega^{0,\bullet}_{\cD^\prime}(\PS)$} for the dynamical field of $5d$ Poisson--Chern--Simons theory in the frame~\eqref{eq:lambda-bar-frame}. The action takes a familiar form
\begin{equation} \label{eq:PCS-PS-prime} S_{\rm PCS}^\prime[\cA] = \frac{\im}{2\pi}\int_\PS\Omega\wedge\biggl(\frac{1}{2}\cA\wedge\dift\cA + \frac{1}{3}\cA\wedge\cL_{\la\lambda\iota\ra^2\xi_0}\cA\wedge\cL_{\la\lambda\iota\ra^2\xi_1}\cA\biggr) . \end{equation}
We must now face the fact that the basis $\bigl\{\tilde{e}^0,\tilde{e}^1\bigr\}$ becomes singular on $\bigl\la\lambda\bar\lambda\bigr\ra=0$. To handle this, we excise the neighbourhood ${\rm Ann}_\eps = \{|\mrm{Re}\,z|\leq\eps\}$ of this locus from $\PS$ and impose suitable boundary conditions. Varying the kinetic term on $\PS\setminus{\rm Ann}_\eps$ yields
\begin{align*}
\delta S_{\rm kin.}^\prime[\cA]={}& - \frac{\im}{4\pi}\int_{\PS\setminus{\rm Ann}_\eps}\Omega\wedge\dift(\cA\wedge\delta\cA) \\
&{+}\, \frac{\im}{2\pi}\int_{\PS\setminus{\rm Ann}_\eps}\delta\cA\wedge(\text{linearised equations of motion}) .
\end{align*}
On-shell we recover the boundary contribution
\begin{displaymath} \delta S_{\rm kin.}^\prime = \frac{\im}{8\pi}\int_{\bbR^{1,2}\times\{\mrm{Re}\,z = \eps\}}\Omega\wedge\tilde {\rm e}^\al\wedge\tilde e_\al\,\cA^\beta\delta\cA_\beta - \frac{\im}{4\pi}\int_{\bbR^{1,2}\times\{\mrm{Re}\,z = -\eps\}}\Omega\wedge\tilde {\rm e}^\al\wedge\tilde e_\al\,\cA^\beta\delta\cA_\beta . \end{displaymath}
Now \smash{$\dif x^{\gamma\delta}\lambda_\gamma\lambda_\delta\,\tilde {\rm e}^\al\,\tilde e_\al = - \sqrt{2}\dif^3x$}, so that
\begin{equation} \label{eq:holomorphic-bdry-term}
\delta S_{\rm kin.}^\prime = - \frac{\im}{4\sqrt{2}\pi}\int_{\bbR^{1,2}}\dif^3x\biggl(\oint_{\{\mrm{Re}\,z = \eps\}} - \oint_{\{\mrm{Re}\,z = -\eps\}}\biggr)\frac{\la\lambda\dif\lambda\ra}{\la\lambda\iota\ra^4}\,\cA^\beta\delta\cA_\beta .\end{equation}
The boundary conditions we impose on $\cA_\al$ are engineered so that they are compatible with the constant frame, at least at the linearised level. Whilst requiring that $\lambda^\gamma\cA_\gamma$ remains finite, we tolerate a singularity in the components $\cA_\al$ of the form
\begin{equation} \label{eq:real-bc} \cA_\al = - \frac{\bar\lambda_\al \lambda^\gamma}{\bigl\la\lambda\bar\lambda\bigr\ra}\cA_\gamma + \text{finite} . \end{equation}
The boundary contribution \eqref{eq:holomorphic-bdry-term} vanishes essentially by the same argument as in Section~\ref{subsec:bcs}.

The only remaining danger occurs at $\lambda\sim\iota$. The boundary condition we impose here is complicated by the fact that $\iota$ is real, so both $\la\lambda\iota\ra$ and $\bigl\la\lambda\bar\lambda\bigr\ra$ vanish there. We continue to require that a version of equation \eqref{eq:real-bc} holds, where now $\lambda^\gamma\cA_\gamma = \cO\bigl(\la\lambda\iota\ra^2\bigr)$ and
\begin{equation} \label{eq:mixed-bc} \cA_\al = - \frac{\bar\lambda_\al \lambda^\gamma}{\bigl\la\lambda\bar\lambda\bigr\ra}\cA_\gamma + \cA^\text{fin.}_\al \end{equation}
for $\cA^\text{fin.}_\al = \cO(\la\lambda\iota\ra)$. The terms of order $\la\lambda\iota\ra$ in the above expression are further required to obey the condition \eqref{eq:constant-bc}, that is
\begin{equation} \label{eq:iota-bc} 2\la\lambda\iota\ra o^\al\cA^\text{fin.}_\al + \la o\lambda\ra\iota^\al\cA^\text{fin.}_\al = \cO\bigl(\la\lambda\iota\ra^2\bigr) . \end{equation}
In order for infinitesimal gauge transformations to respect this condition, they should vanish second order at $\lambda\sim\iota$. It is natural to similarly require that $\tilde {\rm e}^0\cA_0$ vanishes to second order also.

\subsection{Comparing the constant and holomorphic frames} 

We now show that the Poisson--Chern--Simons action in the holomorphic frame \eqref{eq:PCS-PS-prime} is equivalent to that in the constant frame \eqref{eq:PCS-PS} used in the main text. The general argument that this should be true follows from a generalisation of a~lemma of Costello and Gwilliam~\cite[Section~9.2, Lemma~9.2.1]{Costello:2021jvx}. Here we will illustrate this by constructing a field redefinition that moves between the two theories. We will content ourselves to work just to cubic order in the action. Thus, we will explicitly find the relation between the field $a$ defined in the constant distribution $\cD$ and the field $\cA$ introduced in the previous appendix, correct to quadratic order. We also outline how, in principle, this construction could be continued to higher order.

Let us first understand how the gauge fields $a$ and $\cA$ are related at linear order, which follows simply by comparing the two frames. One finds
\begin{displaymath}
a = \frac{1}{\la\lambda\iota\ra}\cA + \frac{\bigl\la\bar\lambda\iota\bigr\ra}{\la\lambda\iota\ra^2\bigl\la\lambda\bar\lambda\bigr\ra}\iota^\gamma\cA_\gamma {\rm e}^1 + \cO\bigl(\cA^2\bigr) ,
\end{displaymath}
which follows since they agree modulo terms wedging to zero against $\Omega$, and $\p_1\ip a=0$. In~component form, this reads
\begin{displaymath}
a_\al = \frac{1}{\la\lambda\iota\ra}\cA_\al + \frac{\bigl\la\bar\lambda\iota\bigr\ra}{\la\lambda\iota\ra^2\bigl\la\lambda\bar\lambda\bigr\ra}\lambda_\al\lambda^\beta\cA_\beta + \cO\bigl(\cA^2\bigr) ,
\end{displaymath}
or dually
\begin{displaymath} \cA_\al = \la\lambda\iota\ra a_\al - \frac{\bigl\la\bar\lambda\iota\bigr\ra}{\bigl\la\lambda\bar\lambda\bigr\ra}\lambda_\al\lambda^\beta a_\beta + \cO\bigl(a^2\bigr) . \end{displaymath}
The boundary conditions \eqref{eq:real-bc}--\eqref{eq:iota-bc} follow from this linearised transformation.

We will also find it useful to introduce a frame for $\ell$ dual to the frame $\bigl\{\tilde {\rm e}^0,\tilde {\rm e}^\al\bigr\}$ for $\cD^\prime_\perp$,
\begin{displaymath} \tilde\xi_0 = - \bigl\la\lambda\bar\lambda\bigr\ra\lambda^\gamma\p_{\bar\lambda^\gamma}\big|_x ,\qquad \tilde\xi_\al = \frac{\bigl(\lambda^\beta\bar\lambda_\al - 2\bar\lambda^\beta\lambda_\al\bigr)}{\bigl\la\lambda\bar\lambda\bigr\ra}\lambda^\gamma\p_{\beta\gamma} . \end{displaymath}
Then $\p_0$, $\p_1$ and $\xi_0$, $\xi_1$ are related by
\begin{displaymath} \p_0\sim \la\lambda\iota\ra^2\xi_0 ,\qquad\p_1 = \la\lambda\iota\ra^2\xi_1 - \frac{\bigl\la\bar\lambda\iota\bigr\ra}{\bigl\la\lambda\bar\lambda\bigr\ra}\iota^\gamma\tilde\xi_\gamma = \la\lambda\iota\ra^2\xi_1 - r \end{displaymath}
in terms of which
\begin{displaymath} a = \cA + (r\ip\cA) \frac{{\rm e}^1}{\la\lambda\iota\ra^2} + \cO\bigl(\cA^2\bigr) . \end{displaymath}

To construct the higher-order terms, we must consider the Lagrangians of the two theories. Note that since $a$ and $\cA$ are both 1-forms, any higher-order corrections must involve contractions of $(0,1)$ vector fields into the gauge field $a$, since the form degree of the Chern--Simons Lagrangian is saturated by three $a$s. These will also be removable by a field redefinition: the Cartan formula \smash{$\bigl[\dift^\prime,\til V\ip\bigr] = \cL^{0,1}_{\til V}$} tells us that contraction and differentiation along $(0,1)$ vector fields cancel in $\dift$-cohomology. Therefore, by a field redefinition the Lagrangian can be brought into a form in which the vertices involve neither and so terminate at cubic order.

The kinetic terms are manifestly equivalent, so it suffices to compare the cubic vertices. For the sake of brevity, we will suppress wedge products. In the constant frame, the interaction is
\begin{displaymath} \begin{aligned}
S_{\rm int.}[a] &= \frac{\im}{6\pi}\int_\PS\Omega\,a\cL_{\p_0}a\cL_{\p_1}a \\
&= \frac{\im}{6\pi}\int_\PS\Omega\,\cA\cL_{\p_0}\biggl(\cA + (r\ip \cA)\frac{{\rm e}^1}{\la\lambda\iota\ra^2}\biggr)\cL_{\p_1}\biggl(\cA + (r\ip\cA)\frac{{\rm e}^1}{\la\lambda\iota\ra^2}\biggr) + \cO\bigl(\cA^4\bigr) .
\end{aligned} \end{displaymath}
In order for the shift from $a$ to $\cA$ to contribute, the Lie derivatives must act on the 1-forms~${\rm e}^1$, since left untouched they will wedge to zero against $\Omega$. These Lie derivatives are currently written in the constant frame $\cD$. To rewrite the interaction using the holomorphic frame, it will be useful to express the vertex in terms of ordinary Lie derivatives. The relevant structure equation is
\begin{displaymath} \dif\biggl(\frac{{\rm e}^1}{\la\lambda\iota\ra^2}\biggr) = 2\frac{\la\lambda\dif\lambda\ra}{\la\lambda\iota\ra^3}\dif x^{\al\beta}\lambda_\al\iota_\beta = 2\frac{\la\lambda\dif\lambda\ra}{\la\lambda\iota\ra^3}\tilde\sigma^\al\lambda_\al , \end{displaymath}
therefore
\begin{displaymath} \cL_{\p_0}\biggl(\frac{{\rm e}^1}{\la\lambda\iota\ra^2}\biggr) = \p_0\ip\dif\biggl(\frac{{\rm e}^1}{\la\lambda\iota\ra^2}\biggr) = \frac{2\tilde\sigma^\al\lambda_\al}{\la\lambda\iota\ra} ,\qquad \cL_{\p_1}\biggl(\frac{{\rm e}^1}{\la\lambda\iota\ra^2}\biggr) = \p_1\ip\dif\biggl(\frac{{\rm e}^1}{\la\lambda\iota\ra^2}\biggr) = 0 . \end{displaymath}
Wedging against $\Omega$,
\begin{displaymath}
\Omega\,\frac{\tilde\sigma^\al\lambda_\al}{\la\lambda\iota\ra} = \Omega\,\tilde {\rm e}^\al\lambda_\al,
 \end{displaymath}
which allows us to recast the vertex as
\begin{displaymath}
\frac{\im}{6\pi}\int_\PS\Omega\,\cA\bigl(\cL_{\la\lambda\iota\ra^2\xi_0}\cA + 2\tilde {\rm e}^\al\lambda_\al r\ip\cA\bigr)\bigl(\cL_{\la\lambda\iota\ra^2\xi_1}\cA - \cL_r\cA\bigr)
\end{displaymath}
to cubic order. At this stage, we can split the Lie derivatives into their $(1,0)$ and $(0,1)$ parts with respect to the \emph{holomorphic frame} $\cD^\prime$, which we write as \smash{$\cL_V = \cL^{1,0}_V + \cL^{0,1}_V$} with \smash{$\cL^{1,0}_V = \{V\ip,\p^\prime\}$} and \smash{$\cL^{0,1}_V = \bigl\{V\ip,\dift^\prime\bigr\}$}. If $V$ is $(1,0)$ vector field in the holomorphic frame, for example, $\xi_0$ or $\xi_1$, then
\begin{displaymath} \cL^{0,1}_V\cA = V\ip\dift^\prime\cA + \dift^\prime V\ip\cA = 0 \end{displaymath}
since $V$ contracts to zero against both $\cA$ and \smash{$\dift^\prime\cA$}. On the other hand, if $V$ is a $(0,1)$ vector field, for example,~$r$, then
\begin{displaymath} \bigl\{r\ip,\p^\prime\bigr\}\cA = \bigl(\p^\prime r\bigr)\ip\cA . \end{displaymath}
We can evaluate $\p^\prime r$ using the structure equations
\begin{displaymath} \begin{aligned}
&\dif\tilde {\rm e}^\al = \dif x^{\al\beta}\lambda_\beta\tilde {\rm e}^0 + \frac{\bigl\la\bar\lambda\iota\bigr\ra}{\la\lambda\iota\ra\bigl\la\lambda\bar\lambda\bigr\ra}{\rm e}^0\tilde {\rm e}^\al = - \lambda^\al\lambda_\beta\tilde {\rm e}^0\tilde {\rm e}^\beta - \frac{\bar\lambda^\al}{\bigl\la\lambda\bar\lambda\bigr\ra}{\rm e}^1\tilde {\rm e}^0 + \frac{\bigl\la\bar\lambda\iota\bigr\ra}{\la\lambda\iota\ra\bigl\la\lambda\bar\lambda\bigr\ra}{\rm e}^0\tilde {\rm e}^\al , \\
&\dif\tilde {\rm e}^0 = - 2\frac{\bigl\la\bar\lambda\iota\bigr\ra}{\la\lambda\iota\ra\bigl\la\lambda\bar\lambda\bigr\ra}{\rm e}^0\tilde {\rm e}^0 ,
\end{aligned} \end{displaymath}
where $\dif$ acts on fields with weight under scaling by trivialising with $\la\lambda\iota\ra$. (Objects invariant under scaling are independent of this choice.) We infer
\begin{displaymath} \begin{aligned}
&\p^\prime r\ip\tilde {\rm e}^\gamma = \p^\prime(r\ip\tilde {\rm e}^\gamma) + r\ip\p^\prime\tilde {\rm e}^\gamma = \p^\prime\biggl(\frac{\bigl\la\bar\lambda\iota\bigr\ra}{\bigl\la\lambda\bar\lambda\bigr\ra}\iota^\gamma\biggr) - \frac{\bigl\la\bar\lambda\iota\bigr\ra^2}{\la\lambda\iota\ra\bigl\la\lambda\bar\lambda\bigr\ra^2}\iota^\gamma {\rm e}^0 = 0 , \\
&\p^\prime r\ip\tilde {\rm e}^0 = \p^\prime\bigl(r\ip\tilde {\rm e}^0\bigr) + r\ip\p^\prime\tilde {\rm e}^0 = 0 .
\end{aligned} \end{displaymath}
As a result, the interaction vertex in the holomorphic frame reads
\begin{gather}
S_{\rm int.}[\cA] = \frac{\im}{6\pi}\int_\PS\Omega\,\cA\bigl(\cL_{\la\lambda\iota\ra^2\xi_0}\cA + 2\tilde {\rm e}^\al\lambda_\al r\ip\cA\bigr)\bigl(\cL_{\la\lambda\iota\ra^2\xi_1}\cA - \cL^{0,1}_r\cA\bigr)\nonumber \\
\hphantom{S_{\rm int.}[\cA]}{}
= \frac{\im}{6\pi}\int_\PS\Omega\,\cA\bigl(\cL_{\la\lambda\iota\ra^2\xi_0}\cA\cL_{\la\lambda\iota\ra^2\xi_1}\cA - \cL_{\la\lambda\iota\ra^2\xi_0}\cA\cL^{0,1}_r\cA \nonumber\\
\hphantom{S_{\rm int.}[\cA]= \frac{\im}{6\pi}\int_\PS\Omega\,\cA\bigl(}{}
+ 2\tilde {\rm e}^\al\lambda_\al(r\ip \cA)\cL_{\la\lambda\iota\ra^2\xi_1}\cA - 2\tilde {\rm e}^\al\lambda_\al(r\ip\cA)\cL^{0,1}_r\cA\bigr) .\label{eq:int-hol}
\end{gather}
The first term has exactly the form we expect for Poisson--Chern--Simons in the holomorphic frame. We would like to show that there exists a quadratic field redefinition which, when inserted into the kinetic term, eliminates the remaining three.

A natural first guess is
\begin{displaymath} \delta\cA = f\cA + B(r\ip\cA)\cL_{\la\lambda\iota\ra^2\xi_0}\cA + C(r\ip\cA)^2\tilde {\rm e}^\al\lambda_\al \end{displaymath}
for constants $B$, $C$ and some function $f$ linear in the field $\cA$. The variation of the kinetic term~is
\begin{displaymath} \delta S_{\rm kin.} = \frac{\im}{2\pi}\int_\PS\Omega \bigl(f\cA + B(r\ip\cA)\cL_{\la\lambda\iota\ra^2\xi_0}\cA + C(r\ip\cA)^2\tilde {\rm e}^\al\lambda_\al\bigr)\dift^\prime\cA . \end{displaymath}
Contracting by parts with some amount of $r$ in the second and third terms, we get
\begin{gather}
\delta S_{\rm kin.} = \frac{\im}{2\pi}\int_\PS\Omega \bigl(f\cA\dift^\prime\cA + (r\ip\cA)\bigl((B-D)\cL_{\la\lambda\iota\ra^2\xi_0}\cA + (C-E)(r\ip\cA)\tilde {\rm e}^\al\lambda_\al\bigr)\dift^\prime\cA \nonumber\\
\hphantom{\delta S_{\rm kin.} = \frac{\im}{2\pi}\int_\PS}{}
+ D\cA(r\ip\cL_{\la\lambda\iota\ra^2\xi_0}\cA)\dift^\prime\cA - D\cA\cL_{\la\lambda\iota\ra^2\xi_0}\cA\bigl(r\ip\dift^\prime\cA\bigr)
\nonumber\\
\hphantom{\delta S_{\rm kin.} = \frac{\im}{2\pi}\int_\PS}{}
+ E\cA(r\ip\cA)(r\ip\tilde {\rm e}^\al\lambda_\al)\dift^\prime\cA - E\cA(r\ip\cA)\tilde {\rm e}^\al\lambda_\al\bigl(r\ip\dift^\prime\cA\bigr)\bigr)\label{eq:kinetic-variation}
\end{gather}
for constants $D$, $E$ to be determined. Let us write $h = - r\ip\tilde {\rm e}^\al\lambda_\al = \la\lambda\iota\ra\bigl\la\bar\lambda\iota\bigr\ra/\bigl\la\lambda\bar\lambda\bigr\ra$. A couple of the new terms can be accounted for by shifting \smash{$f\mapsto f - D\bigl(r\ip\cL_{\la\lambda\iota\ra^2\xi_0}\cA\bigr) + Eh(r\ip\cA)$} and in the remaining two we write \smash{$r\ip\dift^\prime = \cL^{0,1}_r - \dift^\prime r\ip$} to get
\begin{displaymath} \begin{aligned}
\delta S_{\rm kin.} = \frac{\im}{2\pi}\int_\PS\Omega \biggl(&f\cA\dift^\prime\cA + (r\ip\cA)\bigl((B-D)\cL_{\la\lambda\iota\ra^2\xi_0}\cA + (C-E)(r\ip\cA)\tilde {\rm e}^\al\lambda_\al\bigr)\dift^\prime\cA \\
&- D\cA\cL^\prime_{\la\lambda\iota\ra^2\xi_0}\cA\cL^{0,1}_r\cA - E\cA(r\ip\cA)\tilde {\rm e}^\al\lambda_\al\cL^{0,1}_r\cA \\
&+ D\cA\cL_{\la\lambda\iota\ra^2\xi_0}\cA\dift^\prime(r\ip\cA)+ \frac{E}{2}\cA\tilde {\rm e}^\al\lambda_\al\dift^\prime(r\ip\cA)^2\biggr) .
\end{aligned} \end{displaymath}
The terms involving \smash{$\cL^{0,1}_r$} are exactly of the form required to cancel two of the unwanted contributions in \eqref{eq:int-hol}. Concentrating on the final two terms, we follow our nose and integrate by parts with respect to $\dift^\prime$
\begin{displaymath} \begin{aligned}
&\frac{\im}{2\pi}\int_\PS\Omega \biggl(D\cA\cL_{\la\lambda\iota\ra^2\xi_0}\cA\dift^\prime(r\ip\cA) + \frac{E}{2}\cA\tilde {\rm e}^\al\lambda_\al\dift^\prime(r\ip\cA)^2\biggr) \\
&\qquad{}
= \frac{\im}{2\pi}\int_\PS\Omega \biggl(D\cA\dift^\prime\cL_{\la\lambda\iota\ra^2\xi_0}\cA(r\ip\cA) + \frac{E}{2}\cA\dift^\prime(\tilde {\rm e}^\al\lambda_\al)(r\ip\cA)^2 \\
&\qquad\hphantom{= \frac{\im}{2\pi}\int_\PS\Omega \biggl(\,}{}
- D\dift^\prime \cA(r\ip\cA)\cL_{\la\lambda\iota\ra^2\xi_0}\cA- \frac{E}{2}\dift^\prime\cA(r\ip\cA)^2\tilde {\rm e}^\al\lambda_\al\biggr) .
\end{aligned} \end{displaymath}
The final two terms can be collected with those in the first line of equation \eqref{eq:kinetic-variation}. From the structure equation for $\tilde {\rm e}^\al$, we learn that
\begin{displaymath} \dift^\prime\tilde {\rm e}^\al = - \lambda^\al\lambda_\beta\tilde {\rm e}^0\tilde {\rm e}^\beta , \end{displaymath}
therefore, \smash{$\dift^\prime(\tilde {\rm e}^\al\lambda_\al) = 0$}. This leaves
\begin{align}
\frac{\im}{2\pi}\int_\PS\Omega\,D\cA\dift^\prime\cL_{\la\lambda\iota\ra^2\xi_0}\cA(r\ip\cA) ={}& \frac{\im}{2\pi}\int_\PS\Omega\,D\cA\bigl[\dift^\prime,\cL_{\la\lambda\iota\ra^2\xi_0}\bigr]\cA(r\ip\cA) \nonumber\\
&{+}\, \frac{\im}{2\pi}\int_\PS\Omega\,D\cA\cL_{\la\lambda\iota\ra^2\xi_0}\dift\cA(r\ip\cA) .\label{eq:crucial-kinetic-variation}
\end{align}
The first of these terms can be evaluated using \smash{$\bigl[\dift^\prime,\cL^{1,0}_{\xi_0}\bigr] = \cL^{1,0}_{\dif^\prime\xi_0}$} and the structure equations for~${\rm e}^0$,~${\rm e}^1$ to infer that
\begin{gather}
\bigl(\dift\xi_0\bigr) \ip {\rm e}^0 = \dift\bigl(\xi_0\ip {\rm e}^0\bigr) + \xi_0\ip\dift {\rm e}^0 = 0 ,\nonumber\\
\bigl(\dift\xi_0\bigr) \ip {\rm e}^1 = \dift\bigl(\xi_0\ip {\rm e}^1\bigr) + \xi_0\ip\dift {\rm e}^1 = 2\tilde {\rm e}^\al\lambda_\al,\label{eq:dift-xi0}
\end{gather}
where we have used
\begin{displaymath}
\dif {\rm e}^1 = - 2\frac{\dif x^{\al\beta}\iota_\al\lambda_\beta}{\la\lambda\iota\ra}{\rm e}^0 = - \frac{2\bigl\la\bar\lambda\iota\bigr\ra}{\la\lambda\iota\ra\bigl\la\lambda\bar\lambda\bigr\ra}{\rm e}^0{\rm e}^1 + 2{\rm e}^0\tilde {\rm e}^\al\lambda_\al .
\end{displaymath}
The equations \eqref{eq:dift-xi0} imply that \smash{$\dift\xi_0 = 2\tilde {\rm e}^\al\lambda_\al\otimes\xi_1$}, so that
\begin{displaymath}
\bigl[\dift^\prime,\cL^{1,0}_{\la\lambda\iota\ra^2\xi_0}\bigr] = 2\cL^{1,0}_{\tilde {\rm e}^\al\lambda_\al\otimes\la\lambda\iota\ra^2\xi_1} .
\end{displaymath}
In the second term of equation \eqref{eq:crucial-kinetic-variation}, we integrate by parts with respect to \smash{$\cL^{1,0}_{\la\lambda\iota\ra^2\xi_0}$}. Taking care to note that \smash{$\cL^{1,0}_{\xi_0}\Omega = -2h\Omega$} equation \eqref{eq:crucial-kinetic-variation} becomes
\begin{displaymath} \begin{aligned}
\frac{\im}{2\pi}\int_\PS\Omega \bigl(&2D\cA(r\ip\cA)\tilde {\rm e}^\al\lambda_\al\cL_{\la\lambda\iota\ra^2\xi_1}\cA + 2Dh\cA\dift^\prime\cA(r\ip\cA) \\
&- D\cL_{\la\lambda\iota\ra^2\xi_0}\cA\dift^\prime \cA(r\ip\cA) - D\cA\dift^\prime\cA\cL_{\la\lambda\iota\ra^2\xi_0}(r\ip\cA)\bigr) .
\end{aligned} \end{displaymath}
The second and final term can safely be absorbed into $f$, and the third natural combines with one of the terms in the first line of \eqref{eq:kinetic-variation}. The first term has the form required to cancel the final unwanted contribution in \eqref{eq:int-hol}.

Putting this all together, the variation of the kinetic term is
\begin{align}
\frac{\im}{2\pi}\int_\PS&\Omega \bigl(g\cA\dift^\prime\cA + (r\ip\cA)\bigl((B-3D)\cL_{\la\lambda\iota\ra^2\xi_0}\cA + \bigl(C-\tfrac{3}{2}E\bigr)(r\ip\cA)\tilde {\rm e}^\al\lambda_\al\bigr)\dift^\prime\cA\nonumber\\
&{}- D\cA\cL_{\la\lambda\iota\ra^2\xi_0}\cA\cL^{0,1}_r\cA - E\cA(r\ip\cA)\tilde {\rm e}^\al\lambda_\al\cL^{0,1}_r\cA + 2D\cA(r\ip\cA)\tilde {\rm e}^\al\lambda_\al\cL_{\la\lambda\iota\ra^2\xi_1}\cA\bigr),\!\!\!
\end{align}\label{eq:final-kinetic-variation}
where
\begin{displaymath} g = f + D\bigl[r\ip,\cL_{\la\lambda\iota\ra^2\xi_0}\bigr]\cA + (2D-E)h(r\ip\cA) . \end{displaymath}
In order for \eqref{eq:final-kinetic-variation} to cancel against \eqref{eq:int-hol}, we should take
\begin{displaymath} B = -3 ,\qquad C = -3 ,\qquad D = -1 ,\qquad E = -2 . \end{displaymath}
Furthermore, we have $g=0$ implying $f=0$ since
\begin{displaymath}
\bigl[r\ip,\cL_{\la\lambda\iota\ra^2\xi_0}\bigr]\cA = \bigl[r,\la\lambda\iota\ra^2\xi_0\bigr]\ip\cA = \bigl[\la\lambda\iota\ra^2\xi_1,\la\lambda\iota\ra^2\xi_0\bigr]\ip\cA - \bigl[\p_1,\la\lambda\iota\ra^2\xi_0\bigr]\ip\cA = 0 .
\end{displaymath}
Here we have used the fact that $\bigl[\p_1,\la\lambda\iota\ra^2\xi_0\bigr] = 0$ and $\bigl[\la\lambda\iota\ra^2\xi_1,\la\lambda\iota\ra^2\xi_0\bigr]$ is a $(1,0)$ vector field in the holomorphic frame and therefore contracts to zero against $\cA$.

Therefore, performing the field redefinition
\begin{equation} \label{eq:non-linear-redefinition} \cA \mapsto \cA - 3(r\ip\cA)\cL_{\la\lambda\iota\ra^2\xi_0}\cA - 3(r\ip\cA)^2\tilde {\rm e}^\al\lambda_\al + \cO\bigl(\cA^3\bigr), \end{equation}
we recover the Poisson--Chern--Simons Lagrangian in the holomorphic frame up to cubic order~\eqref{eq:PCS-PS-prime}. Inserting this field redefinition in the vertex will create a new term quartic in $\cA$ which can be cancelled by including the cubic field correction in the kinetic term, and so on.

The non-linear field redefinition \eqref{eq:non-linear-redefinition} will necessitate modifications to our boundary conditions on the locus $\lambda\sim\bar\lambda$, and it appears these can generate more severe singularities. Of course, by definition the theory will be equivalent to that in the constant frame and the action will be finite. Nevertheless, we find it unsatisfying that using a distribution adapted to the geometry of the minitwistor correspondence yields such complicated boundary conditions.

Following \cite{Costello:2021bah}, we believe the best way around this is to define non-commutative $5d$ Chern--Simons theory in Euclidean signature from the beginning. An advantage of this perspective is that the correspondence space $\PS_\mathbb{E}$ of Euclidean $\bbR^3$ is honestly transversally holomorphically foliated over $\mT$, so none of the issues generated by the locus $\lambda\sim\bar\lambda$ arise. Whilst it may seem very strange to define KP theory in Euclidean signature, the minitwistor correspondence allows the recovery of Lorentzian data: to compute amplitudes, we can lift minitwistor representatives for scatting states to $\PS_\mathbb{E}$; Lorentzian correlators are obtained by lifting holomorphic surface defects wrapping minitwistor lines to two-dimensional defects on correspondence space. Topological invariance along the fibres of $\PS_\mathbb{E}\to\mT$ ensures that the choice of pullback is immaterial, except where defects and states meet. We we leave a more thorough analysis of this perspective to future work.

\section{Vanishing of the four-point tree amplitude in KP} \label{app:four-point-KP}

In this appendix, we verify that the four-point tree amplitude in KP theory
\begin{equation} \label{eq:app-four-point-KP}
 A_{\rm KP}^4 = \frac{2\bigl(\la1\iota\ra^2 + \la2\iota\ra^2\bigr)\bigl(\la3\iota\ra^2 + \la4\iota\ra^2\bigr)}{\la12\ra^2 + \frac{\fq^2}{4}\la1\iota\ra^2\la2\iota\ra^2\bigl(\la1\iota\ra^2+\la2\iota\ra^2\bigr)^2} + \text{cyclic permutations of \{2,3,4\}} .
 \end{equation}
vanishes to all orders in the non-commutativity parameter $\fq^2$.

First, we express momentum conservation in terms of the spinor helicity variables $\kappa_i$. Recall that \smash{$k_{i,\al\beta} = \kappa_{i,\al}\kappa_{i,\beta} - \tfrac{\fq^2}{12}\iota_\al\iota_\beta\la\kappa\iota\ra^6$}, so momentum conservation reads
\begin{equation} \label{eq:KP-conservation}
 \sum_{i=1}^4 k_i = \sum_{i=1}^4\biggl(\kappa_{i,\alpha}\kappa_{i,\beta} - \frac{\fq^2}{12}\iota_\alpha\iota_\beta\la i\iota\ra^6\biggr) = 0 .
 \end{equation}
It will be useful to introduce some notation for the kinematic invariants on which the amplitude depends. We write $r_i = \la i\iota\ra^2$ for the Lorentz violating invariants. Contracting equation \eqref{eq:KP-conservation} with $\iota^\al\iota^\beta$ and \smash{$\iota^\al\kappa_4^\beta$} gives two constraints involving the $r_i$
\begin{equation} \label{eq:Lorentz-non-inv}
r_1 + r_2 + r_3 + r_4 = 0 ,\qquad\la1\iota\ra\la14\ra + \la2\iota\ra\la24\ra + \la3\iota\ra\la34\ra = 0 .
\end{equation}
These are the same as we'd find in the case of dKP; however, the Lorentz invariant constraints are modified. In particular,
\begin{equation} \label{eq:Lorentz-inv}
(k_1+k_2+k_3)^2 = k_4^2 = - \frac{\fq^2}{6}k_{4,+}^4 = - \frac{\fq^2}{6}(k_1+k_2+k_3)_+^4 .
\end{equation}
Defining deformed `Mandelstam' variables
\begin{displaymath} \begin{aligned}
&s_\fq = \langle12\rangle^2 + \frac{\fq^2}{4}r_1r_2(r_1+r_2)^2 ,\qquad t_\fq = \langle31\rangle^2 + \frac{\fq^2}{4}r_3r_1(r_3+r_1)^2, \\
&u_\fq = \langle23\rangle^2 + \frac{\fq^2}{4}r_2r_3(r_2+r_3)^2 ,
\end{aligned} \end{displaymath}
we can rewrite the constraint \eqref{eq:Lorentz-inv} as
\begin{equation} \label{eq:deformed-constraint} s_\fq + t_\fq + u_\fq = \fq^2r_1r_2r_3r_4 . \end{equation}
Momentum conservation implies that we could, for example, have defined $s_\fq$ through the momenta $k_3$, $k_4$
\begin{align*} s_\fq = \frac{1}{2}(k_1+k_2)^2 + \frac{\fq^2}{12}(k_1+k_2)_+^4 &{}= \frac{1}{2}(k_3+k_4)^2 + \frac{\fq^2}{12}(k_3+k_4)_+^4\\
&{} = \la34\ra^2 + \frac{\fq^2}{4}r_3r_4(r_3+r_4)^2 .
\end{align*}
Expressed in terms of these variables, the amplitude \eqref{eq:app-four-point-KP} is
\begin{align}
A_{\rm KP}^4 ={}& - \frac{2}{s_\fq t_\fq u_\fq}\bigl(t_\fq u_\fq(r_1+r_2)^2 + s_\fq u_\fq(r_3+r_1)^2 + s_\fq t_\fq(r_2+r_3)^2\bigr)\nonumber\\
={}& \frac{2}{s_\fq t_\fq u_\fq}\bigl(t_\fq^2(r_1+r_2)^2 + s_\fq^2(r_3+r_1)^2 - 2s_\fq t_\fq(r_1r_4 + r_2r_3)\bigr) \nonumber\\
&{-}\, \frac{2\fq^2}{s_\fq t_\fq u_\fq}r_1r_2r_3r_4\bigl(t_\fq(r_1+r_2)^2+s_\fq(r_3+r_1)^2\bigr) ,\label{eq:four-point-KP-Mandelstams}
\end{align}
where we have eliminated $u_\fq$ in the numerator using \eqref{eq:deformed-constraint}. We could also eliminate $r_4$, but for the moment retain it for compactness. To simplify this further, we make use of the second constraint in equation \eqref{eq:Lorentz-non-inv}; rearranging and squaring once gives
\begin{displaymath} r_2\la24\ra^2 + r_3\la34\ra^2 + 2\la2\iota\ra\la3\iota\ra\la24\ra\la34\ra = r_1\la14\ra^2 , \end{displaymath}
or in terms of our modified Mandelstam variables
\begin{displaymath}
2\la2\iota\ra\la3\iota\ra\la24\ra\la34\ra = r_1u_\fq - r_2t_\fq - r_3s_\fq - \frac{\fq^2}{4}r_4\bigl(r_1^2(r_1+r_4)^2 -r_2^2(r_2+r_4)^2 - r_3^2(r_3+r_4)^2\bigr) .
\end{displaymath}
Eliminating $u_\fq$ using \eqref{eq:deformed-constraint}, this is
\begin{displaymath} 2\la2\iota\ra\la3\iota\ra\la24\ra\la34\ra = - (r_1+r_2)t_\fq - (r_3+r_1)s_\fq - \frac{\fq^2}{2}r_2r_3r_4(r_1r_4 - r_2r_3) . \end{displaymath}
Squaring a second time,
\begin{displaymath} \begin{aligned}
&4r_2r_3s_\fq t_\fq - \fq^2r_2r_3r_4\bigl(s_\fq r_2(r_2+r_4)^2 + t_\fq r_3(r_3+r_4)^2\bigr) + \frac{\fq^4}{4}r_2^2r_3^3r_4^2(r_2+r_4)^2(r_3+r_4)^2 \\
&\qquad{}= \bigl((r_1+r_2)t_\fq + (r_3+r_1)s_\fq\bigr)^2 + \fq^2r_2r_3r_4(r_1r_4-r_2r_3)((r_1+r_2)t_\fq+(r_3+r_1)s_\fq) \\
&\qquad\quad{}+ \frac{\fq^4}{4}r_2^2r_3^2r_4^2(r_1r_4-r_2r_3)^2 .
\end{aligned} \end{displaymath}
The terms of order $\fq^4$ cancel, since $(r_2+r_4)(r_3+r_4) = r_2r_3 - r_1r_4$. Rearranging the leftovers
\begin{gather*} (r_1+r_2)^2t_\fq^2 + (r_3+r_1)^2s_\fq^2 - 2s_\fq t_\fq(r_1r_4+r_2r_3)\\
\qquad{} - \fq^2r_1r_2r_3r_4\bigl(s_\fq(r_1+r_3)^2 + t_\fq(r_1+r_2)^2\bigr) = 0 ,
\end{gather*}
so the two terms in equation \eqref{eq:four-point-KP-Mandelstams} cancel. As claimed in Section~\ref{subsec:tree-amps}, the tree level four-point amplitude in KP theory vanishes.
\end{appendix}

\subsection*{Acknowledgements}

It is a pleasure to thank Kevin Costello, Lionel Mason, Anthony Morales and Ian Strachan for helpful conversations. We would also like to thank the anonymous referees for useful comments which improved the paper. Some results appearing in this work were obtained during the workshops `Twistors in Geometry and Physics' at the Isaac Newton Institute for Mathematical Sciences in 2024 and `From Good Cuts to Celestial Holography' at St Antony's College in 2025. We thank the organisers for financial assistance and their warm hospitality.

This work was supported in part by the Simons Collaboration on Celestial Holography. Research of RB at Perimeter Institute is supported in part by the Government of Canada through the Department of Innovation, Science and Economic Development and by the Province of Ontario through the Ministry of Colleges and Universities. SH is partly supported by St. John’s College Cambridge, by the Gordon and Betty Moore Foundation, and by the John Templeton Foundation via the Black Hole Initiative. The work of SR is supported by NSF grant DMS-2503353. DS is supported by the STFC (UK) HEP Theory Consolidated grant ST/X000664/1.

\pdfbookmark[1]{References}{ref}
\LastPageEnding

\end{document}